\documentclass[a4paper,11pt]{article}
\pdfoutput=1 

\usepackage{jheppub} 

\usepackage[T1]{fontenc}
\usepackage{dcolumn}
\newcolumntype{d}{D{.}{.}{2.3}}
\usepackage{xspace}
\usepackage[export]{adjustbox}
\usepackage{subcaption}
\usepackage{graphicx}
\usepackage{float}
\usepackage{amsmath}
\usepackage{array}
\usepackage{microtype}
\usepackage{changes}

\graphicspath{{figures/}}

\newcommand{\HEJ}{{\tt HEJ}\xspace}

\hypersetup{pdftitle=Higgs Boson plus Quark Pair Production at High Energies as di-Higgs Background}
\title{\boldmath Higgs Boson plus Quark Pair Production at High Energies as di-Higgs Background \unboldmath}
\author[a,b]{Andreas Maier,}
\author[c]{Giacomo Ruisi}

\affiliation[a]{%
  The Henryk Niewodniczański Institute of Nuclear Physics,
  ul. Radzikowskiego 152, 31-342 Krakow, Poland
}
\affiliation[b]{Institut de F{\'i}sica d'Altes Energies (IFAE), The Barcelona Institute of Science and Technology, Campus UAB, 08193 Bellaterra (Barcelona), Spain}
\affiliation[c]{Deutsches Elektronen-Synchrotron DESY, Platanenallee 6, 15738 Zeuthen, Germany}

\emailAdd{andreas.maier@ifj.edu.pl}
\emailAdd{giacomo.ruisi@desy.de}

\preprint{%
\begin{minipage}[t]{\widthof{COMETA-2026-32}}
  DESY-26-090 \\
  COMETA-2026-32
\end{minipage}
}

\abstract{
We investigate the impact of high-energy QCD resummation on Higgs-boson production processes relevant for Higgs-pair studies at the Large Hadron Collider.
We present the first dedicated study of Higgs boson production in association with a bottom-quark pair and two additional hard jets in the gluon-fusion channel at high energies as a background to vector-boson-fusion Higgs-pair production.
The resummation can significantly modify the QCD contribution in VBF-like regions of phase space, indicating that fixed-order calculations overestimate the contamination from gluon-fusion backgrounds.
We then extend existing High Energy Jets (\HEJ) predictions for Higgs boson production in association with jets by including quark-pair resummation effects for the first time.
}

\begin{document}
\maketitle

\section{Introduction}

After the discovery of the Higgs boson at the CERN Large Hadron Collider (LHC) by the ATLAS~\cite{ATLAS:2012yve} and CMS~\cite{CMS:2012qbp} collaborations, Higgs physics has entered a precision era.
A central objective of this programme is the determination of the Higgs self-couplings.
These parameters are essential for reconstructing the shape of the Higgs potential, which is intimately connected to the electroweak (EW) vacuum stability of the Standard Model (SM)~\cite{Isidori:2001bm, Elias-Miro:2011sqh, Degrassi:2012ry, Buttazzo:2013uya}.
This connection provides a link between particle physics and cosmology and leaves room for a variety of Beyond the Standard Model (BSM) scenarios~\cite{Anderson:1991zb, Bezrukov:2007ep, Bezrukov:2008ej, Bezrukov:2009db, Bezrukov:2010jz, Markkanen:2018pdo, Hiller:2024zjp, Bosse:2026bdk}.

Current experimental measurements across multiple Higgs production modes show no significant deviations from SM predictions, consistently supporting the interpretation of the Higgs boson as a spin-zero, CP-even scalar.
However, a precise determination of the Higgs self-couplings remains elusive.
Direct access to these couplings is primarily achieved through processes involving Higgs boson pair production.
The dominant production mechanism proceeds via the fusion of a pair of gluons (ggF), followed by the subleading electroweak vector-boson fusion (VBF) channel.
The latter is particularly valuable, as it is sensitive to the quartic interactions between Higgs and electroweak gauge bosons.
Given the small total cross section for di-Higgs production, no single golden channel exists, and the overall sensitivity relies on the combination of multiple search channels~\cite{CMS:2018ipl, ATLAS:2024ish, CMS:2026nuu}.
This places equal importance on the accurate modelling of both the signal and the dominant backgrounds in double-Higgs searches.

Indeed, on the theory side, substantial effort has been devoted to improving the precision description of gluon-fusion Higgs boson pair production (ggF HH).
The most advanced calculations include next-to-leading order (NLO) QCD corrections with exact top-mass dependence, higher-order QCD corrections in the heavy-top approximation, electroweak effects, and matching to parton showers~\cite{Borowka:2016ehy, Borowka:2016ypz, Heinrich:2017kxx, Baglio:2018lrj,  Grazzini:2018bsd, Bizon:2018syu, Borowka:2018pxx, Davies:2019dfy, Chen:2019lzz, Bi:2023bnq, Li:2024iio, Heinrich:2024dnz, Jaskiewicz:2024xkd, Davies:2025wke, Alioli:2025xcu, Chen:2026zmi}.
Together, these advances have considerably reduced the theoretical uncertainties associated with ggF HH production.

Comparable effort has been dedicated to single-Higgs production via gluon fusion, for which inclusive and exclusive predictions are available at very high perturbative accuracy~\cite{DelDuca:2001eu, DelDuca:2001fn, Jones:2018hbb, Czakon:2021yub, Chen:2021azt}.
The heavy-top approximation has enabled the computation of one additional perturbative order for Higgs production with up to two associated jets~\cite{Campbell:2006xx, Campbell:2010cz, Boughezal:2013uia, Chen:2014gva, Boughezal:2015dra, Anastasiou:2015vya, Dulat:2017prg, Mistlberger:2018etf, Cieri:2018oms, Chen:2021isd}.

Among the various backgrounds affecting di-Higgs searches, single-Higgs production in association with bottom quarks represents a particularly challenging contribution, especially in channels where at least one Higgs boson decays to bottom quarks.
Reliable modelling of this process therefore plays a crucial role in improving the sensitivity of Higgs pair production analyses.
In recent years, state-of-the-art predictions for Higgs production in association with bottom quarks have been achieved through the inclusion of higher-order QCD corrections and parton-shower matching~\cite{Zhang:2017mdz, Deutschmann:2018avk, Duhr:2019kwi, Pagani:2020rsg, Grojean:2020ech, Manzoni:2023qaf, Biello:2024vdh, Biello:2024pgo, Gavardi:2025zpf, Biello:2025ksu}.

In the context of vector-boson-fusion di-Higgs (VBF HH) searches, the production of a Higgs boson in association with a bottom-quark pair and two additional hard jets represents an important and difficult-to-suppress background.
While VBF HH production is suppressed by roughly an order of magnitude with respect to the gluon-fusion Higgs boson pair channel, it benefits from a clean experimental topology featuring two energetic forward jets in association with the Higgs boson pair.
Recent studies have advanced the precision description of VBF Higgs boson pair production through the inclusion of next-to-leading-order QCD corrections matched to parton showers~\cite{Kilian:2018bhs, Dreyer:2018rfu, Dreyer:2018qbw, Dreyer:2020urf, Jager:2025isz, Braun:2025hvr, Braun:2025fpt}.

Both gluon-fusion Higgs-pair production and gluon-fusion single-Higgs production constitute particularly relevant backgrounds in regions of phase space characterised by VBF-like selections~\cite{Greiner:2015jha, Chen:2025whf}, where large rapidity separations and high dijet invariant masses enhance logarithmic contributions arising from large ratios of the partonic centre-of-mass energy to the characteristic transverse-momentum scale associated with multiple hard emissions~\cite{DelDuca:2003ba}.
As a result, these corrections tend to reduce the ggF component, suggesting that the level of ggF contamination may be overestimated by fixed-order calculations~\cite{Andersen:2018tnm, Andersen:2018kjg}.
Accurate predictions of these backgrounds in this regime are therefore essential for reliable interpretations of VBF Higgs-pair searches.
Motivated by this, we present the first dedicated analysis of gluon-fusion single-Higgs plus jets production as a background to VBF Higgs-pair production within the High Energy Jets (\HEJ) framework, and quantify the impact of high-energy logarithmic resummation on the corresponding predictions.

This work is organised as follows.
Section~\ref{sec:HEJFORHIGGS} outlines the main ingredients required for the inclusion of high-energy QCD resummation within \HEJ in single-Higgs plus jets production, with a specific focus on quark-pair resummation which contributes at next-to-leading logarithmic (NLL) accuracy.
In section~\ref{sec:HHBACKGROUND}, we present the first estimate of Higgs boson production in association with a bottom-quark pair and two additional hard jets in the ggF channel as background to VBF HH production including the effects of high-energy QCD resummation, and compare the resulting predictions with leading-order calculations.
In section~\ref{sec:HJETS}, we update existing predictions for Higgs boson plus jets production in the ggF channel~\cite{Andersen:2022zte} by including quark-pair resummation effects for the first time, and compare the resulting predictions with both experimental measurements and fixed-order NLO calculations.
The paper concludes in section~\ref{sec:conclusions}, where we summarise our findings and discuss their implications.

\section{High Energy Resummation for Higgs Boson plus Jets production}\label{sec:HEJFORHIGGS}

In the following, we present the key components required to calculate the high-energy corrections associated with Higgs boson plus jets production in proton-proton collisions.
We provide a summary of the \HEJ framework for the all-order resummation of high-energy logarithms of the form $(\alpha_{\text{s}} \log s_{ij}/p_{\perp}^2)^N$ and introduce the new ingredients for the production of a Higgs boson together with a quark-antiquark pair.

\subsection{High-energy Scaling of the Amplitudes}\label{subsec:HESCALING}

We are generally interested in the behaviour of scattering amplitudes in the \emph{Multi-Regge Kinematics} (MRK) regime.
This kinematic region is characterised by a large centre-of-mass energy and large invariant masses between all pairs of outgoing particles, each carrying finite transverse momentum.
This setup corresponds to a strong ordering in rapidities.
Specifically, for a $2 \to n$ process, the MRK limit requires
\begin{equation}\label{eq:MRKLIMIT}
y(p_n) \gg \cdots \gg y(p_1), \quad |\mathbf{p}_{i\perp}| \sim \text{finite} \quad \forall i \in \{1, \ldots, n\},
\end{equation}
where the momentum of the $i$-th outgoing particle is $p_i$, with rapidity $y_i \equiv y(p_i)$ and transverse momentum magnitude $|\mathbf{p}_{i\perp}|$.
In this region, Regge theory~\cite{Fadin:2006bj} predicts that the scattering amplitude scales as
\begin{equation}\label{eq:REGGESCALING}
\mathcal{M} \sim s_{12}^{\alpha_1(t_1)} \cdots s_{n-1\,n}^{\alpha_n(t_{n-1})},
\end{equation}
where $s_{i\,i+1}$ denotes the invariant mass between particles $i$ and $i+1$, and $\alpha_i(t_i)$ represents the maximum spin of particles that can be exchanged in the $t$-channel between those particles.
From this scaling behavior, it follows that the leading contributions to QCD amplitudes arise from configurations that maximize the number of gluon exchanges in the $t$-channel.

These configurations define the regions of phase space where leading high-energy logarithms emerge and are thus referred to as \emph{leading-logarithmic} (LL) or \emph{Fadin-Kuraev-Lipatov} (FKL) configurations.
This scaling behaviour has been also observed for Higgs boson plus jets amplitudes~\cite{Andersen:2017kfc, Andersen:2018kjg, Andersen:2022zte}, and holds both in full QCD, with Higgs boson to gluons couplings through quark loops, and in the effective theory obtained by integrating out the heavy top quark (sometimes called Higgs Effective Field Theory, HEFT).
For a single jet production together with a Higgs boson, we can have only LL configurations provided by the two following channels: $gq \to Hq$ and $gg \to Hg$.
In both cases, only a single gluon exchange is allowed in the \textit{t}-channel, and thus we can state that the scaling of the amplitudes are $\mathcal{M} \sim s_{Hq}$ and $\mathcal{M} \sim s_{Hg}$, respectively.

For higher jet multiplicities, we can use the scaling argument to distinguish a richer set of configurations.
If we consider $gq \to Hgq$, where the outgoing particles are organized in ascending rapidity order, two gluon exchanges are allowed in the \textit{t}-channel which results in $\mathcal{M} \sim s_{Hg} s_{gq}$.
The identical scaling behaviour is obtained when we ask for a central Higgs emission: $gq \to gHq$.
On the other hand, when rearranging the outgoing parton flavours $gq \to Hqg$, we end up with only a single gluon exchange and a single quark exchange in the $t$-channel.
Therefore, the resulting amplitude scales as $\mathcal{M} \sim s_{Hq} s_{qg}^{1/2}$ and contributes at~\emph{next-to-leading-logarithmic} (NLL) accuracy.

It is crucial to emphasise that these scaling arguments should not be interpreted in terms of individual underlying Feynman diagrams -- whose contributions may vanish or not depending on the gauge choice -- but rather at the level of the full amplitude in the MRK limit.
In this regime, the full scattering amplitude calculated at Born level factorises into subcomponents of reduced analytic complexity.
These elementary building blocks depend only on external momenta and are contracted in colour and momentum space as if they exchanged either gluon or quark quantum numbers.
Configurations that maximise $t$-channel gluon exchanges are therefore leading in the MRK limit.

The factorisation into simpler subcomponents extends beyond leading logarithmic accuracy.
In particular, factorised amplitudes can still be obtained when the strong rapidity hierachy of equation~\eqref{eq:MRKLIMIT} is relaxed for one or more pairs of coloured particles.
The case in which the hierarchy is relaxed for a single pair is known as the~\textit{Quasi-Multi-Regge kinematic limit} (QMRK).

In section~\ref{subsec:AMPCLASS}, we first review the classification of leading configurations in the MRK limit for Higgs production in association with two jets.
We then discuss the classification of configurations that are leading in the QMRK and therefore subleading in the MRK limit.

\subsection{Amplitude Classification in HEJ}\label{subsec:AMPCLASS}

In pure QCD all the LL configurations have the form $f_a f_b \to f_a \cdots f_b$, where $f_a$, $f_b$ indicate the incoming parton flavours and the ellipsis denotes an arbitrary number of gluon emissions.
Again the particles are written in order of increasing rapidity, namely $f_a$ is the most backward particle and $f_b$ is the most forward one.
Following the scaling argument introduced in section~\ref{subsec:HESCALING}, we can find a first source of subleading logarithmic corrections (NLL) when the most backward or forward outgoing particle is a gluon, but the corresponding incoming parton is a quark or antiquark, $q f_b \to g q \cdots f_b$ and $f_a q \to f_a \cdots f_b q$, respectively.

We generally refer to these subleading configurations as \emph{unordered gluon emissions}, and they allow one $t$-channel gluon exchange less than their FKL counterpart in which the unordered gluon is flipped with the neighbouring (anti-)quark.
A second class of subleading configurations is obtained when we ask for a \emph{quark-antiquark pair emission}.
They can be either produced centrally, $f_a f_b \to f_a \cdots q \bar{q} \cdots f_b$, or at the edges of the rapidity chain when the corresponding incoming parton is a gluon, for instance in the most forward region these configurations have the form $f_a g \to f_a \cdots q \bar{q}$.

These are non-FKL configurations, as in both scenarios the $q \bar{q}$ pair can only exchange a $t$-channel quark, and thus lowering the scaling of the amplitude in the MRK limit.
In principle, every time we ask for an extra $q\bar{q}$ pair with $y_q \sim y_{\bar{q}}$ in the rapidity chain, we force a $t$-channel quark exchange.

In contrast with the BFKL formalism~\cite{Fadin:1975cb, Kuraev:1976ge, Kuraev:1977fs} in which the impact factors would depend on transverse momenta only~\cite{DelDuca:1995zy, DelDuca:1999iql}, the factorised framework developed for \HEJ~\cite{Andersen:2009nu, Andersen:2009he, Andersen:2011hs} preserve full momentum dependence, and thus ensuring the analytic properties of the tree-level amplitude of crossing symmetry, gauge invariance and Lorentz invariance.
In this regard, it is worth to highlight that the \emph{extremal quark-antiquark current} $g \to q \bar{q}$ can be obtained via crossing symmetry from the \emph{unordered gluon current} $q \to g q$~\cite{Gunion:1985vca}.
Because of the much milder approximations in \HEJ with respect to BFKL, the corrections when matching to fixed-order amplitudes are also found to be smaller~\cite{Andersen:2009nu, Andersen:2016vkp, Andersen:2020yax}.

As already mentioned in a few examples in section~\ref{subsec:HESCALING}, most of the observations above hold also for Higgs boson plus jets processes, in which the production of the Higgs proceeds via an effective coupling to two or more gluons.
The exact dependence on the top-quark mass must be included in this effective coupling, as the outgoing invariant masses are large in the high-energy region~\cite{Andersen:2018kjg}.

Since a final-state Higgs boson at an intermediate rapidity $y_j$, such that $y_{j-1} \ll y_j \ll y_{j+1}$, can always exchange $t$-channel gluons with the outgoing partons $j-1$ and $j+1$, then all configurations of the type $f_a f_b \to f_a \cdots H \cdots f_b$ are FKL, so contributing at LL accuracy.
This can be also understood by observing that for fixed transverse momenta and $s_{cd} \gg s_{cH}, s_{Hd} \gg |s_{bd}|$ the scattering amplitude for $g_a q_b \to g_c H q_d$ calculated at Born level factorises into three components depending on the momenta of $(p_a, p_c)$, $(p_a - p_c , p_H)$ and $(p_b, p_d)$ only, contracted in colour and momentum space as if they exchanged two $t$-channel gluons, as shown in the following
\begin{align}
  \includegraphics[valign=c,width=0.3\textwidth]{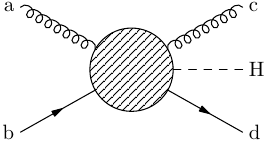}\ \quad \overset{\longrightarrow}{\substack{s_{cd} \gg s_{cH}, s_{Hd} \gg |s_{bd}|\\\mathrm{fixed}\ p_t}}\ \quad \includegraphics[valign=c,width=0.3\textwidth]{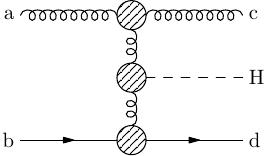}
\label{eq:amplitudesplit_FKL_cenh}
\end{align}
The analogous conclusion can be drawn when the Higgs is the most backward (forward) outgoing particle and the corresponding incoming backward (forward) parton is a gluon.
In this case, the Higgs can always exchange a $t$-channel gluon with the neighbouring outgoing parton.
Therefore, all configurations of the form $g f_b \to H \cdots f_b$ and $f_a g \to f_a \cdots H$ are FKL, such as
\begin{align}
  \includegraphics[valign=c,width=0.3\textwidth]{figures/asy/higgs2jetsamp.pdf}\ \quad \overset{\longrightarrow}{\substack{s_{Hd} \gg s_{Hc}, s_{cd} \gg |s_{bd}|\\\mathrm{fixed}\ p_t}}\ \quad \includegraphics[valign=c,width=0.3\textwidth]{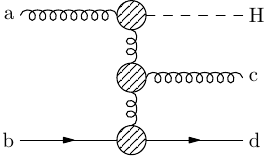}
\label{eq:amplitudesplit_FKL_exh}
\end{align}
Here, it is interesting to note the straightforward generalisation with the pure QCD case: the Higgs basically behaves as a 'massive gluon' when we consider pure QCD amplitudes in the MRK limit.
The reason behind that is due to the similarity between the triple gluon and the effective gluon-gluon-Higgs vertices.
Indeed, we can combine all the QCD NLL configurations described in this section with a Higgs emission to obtain NLL accurate amplitudes for Higgs boson plus jets, for example in the case of an unordered gluon emission
\begin{align}
  \includegraphics[valign=c,width=0.3\textwidth]{figures/asy/higgs2jetsamp.pdf}\ \quad \overset{\longrightarrow}{\substack{s_{Hd}, s_{Hc} \gg s_{dc}, |s_{bc}|, |s_{bd}|\\\mathrm{fixed}\ p_t}}\ \quad \includegraphics[valign=c,width=0.3\textwidth]{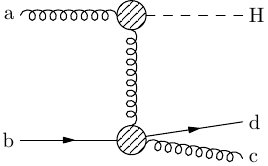}
\label{eq:amplitudesplit_UNO}
\end{align}
It is important to stress that we always require fixed transverse momenta for the outgoing particles to make sure that the large invariant masses $s_{ij}$ are reached not by increasing transverse momenta, but rather by a hierarchy in positive and negative light-cone momenta.
Special care must be taken when the Higgs is not emitted centrally, namely when it is radiated either at the beginning or at the end of the rapidity chain, and the corresponding incoming parton is a quark or an antiquark.
In these cases, we have a second class of subleading logarithmic corrections, which we can call \emph{unordered Higgs emissions}, such as $q f_b \to H q \cdots f_b$ and $f_a q \to f_a \cdots q H$.
They look like
\begin{align}
  \includegraphics[valign=c,width=0.3\textwidth]{figures/asy/higgs2jetsamp.pdf}\ \quad \overset{\longrightarrow}{\substack{s_{cd}, s_{cH} \gg s_{dH}, |s_{bH}|, |s_{bd}|\\\mathrm{fixed}\ p_t}}\ \quad \includegraphics[valign=c,width=0.3\textwidth]{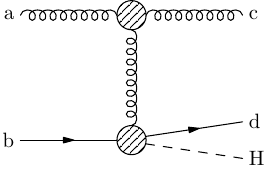}
\label{eq:amplitudesplit_UNOHIGGS}
\end{align}
These configurations permit a $t$-channel gluon exchange less than their FKL counterpart in which the unordered Higgs is swapped with the neighbouring (anti-)quark.
When merging this last class with the QCD NLL configurations, we end up with \emph{next-to-next-to-leading-logarithmic} (NNLL) contributions that are already accounted for in \HEJ.
As an example, we can have the following 'unordered Higgs' and 'unordered gluon' process $Q q \to H Q \cdots q g$ that allows for two $t$-channel exchanges less than the equivalent FKL configuration.

All the Higgs plus jets processes presented up to this point have been considered for phenomenological analyses in~\cite{Andersen:2022zte}.
In this work, we include for the first time the NLL corrections arising from quark-antiquark emissions along with a Higgs boson production and are relevant for $H + \geq 2j$ final states.
For every event in which the Higgs and $q\bar{q}$ are emitted at the opposite edges of the rapidity chain, only one possible configuration is allowed.
In such cases, we would always have two incoming gluons, one splitting into the Higgs and the other one splitting into the $q\bar{q}$ pair, as shown in the following for $H+2j$
\begin{align}
  \includegraphics[valign=c,width=0.3\textwidth]{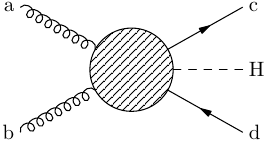}\ \quad \overset{\longrightarrow}{\substack{s_{Hc}, s_{Hd} \gg s_{cd}, |s_{bc}|, |s_{bd}|\\\mathrm{fixed}\ p_t}}\ \quad \includegraphics[valign=c,width=0.3\textwidth]{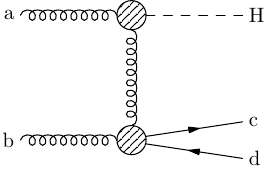}
\label{eq:amplitudesplit_EXQQX}
\end{align}
For higher jet multiplicities, so for $\geq 3j$, the Higgs boson can be emitted in between the $q\bar{q}$ and one additional jet as well.
Therefore, we would need only one incoming gluon splitting into the quark-antiquark pair, and no restrictions for the other incoming particle.
A central $q\bar{q}$ emission does not require any specific incoming parton flavour, unless the Higgs is emitted at the edge.
This kind of configurations only enter for $\geq 3j$, such as
\begin{align}
  \includegraphics[valign=c,width=0.3\textwidth]{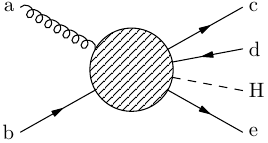}\ \quad \overset{\longrightarrow}{\substack{s_{He} \gg s_{Hi}, s_{ie} \gg s_{cd}, |s_{be}|\\i \in \{c,d\},\ \mathrm{fixed}\ p_t}}\ \quad \includegraphics[valign=c,width=0.3\textwidth]{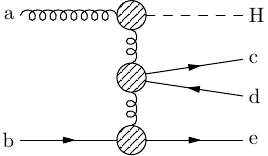}
\label{eq:amplitudesplit_CENQQX}
\end{align}
Thanks to the inclusion of $q\bar{q}$ processes presented in this study, it is also straightforward to account for another source of NNLL contributions at Born level, namely all the combinations of 'unordered Higgs' plus 'quark-antiquark' emissions.

\subsection{HEJ Amplitudes for Higgs Boson plus Quark Pair Production}

The construction of the leading-logarithmic calculation of $p p \to H +\ge 2j$ in the \HEJ framework was described in detail in~\cite{Andersen:2017kfc,Andersen:2018kjg}.
In addition to the LL resummation, gauge-invariant subsets of next-to-leading logarithmic (NLL) corrections originating from non-FKL configurations have also been included in \HEJ.
One source of NLL corrections are the \emph{unordered Higgs emissions} already described in section~\ref{subsec:AMPCLASS}, which only permit $n-2$ $t$-channel gluon exchanges instead of the $n-1$ exchanges found in LL configurations.
In these cases, it is sufficient to adapt the matrix element formula for the corresponding LL configurations to a flipped rapidity order of outgoing (anti-)quark and Higgs boson.
A second source of NLL corrections is given by the \emph{unordered gluon emissions}, also introduced in section~\ref{subsec:AMPCLASS}.
For a derivation and explicit expressions, see~\cite{Andersen:2017kfc}.

The goal of the current work is to investigate for the first time the impact of a further type of NLL configurations, namely \emph{quark-antiquark pair emissions}.
Here we will just provide a coincise summary of this new set of sub-leading corrections.
The full derivation of the Born extremal and central quark-antiquark currents can be found in~\cite{Andersen:2020yax} for W boson plus jets production.
Therefore, no genuine derivation of new \HEJ currents for Higgs boson production with a quark pair is needed.

In the MRK limit, scattering amplitudes factorise into a neat product of three functions, two of which are process-independent.
By exploiting this structure, we can reuse the same building blocks to derive previously unavailable matrix elements for processes involving a single Higgs boson accompanied by a quark-antiquark pair in the final state.
Crucially, this work presents the first complete implementation of $H + q\bar{q}$ production within the \HEJ framework, bringing together all required ingredients and providing the first dedicated phenomenological studies of these processes.

\begin{figure}[btp]
    \centering
    \begin{subfigure}[b]{0.42\textwidth}
        \includegraphics[width=\textwidth]{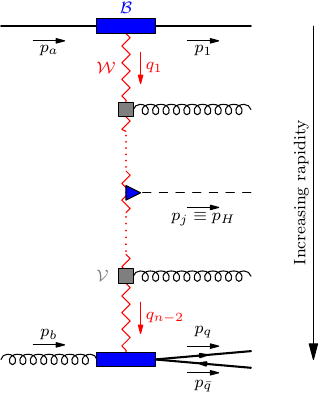}
        \caption{$ f_a g \to f_a \cdots H \cdots q\bar{q}$ process}
    \end{subfigure}
    \hfill
    \begin{subfigure}[b]{0.42\textwidth}
        \includegraphics[width=\textwidth]{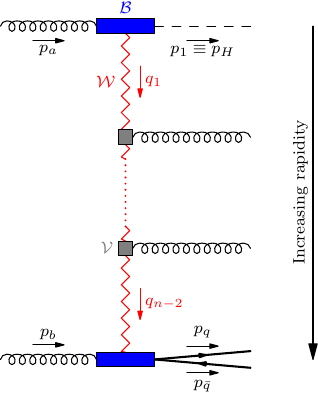}
        \caption{$ gg \to H \cdots q\bar{q}$ process}
    \end{subfigure}

\caption{
On the left panel, the diagrammatic structure of the \HEJ matrix element for central Higgs plus extremal quark-antiquark pair production.
On the right panel, the diagrammatic structure of the \HEJ matrix element for extremal Higgs together with an extremal quark-antiquark emission.
The building blocks of the \HEJ matrix elements are given by the \textit{Born functions} $\mathcal{B}$.
The grey boxes in both figures stand for additional gluon radiation.
The latter is described by the \textit{Lipatov vertex} $\mathcal{V}$.
The virtual corrections, illustrated by the red zigzag lines, are given by the Reggeised $t$-channel colour-octet exchanges.
They are obtained via the \textit{Lipatov Ansatz} $\mathcal{W}$.
}
\label{fig:ME_h_jets_extremal}
\end{figure}

We start with the case in which an incoming gluon splits into a $q\bar{q}$ pair, i.e. the \emph{extremal quark-antiquark configuration}.
For the production of a central Higgs boson together with an extremal quark pair, we obtain the form
\begin{equation}
  \label{eq:ME_fact}
  \begin{split}
    \overline{\left|\mathcal{M}_{\HEJ}^{f_a g \to f_a \cdots H
          \cdots  q \bar{q}}\right|}^2 ={}&\mathcal{B}_{f_a, H, q\bar{q}}(p_a, p_b, p_1, p_q, p_{\bar{q}}, q_j, q_{j+1})\\
    &\cdot \prod_{\substack{i=1\\i \neq j}}^{n-3} \mathcal{V}(p_a,p_b,p_1,p_n,q_i,q_{i+1})\\
    &\cdot \prod_{i=1}^{n-2} \mathcal{W}(q_i, y_i, y_{i+1}),
  \end{split}
\end{equation}
for the modulus square of the matrix element, summed and averaged over helicities and colours.
In this expression, $p_a$ $(p_b)$ is the incoming momentum in the backward (forward) direction and $p_1,\dots,p_{n-2}, p_{n-1} \equiv p_q, p_n \equiv p_{\bar{q}}$ are the outgoing momenta ordered in increasing rapidity, where $p_j \equiv p_H$ with $1 < j < n-1$ is the momentum of the Higgs boson. Configurations where the antiquark rapidity is smaller than the quark one are completely analogous.
The $t$-channel momenta are given by
\begin{equation}
  \label{eq:q_i}
  q_1 =  p_a - p_1,\qquad q_i = q_{i-1} - p_i \; \text{ for } 1 < i < n.
\end{equation}
The diagrammatic representation of this amplitude in \HEJ is illustrated in the left panel of figure~\ref{fig:ME_h_jets_extremal}.
At Born level, the right-hand side of equation~(\ref{eq:ME_fact}) reduces to the function $\mathcal{B}_{f_a, H, q\bar{q}}$, described below.
The object $\mathcal{V}$ comprises the real corrections due to the production of $n-4$ gluons in addition to $f_a,~q\bar{q}~\text{pair}$ and the Higgs boson, where $n$ is the number of outgoing particles.
The function $\mathcal{V}$ is given by the contraction of two Lipatov vertices~\cite{Andersen:2017kfc} that read\footnote{In the numerical analyses in sections~\ref{sec:HHBACKGROUND} and \ref{sec:HJETS}, we adopt a slightly different momentum assignment for the Lipatov function, with $p_n \to p_q + p_{\bar{q}}$ (and, equivalently, $p_1 \to p_q + p_{\bar{q}}$ for a backward quark-antiquark emission).
The numerical impact of this choice has been found to be negligible.}
\begin{align}
  \label{eq:V}
  \mathcal{V}(p_a,p_b,p_1,p_n,q_i,q_{i+1})={}& -\frac{C_A}{t_i t_{i+1}} V_\mu(p_a,p_b,p_1,p_n, q_i, q_{i+1}) V^\mu(p_a,p_b,p_1,p_n,q_i, q_{i+1}),\\
  \label{eq:V_Lipatov}
  V^\mu(p_a,p_b,p_1,p_n, q_i, q_{i+1})={}& -(q_i+q_{i+1})^\mu \nonumber\\
  &+ \frac{p_a^\mu}{2} \left( \frac{q_i^2}{p_{i+1}\cdot p_a} +
  \frac{p_{i+1}\cdot p_b}{p_a\cdot p_b} + \frac{p_{i+1}\cdot p_n}{p_a\cdot p_n}\right) +
p_a \leftrightarrow p_1 \nonumber\\
  &- \frac{p_b^\mu}{2} \left( \frac{q_{i+1}^2}{p_{i+1} \cdot p_b} + \frac{p_{i+1}\cdot
      p_a}{p_b\cdot p_a} + \frac{p_{i+1}\cdot p_1}{p_b\cdot p_1} \right) - p_b
  \leftrightarrow p_n,
\end{align}
where $t_i = q_i^2$ are the squares of the $t$-channel momenta.
The Lipatov Ansatz $\mathcal{W}$ accounts for the all-order finite contribution coming from the sum of the virtual corrections and unresolved real corrections.
It is process-independent and described in detail in~\cite{Andersen:2017kfc}.

The process-dependent Born-level factor is given by
\begin{equation}
  \label{eq:Born_h_jets_central}
 \mathcal{B}_{f_a, H, q\bar{q}} =  \frac {(4\pi\alpha_s)^{n-1}} {4(N_C^2-1)} \frac{N_C}{(N_C^2 - 1)}
    \frac {K_{f_a}(p_1^-, p_a^-)} {t_1}\
 \frac{K_{q}}{t_{n-2}} \frac{\left\|S_{f_a
          g\to f_a H q\bar{q}}\right\|^2}{t_j t_{j+1}}.
\end{equation}
Here, $\alpha_s$ is the strong coupling and $N_c = 3$ is the number of colours.
The difference between incoming gluons and (anti-)quarks is completely absorbed into the \emph{colour acceleration multipliers} $K_f$ with
\begin{align}
  \label{eq:K_g}
K_g(x, y) ={}& \frac{1}{2}\left(\frac{x}{y} + \frac{y}{x}\right)\left(C_A -
  \frac{1}{C_A}\right)+\frac{1}{C_A} &\text{for gluons,}\\
  \label{eq:K_q}
  K_q(x, y) ={}& C_F &\text{for quarks and antiquarks.}
\end{align}
$C_F = \frac{N_C^2-1}{2N_C}$ and $C_A = N_C$ are the usual Casimir invariants.
$S_{f_a g \to f_a H q \bar{q}}$ is a contraction of currents with the Higgs boson production vertex.
The double vertical bars indicate the sum over helicities of the corresponding amplitudes
\begin{equation}
  \label{eq:S_fHqqx}
  \left\|S_{f_a g\to f_a H q \bar{q}}\right\|^2 = \sum_{\substack{\lambda_a =
  +,-\\\lambda_b = +,-}} \left|j^{\lambda_a}_\mu(p_1, p_a) V_H^{\mu\nu}(q_j,q_{j+1})
j_{\nu,q\bar{q}}^{\lambda_b}(p_q, p_{\bar{q}}, p_b)\right|^2.
\end{equation}
$V_H$ is the well-known one-loop effective coupling between the Higgs boson and two gluons in the normalisation of~\cite{Andersen:2018kjg}, including the full quark-mass dependence.
The inclusion of this piece in equation~\eqref{eq:ME_fact} then gives the correct finite quark-mass contributions at LL for \emph{any} number of final state partons/jets.
The current $j^{\lambda}_{\mu, q\bar{q}}$ accounts for the extremal quark-antiquark emission and the current $j$ is given by
\begin{equation}
  \label{eq:current}
  j^{\lambda}_\mu(p, q) = \bar{u}^\lambda(p)\gamma_\mu u^\lambda(q).
\end{equation}
It is worth to highlight that $j^{\lambda}_{\mu, q\bar{q}}$ can be obtained via crossing symmetry from the unordered gluon emission current $j^{\lambda}_{\mu, \text{uno}}$.
The full expression for $j^{\lambda}_{\mu, \text{uno}}$ can be found in~\cite{Andersen:2018kjg}.
Starting from the unordered gluon case, in which an incoming quark $q(p_a)$ splits into the most backward gluon $g(p_g)$, the second most backward quark $q(p_1)$ and a $t$-channel Reggeised gluon $g^\ast$, namely
\begin{equation}
  q(p_a) \to g(p_g) q(p_1) g^\ast(p_1 + p_g - p_a),
\end{equation}
and by replacing
\begin{equation}
\label{eq:crossing}
  p_a \to - p_{\bar{q}}\,, \qquad p_g \to - p_a\,, \qquad  p_1 \to p_q\,,
\end{equation}
one obtains the extremal quark-antiquark current.
After crossing we end up with negative-energy spinors such as $\lvert - p_a \rangle$.
Using analytic continuation these can be rewritten as positive-energy ones by pulling out a phase factor $i$, e.g. $\lvert - p_a \rangle \to i \, \lvert  p_a \rangle$~\cite{Gunion:1985vca}.
Importantly, equation~\eqref{eq:Born_h_jets_central} has an extra $N_C / (N_C^2 - 1)$ factor with respect to the unordered-gluon Born function.
This factor is required to have the correct colour averaging after crossing symmetry.

Also $K_q$ instead of $K_g$ in~\eqref{eq:Born_h_jets_central} is justified by the crossing symmetry.
Alternatively, the current $j_{q\bar{q}}$ in~\eqref{eq:S_fHqqx} can be derived from scratch by calculating the full QCD amplitude for $f_a g \to f_a q \bar{q}$ at LO, and then dropping all terms that are kinematically suppressed in the QMRK limit $s_{q\bar{q}} \ll s_{aq}, s_{a\bar{q}}$.

For the case of a peripheral Higgs-boson emission together with an extremal quark-antiquark pair, as depicted in the right panel of figure~\ref{fig:ME_h_jets_extremal}, it is sufficient to slightly rearrange the ingredients in equation~\eqref{eq:ME_fact}.
Starting from the Born function $\mathcal{B}$ in equation~\eqref{eq:Born_h_jets_central}, we have to drop the first colour acceleration multiplier $K_{f_a}$ and $t_j, t_{j+1}$.
Afterwards, we have to replace $\mathcal{S}_{f_a g \to f_a H q \bar{q}}$ with a different contraction of currents that is given by
\begin{equation}
  \left\|S_{g g\to H q \bar{q}}\right\|^2 = \sum_{\substack{\lambda_a =
   +,-\\\lambda_b = +,-}} \left|\epsilon^{\lambda_a}_\mu(p_a) V_H^{\mu\nu}(p_a,p_a - p_1)
   j_{\nu,q\bar{q}}^{\lambda_b}(p_q, p_{\bar{q}}, p_b)\right|^2,
\end{equation}
with $p_1 \equiv p_H$. Hence, the resulting Born function $\mathcal{B}_{H,q\bar{q}}$ takes the form
\begin{equation}
  \label{eq:Born_h_jets_extremal}
 \mathcal{B}_{H, q\bar{q}} =  \frac {(4\pi\alpha_s)^{n-1}} {4(N_C^2-1)} \frac{N_C}{(N_C^2 - 1)}
 \frac{K_{q}}{t_1 t_{n-2}} \left\|S_{g g \to H q\bar{q} }\right\|^2.
\end{equation}
We stress that in the specific case of no additional gluon radiation, the full matrix element for $gg \to Hq\bar{q}$ reduces to equation~\eqref{eq:Born_h_jets_extremal}.
With additional central gluon emissions between the Higgs boson and the quark-antiquark pair, the Lipatov vertex defined in equation~\eqref{eq:V_Lipatov} is modified such that its third argument becomes $p_a$ instead of $p_1$, as the outer particle -- the peripheral Higgs boson -- is no longer colour-charged.
The validity of this modified Lipatov vertex has been confirmed in~\cite{Andersen:2022zte}.

The last NLL configuration is when we allow for a \emph{central quark-antiquark emission}.
In this process, we have two currents scattering off of each other, but this time, via an effective vertex, which connects together two FKL chains.
Each FKL chain radiates a $t$-channel gluon that in turn splits into a $q\bar{q}$ pair.
As a result, we end up with this pair emitted in between the most backward and forward jets.
\begin{figure}[btp]
    \centering
    \begin{subfigure}[b]{0.42\textwidth}
        \includegraphics[width=\textwidth]{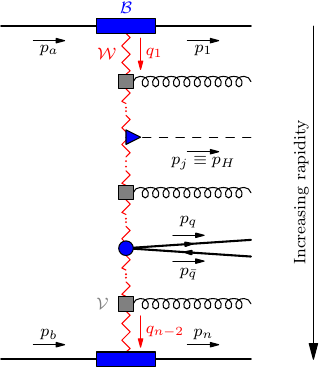}
        \caption{$ f_a f_b \to f_a \cdots H \cdots q\bar{q} \cdots f_b$ process}
    \end{subfigure}
    \hfill
    \begin{subfigure}[b]{0.42\textwidth}
        \includegraphics[width=\textwidth]{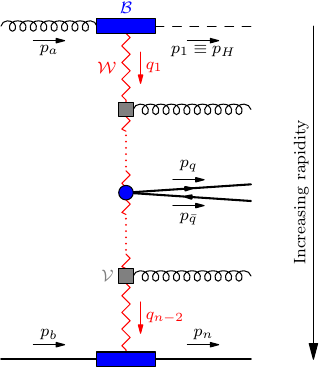}
        \caption{$ g f_b \to H \cdots q\bar{q} \cdots f_b$ process}
    \end{subfigure}

\caption{
On the left panel, the diagrammatic structure of the \HEJ matrix element for central Higgs plus central quark-antiquark pair production.
On the right panel, the diagrammatic structure of the \HEJ matrix element for extremal Higgs together with a central quark-antiquark emission.
The building blocks of the \HEJ matrix elements are given by the \textit{Born functions} $\mathcal{B}$.
The grey boxes in both figures stand for additional gluon radiation.
The latter is described by the \textit{Lipatov vertex} $\mathcal{V}$.
The virtual corrections, illustrated by the red zigzag lines, are given by the Reggeised $t$-channel colour-octet exchanges.
They are obtained via the \textit{Lipatov Ansatz} $\mathcal{W}$.
}
\label{fig:ME_h_jets_central}
\end{figure}
The \HEJ diagrammatic representation of a central quark-pair and an Higgs boson emission is illustrated in figure~\ref{fig:ME_h_jets_central}.
As in the case of extremal quark-pair emission, two distinct configurations arise: either the Higgs is emitted centrally or peripherally.
Both scenarios are illustrated in the left and right panels of figure~\ref{fig:ME_h_jets_central}, respectively.

Since the outgoing $q\bar{q}$ are allowed to exchange colour-singlet $t$-channel propagator, we do not need to impose any strong rapidity ordering between the quark-antiquark pair.
Therefore, we have the following QMRK limit
\begin{equation}
  y_1 \ll y_{j-1} \ll y_q , y_{\bar{q}} \ll y_{j+2}  \ll y_n.
\end{equation}
The derivation of this current depends on the sum of all leading-order Feynman diagrams contributing to the process $f_a f_b \to f_a q \bar{q} f_b$.
Afterwards, we just drop the terms that are kinematically suppressed.
The associated Born function in the case of a central Higgs emission is
\begin{equation}
  \label{eq:Born_cenh_cenqqx}
 \mathcal{B}_{f_a, H, q\bar{q},f_b} =  \frac {(4\pi\alpha_s)^{n-1}} {4(N_C^2-1)}
    \frac {K_{f_a}(p_1^-, p_a^-)} {t_1}\
 \frac{K_{f_b}(p_n^+, p_b^+)}{t_{n-2}} \frac{\left\|S_{f_a
          f_b \to f_a H q\bar{q} f_b}\right\|^2}{t_j t_{j+1}},
\end{equation}
where $S_{f_a f_b \to f_a H q\bar{q} f_b}$ is given by the contraction of the following currents with the Higgs boson production vertex
\begin{equation}
  \label{eq:S_fHqqxf}
  \left\|S_{f_a f_b \to f_a H q \bar{q} f_b}\right\|^2 = \frac{1}{2 q^2_{-} q^2_{+}}\sum_{\substack{\lambda_a =
  +,-\\\lambda_b = +,- \\\lambda_c = +,-}} \left|j^{\lambda_a}_\mu(p_1, p_a) V_H^{\mu\nu}(q_j,q_{j+1})
j_{\nu,q\bar{q}}^{\rho,\lambda_c}(p_q, p_{\bar{q}}, q_{-}) j^{\lambda_b}_{\rho}(p_n, p_b)\right|^2.
\end{equation}
In~\eqref{eq:S_fHqqxf}, $q_{-}~(q_{+})$ refers to the $t$-channel momentum before (after) the quark-pair emission, such that $q_{+} = q_{-} - p_q - p_{\bar{q}}$.
It is worth to stress that $j_{q\bar{q}}$ in~\eqref{eq:S_fHqqxf} is a different quark-antiquark current from the one appearing in~\eqref{eq:S_fHqqx}, see~\cite{Andersen:2020yax}.
The Lipatov vertex $\mathcal{V}$, shown in~\eqref{eq:V} and~\eqref{eq:V_Lipatov}, must be rearranged in such a way its dependence on $p_q,p_{\bar{q}}$ is replaced by $p_n$.

\begin{figure}[htbp]
    \centering
    \begin{subfigure}[b]{0.48\textwidth}
        \includegraphics[width=\textwidth]{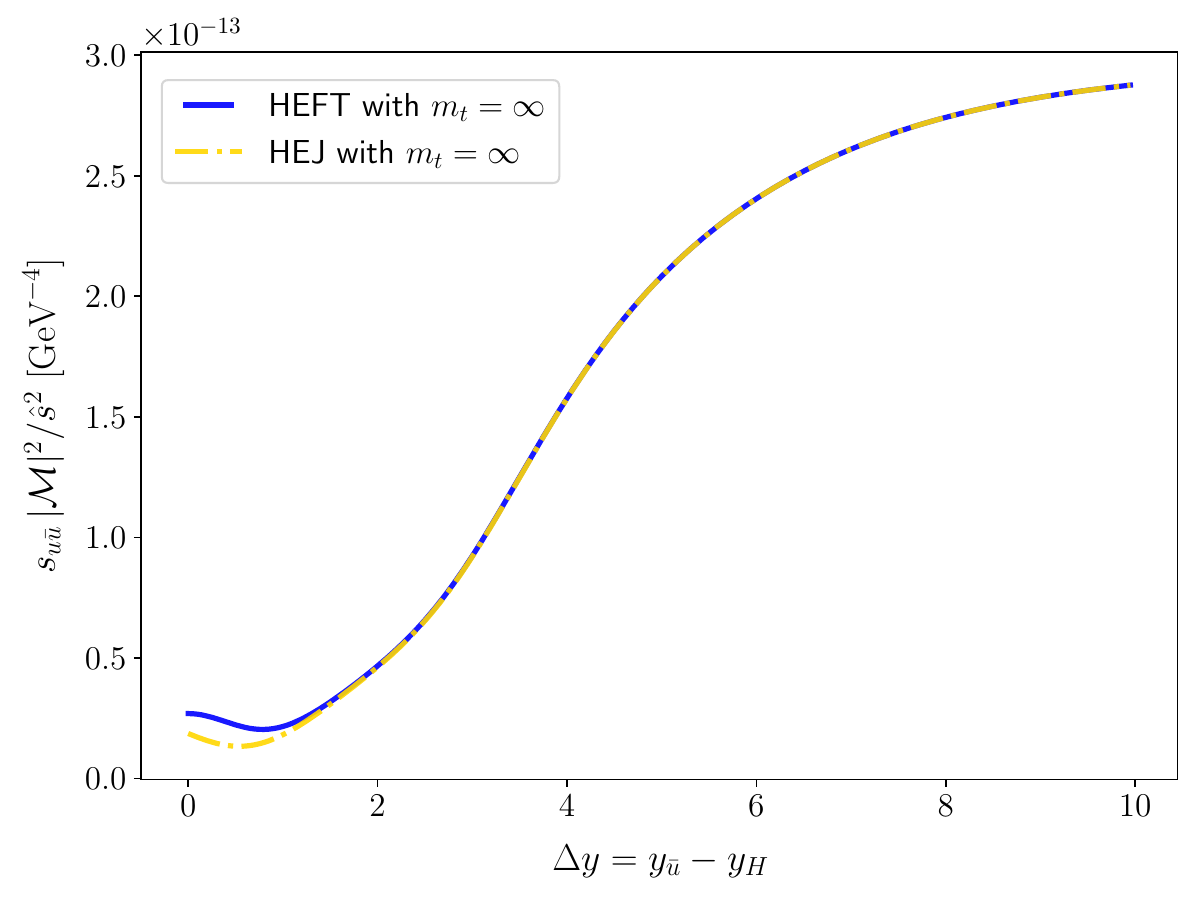}
        \caption{$s_{u\bar{u}}\,\vert\mathcal{M}\vert^2/\hat{s}^2$ for $gg \to Hu\bar{u}$ process}
    \end{subfigure}
    \hfill
    \begin{subfigure}[b]{0.48\textwidth}
        \includegraphics[width=\textwidth]{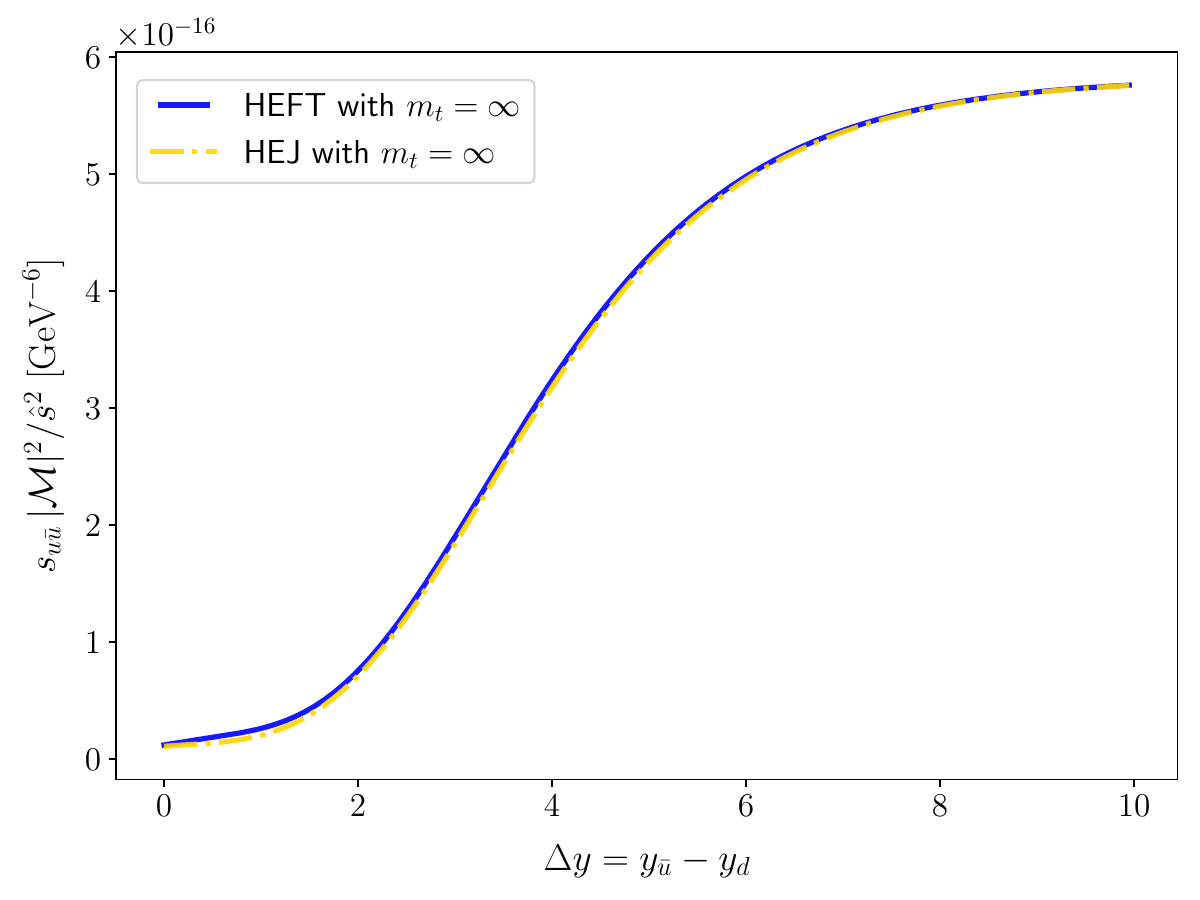}
        \caption{$s_{u\bar{u}}\,\vert\mathcal{M}\vert^2/\hat{s}^2$ for $dg \to dHu\bar{u}$ process}
    \end{subfigure}

    \vspace{1em}

    \begin{subfigure}[b]{0.48\textwidth}
        \includegraphics[width=\textwidth]{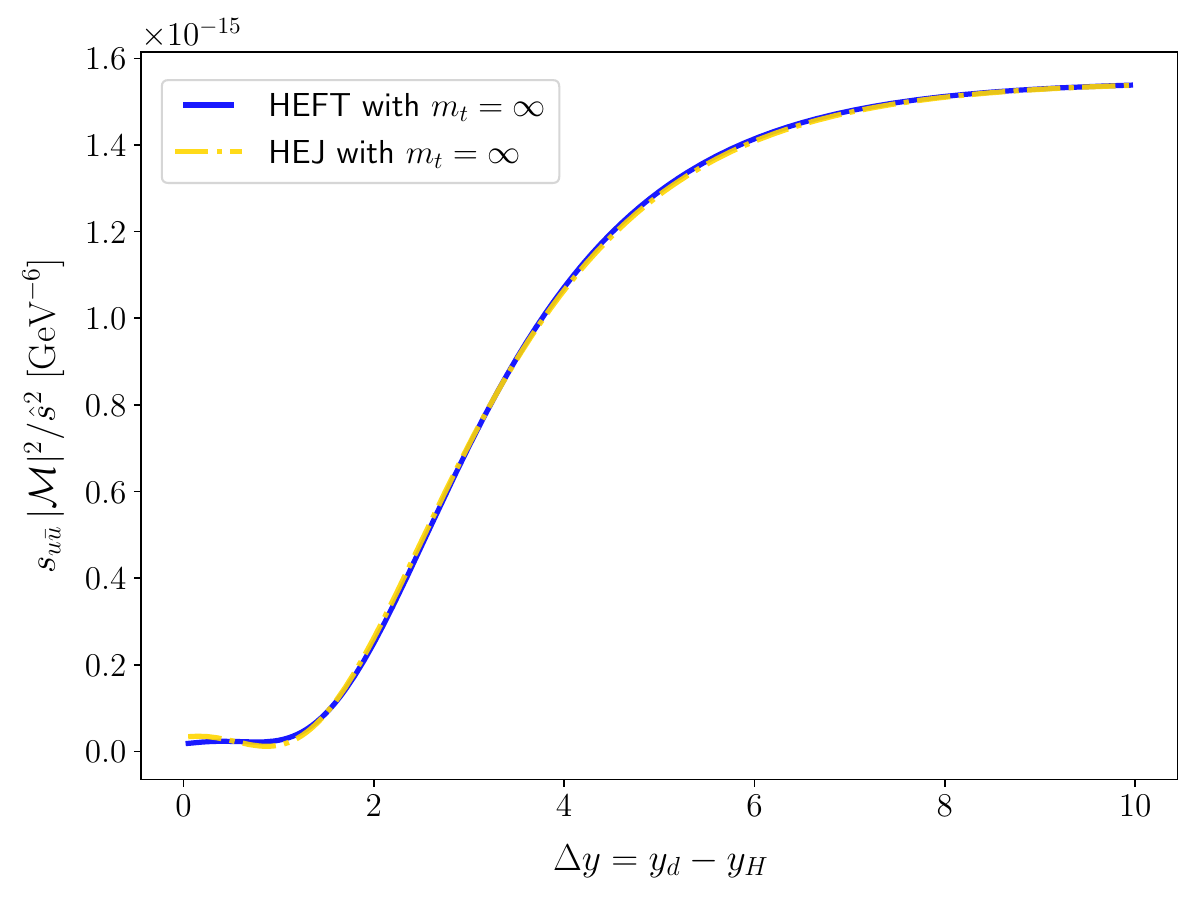}
        \caption{$s_{u\bar{u}}\,\vert\mathcal{M}\vert^2/\hat{s}^2$ for $gd \to Hu\bar{u}d$ process}
    \end{subfigure}
    \hfill
    \begin{subfigure}[b]{0.48\textwidth}
        \includegraphics[width=\textwidth]{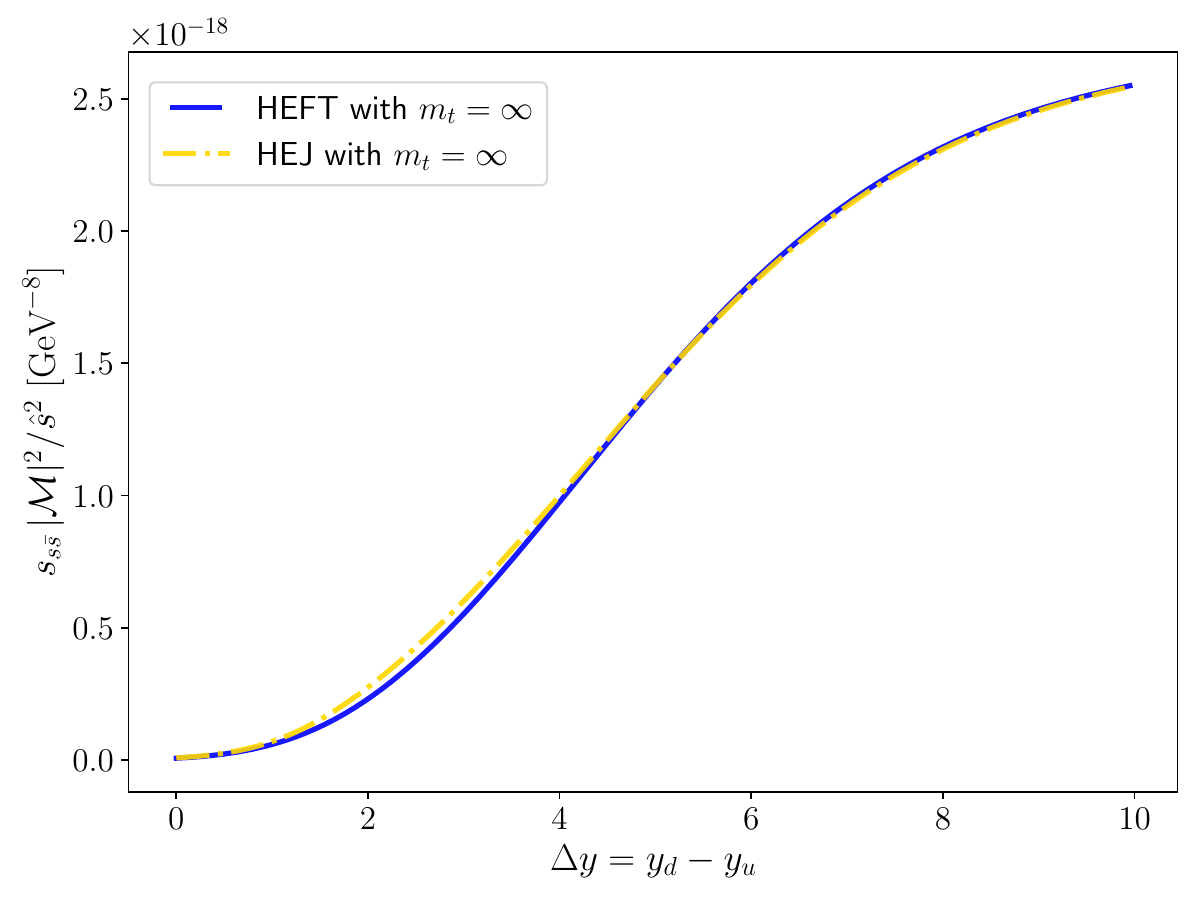}
        \caption{$s_{s\bar{s}}\,\vert\mathcal{M}\vert^2/\hat{s}^2$ for $ud \to us\bar{s}Hd$ process}
    \end{subfigure}

    \caption{Regge scaling of the squared matrix elements given in equation~(\ref{eq:REGGESCALING}) for four different processes.
      On the top part the \emph{extremal quark-antiquark emission} together with an extremal Higgs boson and a central one, respectively.
      On the bottom part the \emph{central quark-antiquark emission} along with an extremal Higgs boson and a central one, respectively.
      The MRK limit is reached by increasing $\Delta y$ on the $x$-axis, see equation~(\ref{eq:MRKLIMIT}).}
\label{fig:scaling_explorers}
\end{figure}
For the case of a peripheral Higgs boson, see the right panel of figure~\ref{fig:ME_h_jets_central} for reference, we can proceed in the same manner as for the extremal quark-pair emission.
First, we replace $j_{\mu}^{\lambda_a}$ in~\eqref{eq:S_fHqqxf} with $\epsilon_{\mu}^{\lambda_a}$, then we rearrange the momenta flowing into the effective gluon-gluon-Higgs vertex.
Therefore, the resulting contraction of currents can be written as
\begin{equation}
  \label{eq:S_Hqqxf}
  \left\|S_{g f_b \to H q \bar{q} f_b}\right\|^2 = \frac{1}{2 q^2_{-} q^2_{+}}\sum_{\substack{\lambda_a =
  +,-\\\lambda_b = +,- \\\lambda_c = +,-}} \left|\epsilon^{\lambda_a}_\mu(p_a) V_H^{\mu\nu}(p_a,p_a - p_1)
j_{\nu,q\bar{q}}^{\rho,\lambda_c}(p_q, p_{\bar{q}}, q_{-}) j^{\lambda_b}_{\rho}(p_n, p_b)\right|^2,
\end{equation}
with $p_1 \equiv p_H$. In turn, the Born function $\mathcal{B}_{H,q\bar{q},f_b}$ takes the form
\begin{equation}
  \label{eq:Born_exh_cenqqx}
 \mathcal{B}_{H, q\bar{q},f_b} =  \frac {(4\pi\alpha_s)^{n-1}} {4(N_C^2-1)}
 \frac{K_{f_b}(p_n^+, p_b^+)}{t_1 t_{n-2}} \left\|S_{g
          f_b \to H q\bar{q} f_b}\right\|^2.
\end{equation}
To account for additional gluon radiation, the Lipatov vertex $\mathcal{V}$ in~\eqref{eq:V_Lipatov} has to be modified by substituting again $p_q,p_{\bar{q}}$ with $p_n$, and more importantly its third argument $p_1$ with $p_a$.
We emphasise once again that this last replacement is necessary, since the outer particle is colourless.

The above matrix elements are validated in the limit of an infinite top-quark mass by comparing to amplitudes extracted from \texttt{Madgraph5\_aMC@NLO}~\cite{Alwall:2014hca}, see figure~\ref{fig:scaling_explorers}.
Both \HEJ-approximated and exact amplitudes are calculated in a one-dimensional phase-space as a function of the rapidity separation between all pairs of particles.
The momentum configurations used for the plots in figure~\ref{fig:scaling_explorers} are listed in table~\ref{tab:explorerPS}.
It is worth to stress that the qualitative behaviour shown is not dependent on specific values of azimuthal angle or transverse momentum, but only on the rapidity assignment of the particles.
\begin{table}[htbp]
\begin{center}
\begin{tabular}{ |c||l| }
\hline
Process & Momenta configuration \\
\hline
$gg \to Hu\bar{u}$ &
\(
\begin{cases}
      y_H = -\Delta, y_u = \frac{\Delta}{2} \text{ and } y_{\bar{u}} = \Delta &\\
      \phi_u = \frac{11}{9}\pi \text{ and } \phi_{\bar{u}} = \frac{\pi}{7} &\\
     p_{u\perp} = 40 \, \text{GeV} \text{ and } p_{\bar{u}\perp} = 40 \, \text{GeV} &
    \end{cases}
\)  \\
$dg \to dHu\bar{u}$ &
\(
\begin{cases}
      y_d = -\Delta, y_H = -\frac{\Delta}{3}, y_u = \frac{\Delta}{3} \text{ and } y_{\bar{u}} = \Delta &\\
      \phi_H = \frac{6}{11}\pi, \phi_u = \frac{11}{9}\pi \text{ and } \phi_{\bar{u}} = \frac{3}{4}\pi &\\
     p_{H\perp} = 40 \, \text{GeV}, p_{u\perp} = 40 \, \text{GeV} \text{ and } p_{\bar{u}\perp} = 40 \, \text{GeV}  &
    \end{cases}
\)     \\
$gd \to Hu\bar{u}d$ &
\(
\begin{cases}
      y_H = -\Delta, y_u=-\frac{\Delta}{3}, y_{\bar{u}}=\frac{\Delta}{3} \text{ and } y_d = \Delta &\\
      \phi_u = \frac{11}{9}\pi, \phi_{\bar{u}} = \frac{6}{11}\pi \text{ and } \phi_d = \frac{3}{4}\phi & \\
     p_{u\perp} = 40 \, \text{GeV}, p_{\bar{u}\perp} = 40 \, \text{GeV} \text{ and } p_d = 40 \, \text{GeV} &
    \end{cases}
\)  \\
$ud \rightarrow us\bar{s}Hd$ &
\(
\begin{cases}
      y_u = -\Delta, y_s=-\frac{\Delta}{2}, y_{\bar{s}} = 0, y_{H} = \frac{\Delta}{2} \text{ and } y_d = \Delta &\\
      \phi_s = \frac{6}{11}\pi, \phi_{\bar{s}} = \frac{3}{4}\pi, \phi_H = \frac{8}{5}\pi \text{ and } \phi_d = \frac{11}{9}\pi & \\
     p_{s\perp} = 40 \, \text{GeV}, p_{\bar{s}\perp} = 40 \, \text{GeV}, p_{H\perp} = 40 \, \text{GeV} \text{ and } p_{d\perp} = 40 \, \text{GeV}  &
    \end{cases}
\)  \\
 \hline
\end{tabular}
\end{center}
\caption{The momentum configurations used in figure~\ref{fig:scaling_explorers}.}
\label{tab:explorerPS}
\end{table}

\section{Higgs Boson plus Quark Pair as VBF Di-Higgs Background}\label{sec:HHBACKGROUND}

In this section, we present the first study of single Higgs boson production via gluon fusion in association with a bottom-quark pair and at least two additional hard jets as a background to VBF Higgs-pair production.
The predictions are obtained within the \HEJ framework, including the effects of high-energy QCD resummation, and are compared with leading-order perturbative calculations.

The analysis considered in this work is inspired by a CMS search for nonresonant Higgs-boson pair production in the $\gamma\gamma b\bar b$ final state using a data sample corresponding to an integrated luminosity of $137~\mathrm{fb}^{-1}$ collected at a centre-of-mass energy of 13~TeV from 2016 to 2018~\cite{CMS:2020tkr}.
The $\gamma \gamma b\bar{b}$ final state has a combined branching fraction of $2.63 \pm 0.06 \cdot 10^{-3}$~\cite{LHCHiggsCrossSectionWorkingGroup:2016ypw} for a Higgs boson mass of 125 GeV.
This channel is one of the most sensitive to HH production because of the large SM branching fraction of Higgs boson decays to bottom quarks, the good mass resolution of the $H \to \gamma \gamma$ channel, and relatively low background rates.

The main goal of the present study is to provide novel theoretical predictions to improve the modelling of backgrounds to di-Higgs searches in the VBF production mode.
This channel is particularly relevant, as it gives access to the trilinear Higgs self-coupling (HHH), as well as to the coupling between two vector bosons and the Higgs boson (VVH) and the coupling between a pair of Higgs bosons and a pair of vector bosons (VVHH).
The dominant background considered in this work arises from single-Higgs plus jets production via gluon fusion.
In VBF topologies, characterised by large dijet invariant masses and sizeable rapidity separations between the hard jets, high-energy logarithmic corrections are expected to become increasingly important~\cite{Andersen:2018kjg, Andersen:2018tnm}.
Accounting for these effects is therefore essential for a reliable prediction of the ggF background in these regions of phase space.

\begin{figure}[htbp]
    \centering

    \begin{subfigure}[c]{0.32\textwidth}
        \centering
        \includegraphics[valign=c,width=\textwidth]{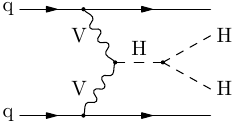}
    \end{subfigure}
    \hfill
    \begin{subfigure}[c]{0.32\textwidth}
        \centering
        \includegraphics[valign=c,width=\textwidth]{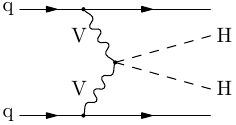}
    \end{subfigure}
    \hfill
    \begin{subfigure}[c]{0.32\textwidth}
        \centering
        \includegraphics[valign=c,width=\textwidth]{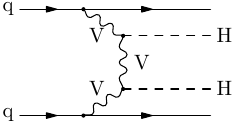}
    \end{subfigure}

\caption{
Feynman diagrams contributing to the pair production of Higgs bosons via VBF at LO.
The left figure displays the diagram sensitive to the HHH vertex, in the middle figure the diagram involving the VVHH vertex, and on the right figure the diagram with two VVH vertices.
}
\label{fig:vbf_process}
\end{figure}

\begin{figure}[htbp]
    \centering

    \begin{subfigure}[c]{0.32\textwidth}
        \centering
        \includegraphics[valign=c,width=\textwidth]{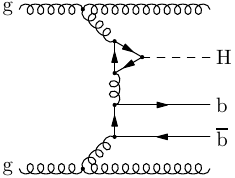}
    \end{subfigure}
    \hfill
    \begin{subfigure}[c]{0.32\textwidth}
        \centering
        \includegraphics[valign=c,width=\textwidth]{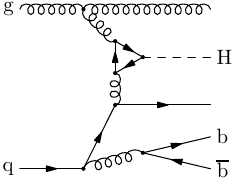}
    \end{subfigure}
    \hfill
    \begin{subfigure}[c]{0.32\textwidth}
        \centering
        \includegraphics[valign=c,width=\textwidth]{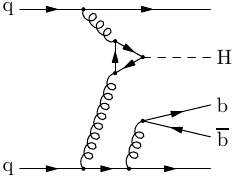}
    \end{subfigure}

\caption{
Examples of Feynman diagrams contributing to the production of a single Higgs boson in association with four jets via ggF at LO.
All the diagrams correspond to SM processes, involving the top and bottom Yukawa couplings, $y_t$ and $y_b$, respectively.
Importantly, these processes are a significant background to di-Higgs searches, both via VBF and ggF production modes, where one of the Higgs bosons decays into a bottom-pair.
}
\label{fig:ggf_process}
\end{figure}

The Feynman diagrams contributing to double-Higgs VBF production mode at LO are shown in figure~\ref{fig:vbf_process}, and some of the ones contributing to the background at LO are shown in figure~\ref{fig:ggf_process}.
While the HHH coupling is mainly constrained from measurements of HH production via ggF, and the VVH coupling is constrained by measurements of vector boson associated production of a single Higgs boson and the decay of the Higgs boson to a pair of bosons, the VVHH coupling is only directly measurable via VBF HH production.
The measurement of the HHH, VVH, and VVHH couplings constitutes a central component of the LHC physics programme, as these interactions directly probe the structure of the electroweak symmetry-breaking sector and the shape of the scalar potential, thereby providing stringent tests of the Standard Model.

\subsection{High Energy Jets Predictions}
\label{sec:HEJ_HH_bkg}

The main focus of this section is to present theoretical predictions for Higgs boson production in association with two $b$-jets and at least two additional hard jets via gluon fusion, including the effects of high-energy logarithmic corrections of the form $(\alpha_s \log(s_{ij}/p_{\perp}^2))^N$ through the \HEJ framework.
The resulting predictions are compared with fixed-order leading-order calculations obtained using \texttt{Sherpa}~\cite{Sherpa:2019gpd} through the matrix-element generators \texttt{Comix}~\cite{Gleisberg:2008fv} and \texttt{Amegic}~\cite{Krauss:2001iv}.

The leading-order partonic events generated with \texttt{Sherpa} are passed to \texttt{HEJ}, where each Born-level phase-space point is supplemented with the corresponding high-energy logarithmic correction, both real and virtual.
The resummation is performed according to the event classification described in the previous section~\ref{subsec:AMPCLASS}, ensuring that the resulting predictions retain leading-order accuracy while including the all-order resummation of high-energy logarithms.
More details about the high-energy resummation within the \HEJ 2 framework are given in~\cite{Andersen:2018tnm, Andersen:2023kuj}.

For the results presented in this section, the fixed-order input is limited to $Hb\bar{b}+2$ jets and $Hb\bar{b}+3$ jets configurations.
A discussion of the truncation of the Born-level multiplicity and its impact on the predictions is given in appendix~\ref{app:hsixjets}.

For accurate predictions in the high-energy regime, finite top-quark mass effects must be taken into account.
In principle, the \HEJ\ framework allows one to retain the full dependence on both the top- and bottom-quark masses, $m_t$ and $m_b$, within the resummed matrix elements.
However, computing the leading-order input with the exact dependence on $m_t$ rapidly becomes computationally prohibitive for increasing jet multiplicities.
For this reason, the event generation is performed in the HEFT in the $m_t \to \infty$ limit, while finite top-quark mass effects are restored through the resummation weight.
For the evaluation of finite top-quark mass corrections within \HEJ, we implement \texttt{QCDLoop}~\cite{Carrazza:2016gav}.

Although the dependence on the bottom-quark mass and Yukawa coupling could also be retained, we do not include these effects in the present study.
This choice is made for consistency with the 5FS adopted throughout the analysis and is supported by the comparison presented in appendix~\ref{app:bmass}, where both the bottom-quark mass and Yukawa contributions are shown to have a negligible impact in the VBF-like phase-space region considered here.
The weight of a leading-order matched resummation event is therefore constructed from the all-order \HEJ\ matrix element $\mathcal{M}_{\HEJ}(m_t, m_b)$ and the corresponding leading-order matrix element $\mathcal{M}_{\rm LO}(m_t, m_b)$ according to
\begin{equation}
 w \propto \vert \mathcal{M}_{\HEJ} (m_t, 0) \vert^2 \frac{\vert \mathcal{M}_{\rm LO}(\infty, 0)\vert^2}{\vert \mathcal{M}_{\HEJ,\rm LO}(\infty, 0)\vert^2} \quad \geq 4 \,{\rm jets.}
\end{equation}
In the above equation $ \mathcal{M}_{\HEJ,\rm LO}$ denotes the leading-order truncation of the \HEJ\ matrix element.
Empirically, the transverse-momentum cuts in the leading-order generation are taken about 10--20\% looser than the final analysis cuts, with a slightly larger difference at high jet multiplicities.
To improve efficiency, the leading-order generation is split into two disjoint samples: a high-statistics sample where all particles satisfy the analysis $p_{T}$ cuts, and a low-statistics sample containing events with at least one particle below these cuts.
\HEJ\ resummation is applied separately to each sample and the results are then combined.

All theoretical predictions are obtained with the central scale choice
\begin{equation}\label{eq:scale_choice}
  \mu_{\rm F} = \mu_{\rm R} = \max(m_{12}, m_{\rm H})
\end{equation}
in which $m_{12}$ stands for the invariant mass between the two hardest jets and $m_{\rm H}$ is the Higgs boson mass.
To provide an estimate of the theory uncertainty due to the scale dependence, a conventional 7-point scale variation with a factor of two is employed.

The cuts employed in this study follow the analysis strategy of the CMS collaboration~\cite{CMS:2020tkr} and are implemented in \texttt{Rivet}~\cite{Bierlich:2024vqo}.
The most relevant experimental cuts are reported in table~\ref{table:hh-baseline}, see~\cite{CMS:2020tkr} for the comprehensive list.
The Higgs boson candidate is identified from the diphoton decay channel, while jets are reconstructed using the anti-$k_t$ algorithm~\cite{Cacciari:2008gp} with a radius parameter of $R = 0.4$, as implemented in \texttt{FastJet}~\cite{Cacciari:2011ma}.
In addition, the PDF4LHC21\_40 PDF set~\cite{PDF4LHCWorkingGroup:2022cjn} is used within the massless 5FS through the \texttt{LHAPDF} interface~\cite{Buckley:2014ana}, following the recommendations of the LHC Higgs Working Group.

The results presented in the following correspond to 4-jet inclusive observables, requiring events with at least four jets, two of which are identified as $b$-jets, while the remaining jets are treated as VBF-jet candidates.
The $b$-jet selection employed in this study differs from that adopted in the experimental analysis~\cite{CMS:2020tkr}.
In the latter, jets originating from bottom quarks are identified using a secondary-vertex tagging algorithm based on a deep neural network (DNN).
The resulting $b$-tagging score is used to rank and classify all $b$-jet candidates in all events with at least two $b$-jets satisfying the baseline selection criteria reported in table~\ref{table:hh-baseline}.
In contrast, our analysis selects the $b$-jet pair whose invariant mass is closest to the Higgs boson mass.
This choice is motivated by the requirement that the reconstructed $b\bar{b}$ system lies within the Higgs boson mass window, $70 < m_{b\bar{b}} < 190~\mathrm{GeV}$, thereby enhancing compatibility with a Higgs boson decay and reducing background contamination.

\begin{figure}[htbp]
    \centering

    \begin{subfigure}[b]{0.49\textwidth}
        \centering
        \includegraphics[width=\textwidth]{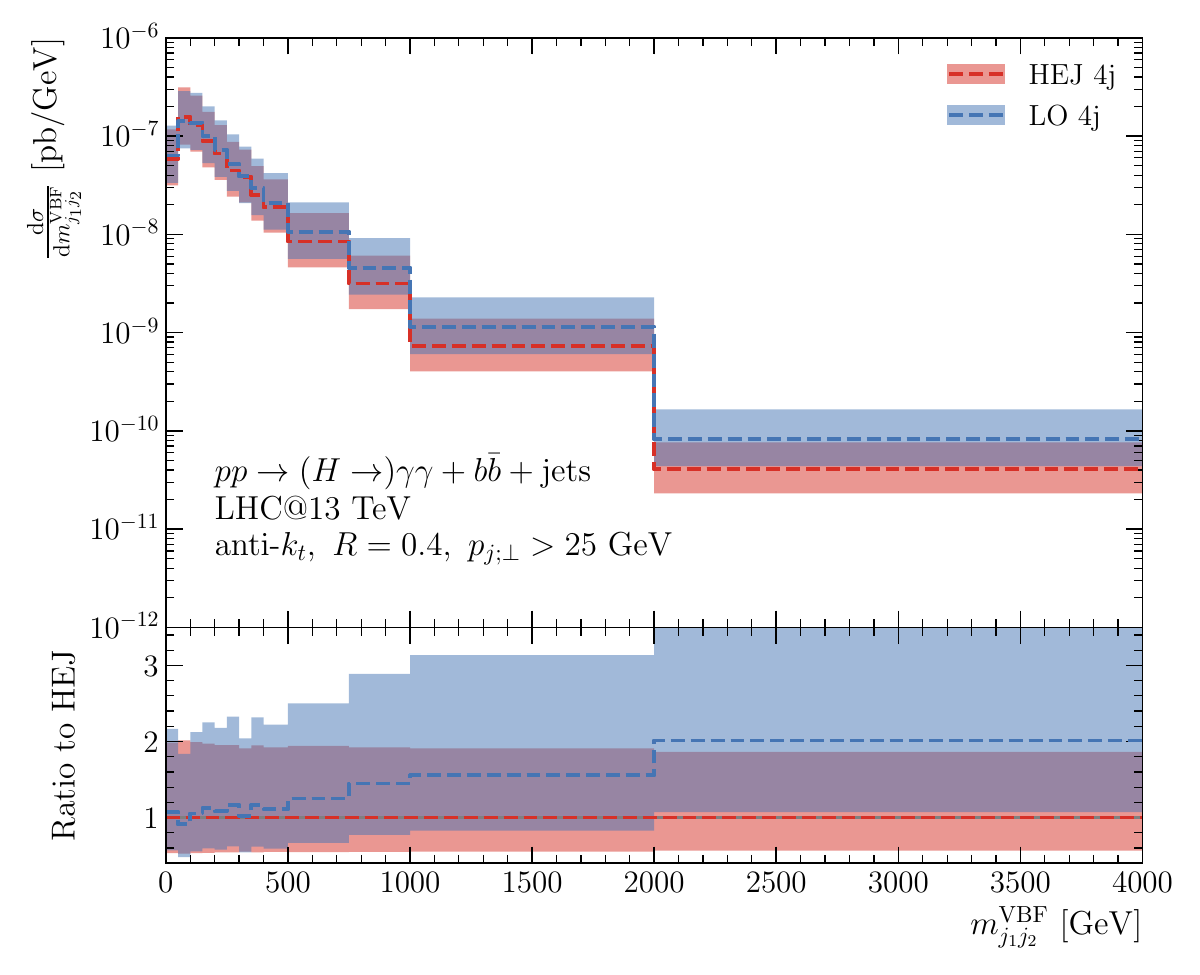}
        \caption{VBF-dijet invariant mass.}
    \label{fig:hh_vbfdijetmass}
    \end{subfigure}
    \hfill
    \begin{subfigure}[b]{0.49\textwidth}
        \centering
        \includegraphics[width=\textwidth]{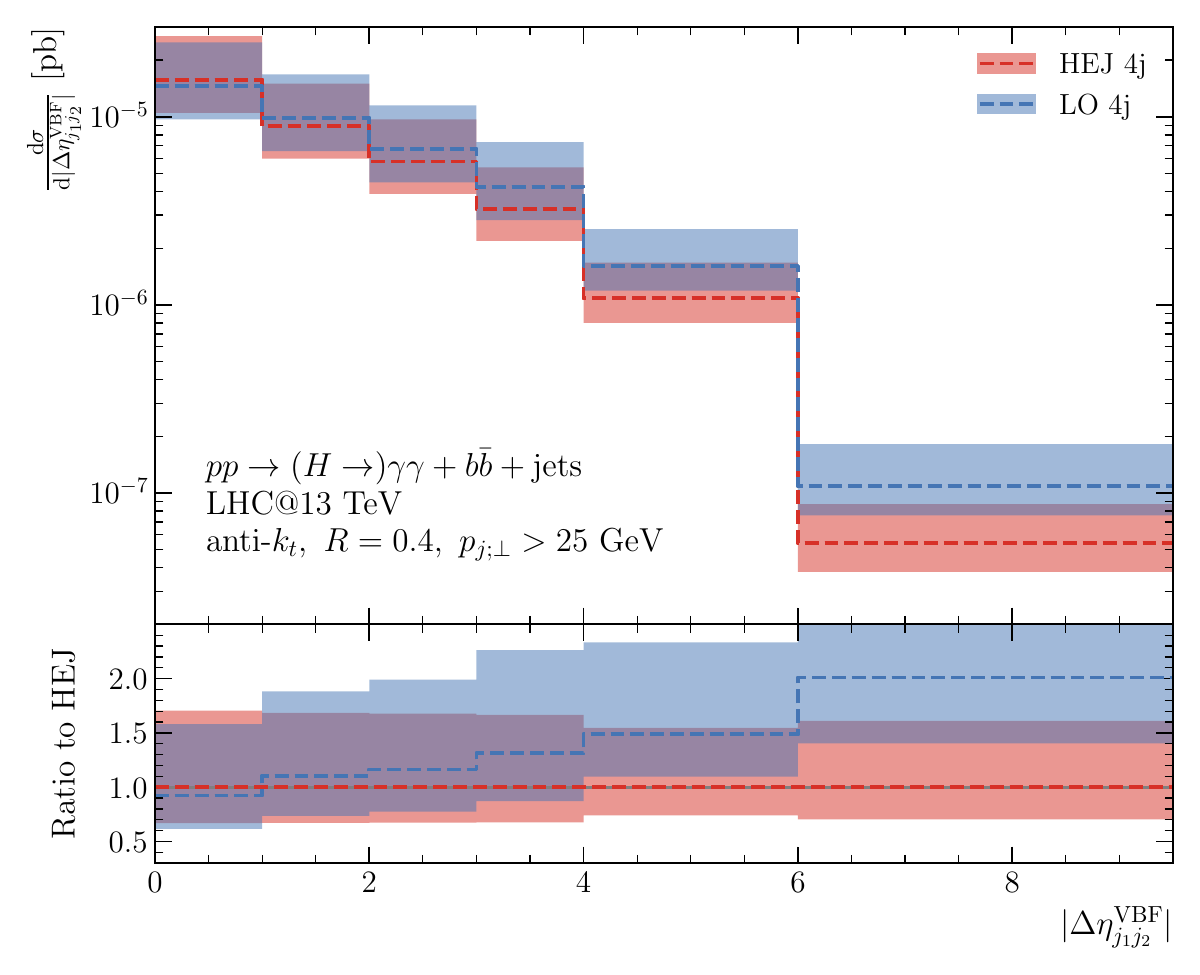}
        \caption{VBF-dijet pseudo-rapidity separation.}
    \label{fig:hh_vbfjetsdeltaeta}
    \end{subfigure}

    \vspace{1em}

    \begin{subfigure}[b]{0.49\textwidth}
        \centering
        \includegraphics[width=\textwidth]{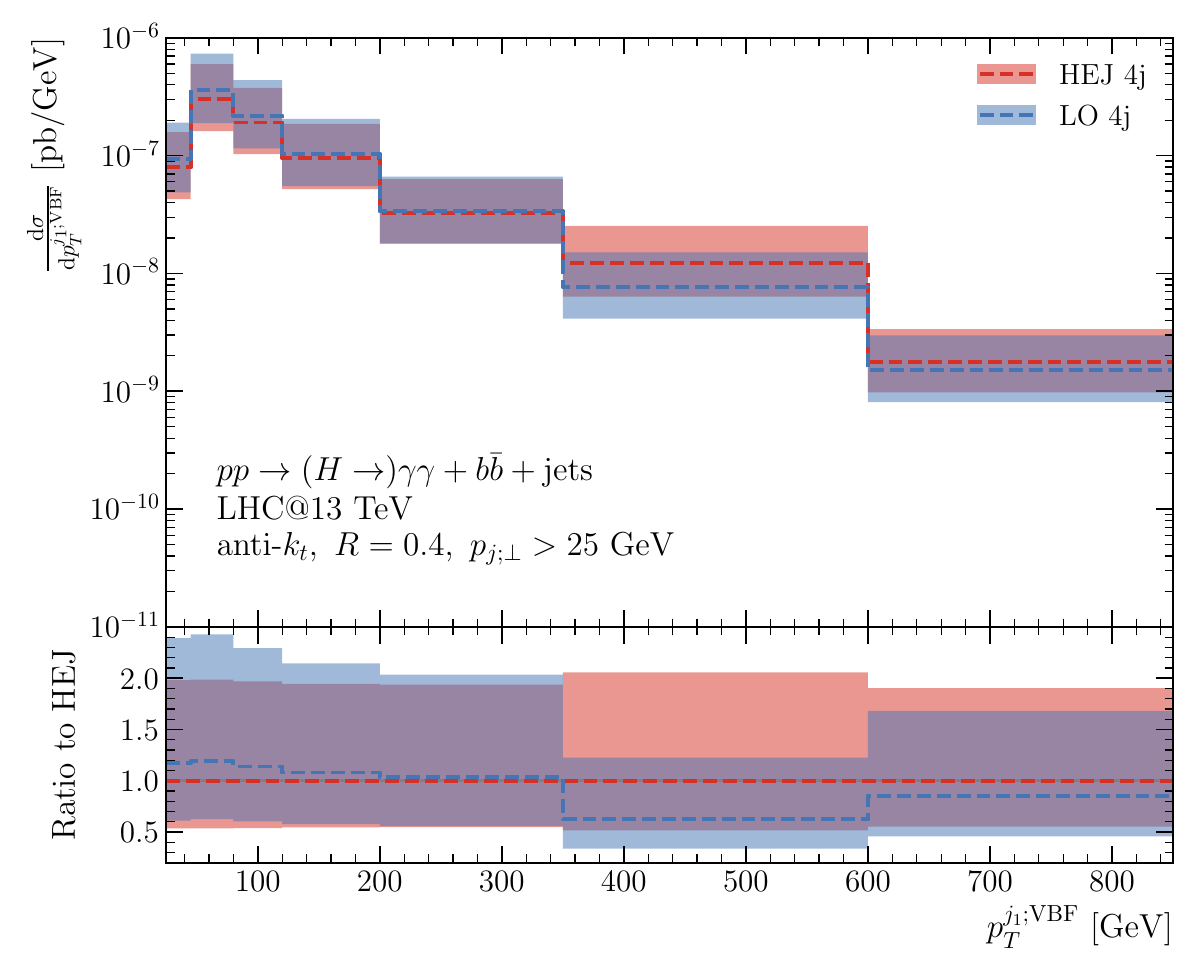}
        \caption{Leading VBF-jet transverse momentum.}
    \label{fig:hh_vbfjet0pt}
    \end{subfigure}
    \hfill
    \begin{subfigure}[b]{0.49\textwidth}
        \centering
        \includegraphics[width=\textwidth]{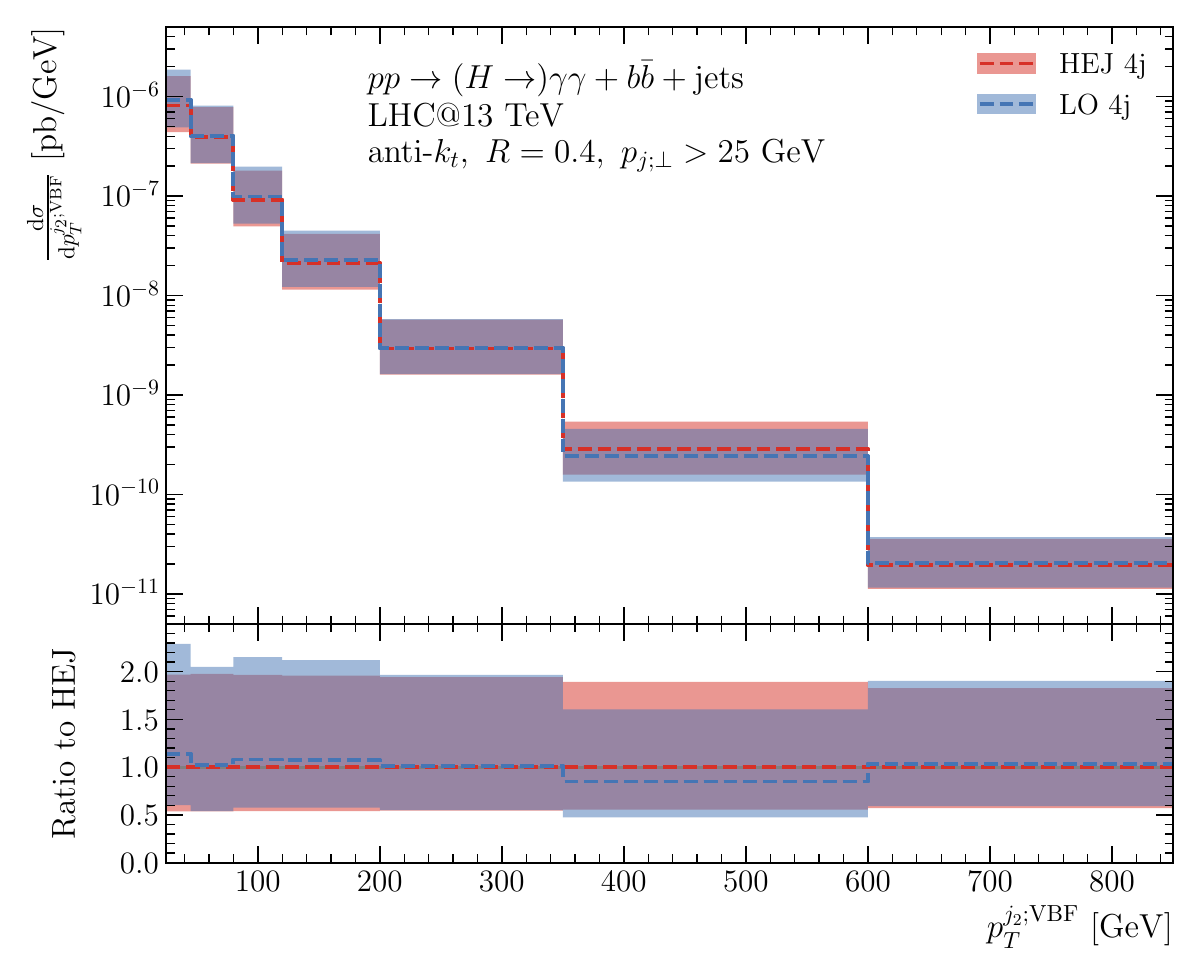}
        \caption{Subleading VBF-jet transverse momentum.}
    \label{fig:hh_vbfjet1pt}
    \end{subfigure}

\caption{
Comparison between \HEJ and LO for 4-jet inclusive observables.
\ref{fig:hh_vbfdijetmass}: Invariant mass of the two VBF jets.
\ref{fig:hh_vbfjetsdeltaeta}: Pseudo-rapidity separation between the two VBF jets.
\ref{fig:hh_vbfjet0pt}: Transverse momentum of the hardest VBF jets.
\ref{fig:hh_vbfjet1pt}: Transverse momentum of the second hardest VBF jets.
}
\label{fig:hh_vbf_observables}
\end{figure}

The selection of the VBF-jet candidates is instead identical to that employed in the experimental analysis.
After applying the requirements listed in table~\ref{table:hh-baseline}, the jet pair with the largest dijet invariant mass $m^{\rm VBF}_{j_1j_2}$  -- excluding the previously identified $b\bar{b}$ system -- is chosen as the VBF-tagged jet pair.
\begin{table}[H]
\begin{center}
\begin{tabular}{|l|c|}
 \hline
Description & Baseline cuts  \\
 \hline
Photon transverse momentum & $p_T(\gamma) > 25 \text{ GeV}$  \\
Diphoton invariant mass & $100 \text{ GeV} < m_{\gamma\gamma} < 180 \text{ GeV} $ \\
Pseudo-rapidity of the photons & $|\eta_{\gamma} |<2.5$ \\ & excluding $1.44< |\eta_\gamma| < 1.57$\\
Ratio of harder photon $p_T$ to diphoton invariant mass & $p_T(\gamma_1)/m_{\gamma \gamma} > 0.35 $  \\
Ratio of softer photon $p_T$ to diphoton invariant mass & $p_T(\gamma_2)/m_{\gamma \gamma} > 0.25 $  \\
\hline
$b$-jet transverse momentum & $p_T(j) > 25 \text{ GeV}$  \\
$b$-jets invariant mass & $70 \text{ GeV} < m_{b\bar{b}} < 190 \text{ GeV} $ \\
$b$-jet rapidity & $|y_j| <2.5$  \\
\hline
Hardest VBF-jet transverse momentum & $p_T(j) > 40 \text{ GeV}$  \\
Softest VBF-jet transverse momentum & $p_T(j) > 30 \text{ GeV}$  \\
VBF-jet rapidity & $|y_j| <4.7$  \\
\hline
\end{tabular}
\caption{Baseline cuts of the CMS analysis~\cite{CMS:2020tkr}.}
\label{table:hh-baseline}
\end{center}
\end{table}
We first discuss observables where high-energy resummation is expected to lead to significantly improved descriptions. Figure~\ref{fig:hh_vbfdijetmass} shows the invariant mass distribution of the two VBF jets for both \HEJ\ and fixed-order LO QCD predictions.
The all-order resummation of high-energy logarithmic corrections of the form $\log((m_{j_1j_2}^{\rm VBF})^2/t)$ leads to a drastic reduction of the central prediction by a factor of up to two.
A similar pattern is observed in figure~\ref{fig:hh_vbfjetsdeltaeta}, which displays the pseudo-rapidity separation between the two VBF jets.
While the large scale uncertainties preclude an exact quantitative assessment, the systematic difference in shapes strongly suggests that high-energy resummation is essential for reliable predictions.

VBF selections requiring large dijet invariant masses and wide rapidity separations are regularly employed to reduce the gluon-fusion background in single-Higgs production.
In contrast, their use in di-Higgs measurements has been more analysis dependent.
For instance, they have not been applied in the $b\bar{b}\gamma\gamma$ analyses of~\cite{CMS:2020tkr, ATLAS:2023gzn,  ATLAS:2025hhd} and $b\bar{b}\tau^-\tau^+$ measurement of~\cite{ATLAS:2024pov}.
They have, however, been used in other studies with $b\bar{b}\tau^-\tau^+$ as final state~\cite{CMS:2022hgz} and in several other measurements involving $b\bar{b}b\bar{b}$~\cite{ATLAS:2020jgy, ATLAS:2023qzf, ATLAS:2024lsk}.

Clearly, such cuts are highly effective in suppressing the background from single-Higgs production with a bottom-quark pair.
According to our analysis, this background suppression is even more effective than expected from fixed-order predictions.
This finding is in line with earlier observations on the enhanced effectiveness of VBF cuts in single-Higgs production~\cite{Andersen:2018kjg}.

In contrast, no sizeable differences between \HEJ and LO are observed for transverse-momentum observables.
This is illustrated in figure~\ref{fig:hh_vbfjet0pt} and figure~\ref{fig:hh_vbfjet1pt}, which show the transverse-momentum distributions of the leading and subleading VBF jets, respectively.

A similarly good agreement is observed for observables involving the $b$-jet pair.
Since the \HEJ\ resummation does not address logarithmic corrections associated with the $b\bar{b}$ system, no sizeable modifications of the LO prediction are expected.
For this reason, these distributions are not discussed further.

\begin{figure}[htbp]
    \centering

    \begin{subfigure}[b]{0.49\textwidth}
        \centering
        \includegraphics[width=\textwidth]{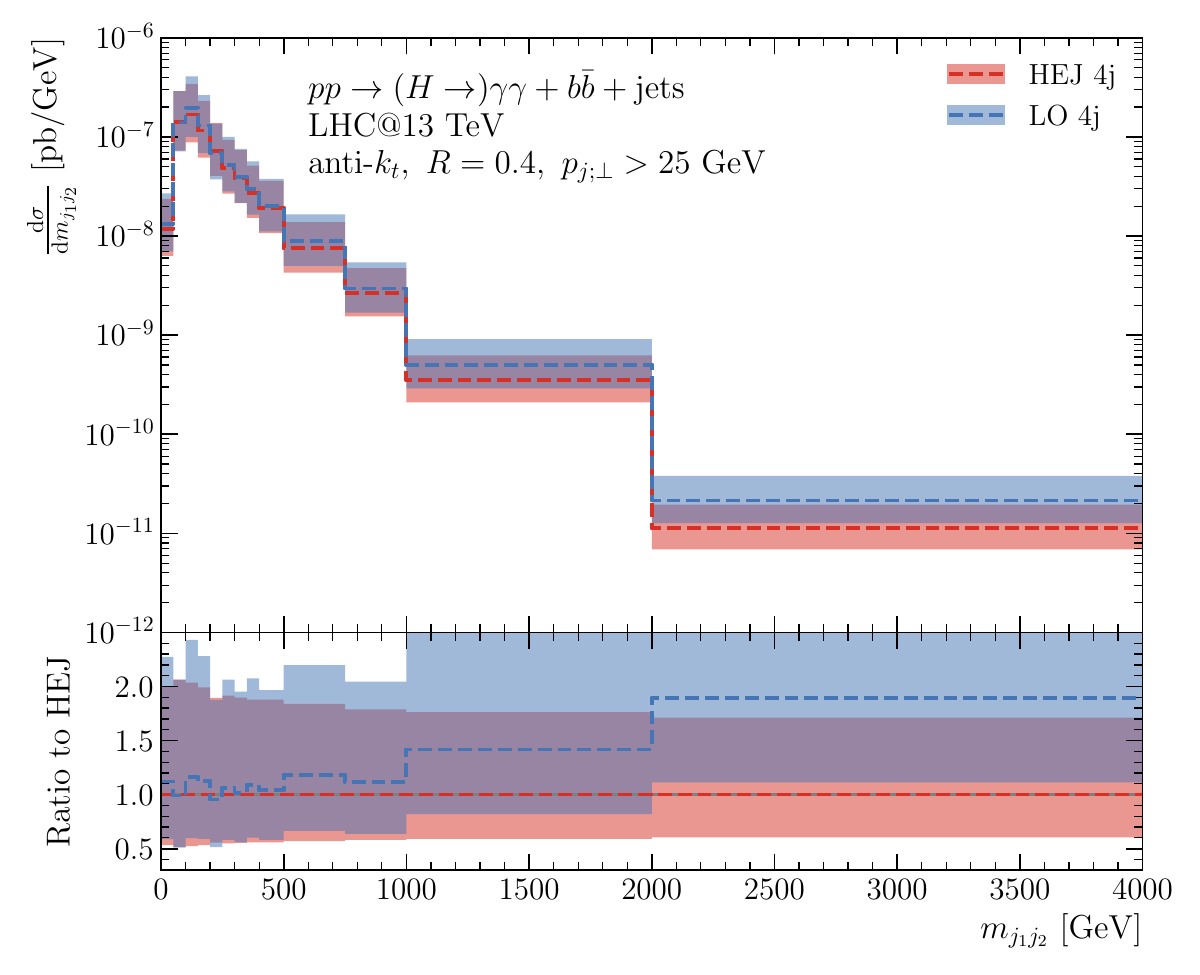}
        \caption{Dijet invariant mass.}
    \label{fig:hh_jjmass}
    \end{subfigure}
    \hfill
    \begin{subfigure}[b]{0.49\textwidth}
        \centering
        \includegraphics[width=\textwidth]{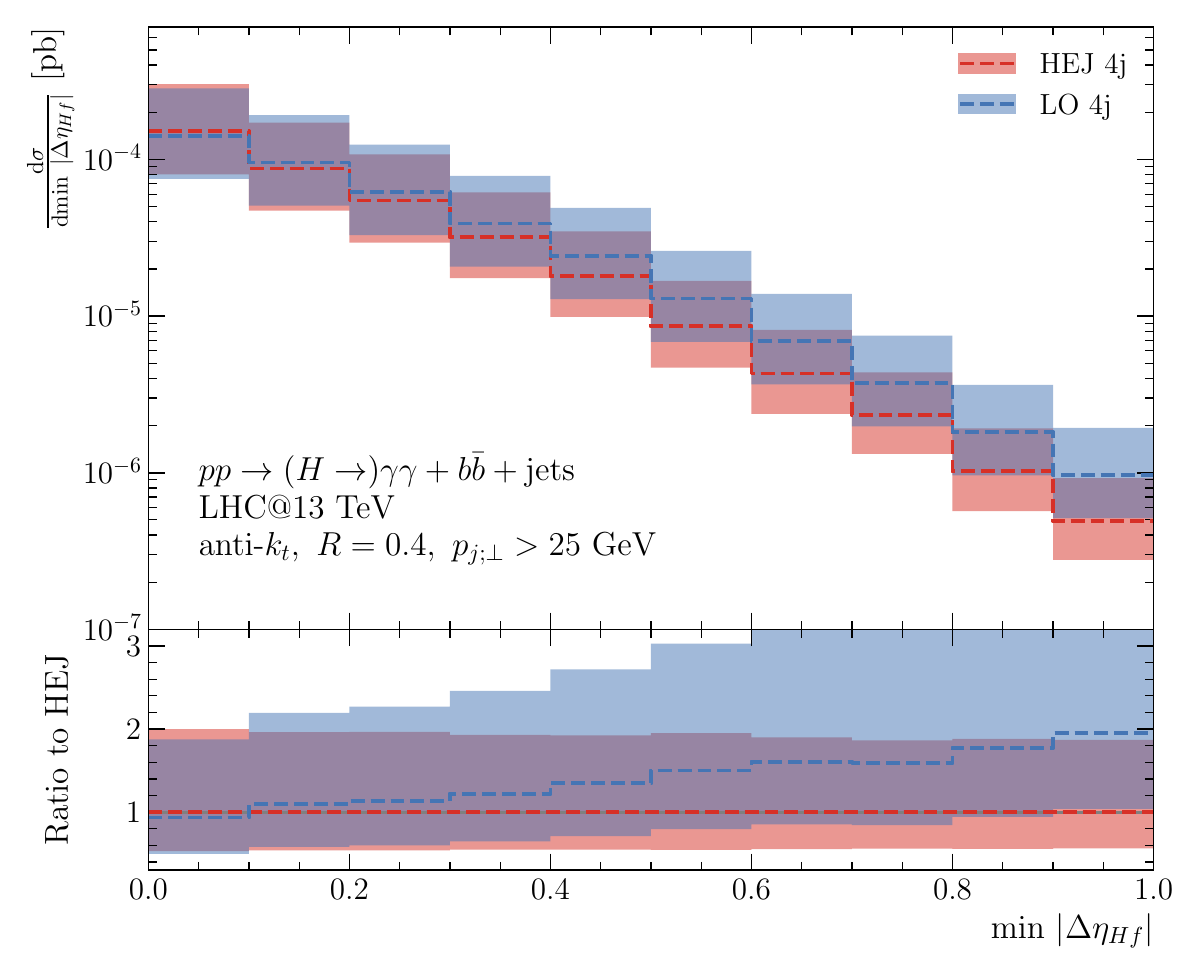}
        \caption{Minimum pseudo-rapidity separation.}
    \label{fig:hh_ggjmindeltaeta}
    \end{subfigure}

\caption{
Comparison between \HEJ and LO for 4-jet inclusive observables sensitive to high-energy effects.
\ref{fig:hh_jjmass}: Invariant mass of the hardest jets.
\ref{fig:hh_ggjmindeltaeta}: Minimum pseudo-rapidity separation between the diphoton system and any other outgoing final-state particle.
}
\label{fig:hh_jj_observables}
\end{figure}

These corrections become sizeable again for observables that probe a broader region of phase space, such as the invariant mass of the two hardest jets, $m_{j_1j_2}$, shown in figure~\ref{fig:hh_jjmass}, where no distinction is made between $b$-jets and VBF-jets.
Unlike the VBF-dijet invariant mass, displayed in figure~\ref{fig:hh_vbfdijetmass}, the observable $m_{j_1j_2}$ also receives contributions from configurations in which one or both of the leading jets are $b$-jets.
As a consequence, the onset of visible differences between the \HEJ\ and LO QCD predictions is shifted towards larger invariant masses, becoming apparent only for $m_{j_1j_2} \gtrsim 1000~\mathrm{GeV}$, compared to $m_{j_1j_2}^{\rm VBF} \gtrsim 500~\mathrm{GeV}$ for the VBF-tagged jet pair.

A similar behaviour is observed for the minimum pseudo-rapidity separation between the diphoton system -- or, equivalently, the reconstructed Higgs boson -- and any other final-state particle, shown in figure~\ref{fig:hh_ggjmindeltaeta}.
As for the invariant mass of the two hardest jets, this observable probes a broad region of phase space and is therefore particularly sensitive to high-energy logarithmic corrections.
In the last two bins, corresponding to configurations in which the Higgs boson is separated by at least $0.8$ units of pseudo-rapidity from every other final-state particle, the LO prediction exceeds the \HEJ\ result by nearly a factor of two.
This corresponds to a relative difference of approximately $100\%$, highlighting the sizeable impact of the high-energy resummation in this region.

To conclude, the inclusion of high-energy resummation effects leads to the largest deviations from the fixed-order prediction precisely in those observables that are expected to be sensitive to large logarithmic corrections, namely observables probing large invariant masses and wide rapidity separations.
In contrast, for observables that are not directly sensitive to the high-energy limit, the \HEJ\ predictions remain remarkably close to the corresponding LO results.
This behaviour further illustrates the perturbative stability of the \HEJ\ framework.
Since the resummation is constructed on top of the fixed-order input, the calculation retains at least leading-order accuracy by construction.
As a consequence, in regions where high-energy logarithms are numerically small, the resummation leaves the fixed-order prediction essentially unchanged, while providing an all-order description in the kinematic regions where such corrections become relevant.

Finally, it should be noted that both the LO and \HEJ\ predictions are affected by sizeable scale uncertainties, reflecting the complexity of the process and the absence of next-to-leading-order calculations for this final state.
In particular, the Born-level contribution already involves a high jet multiplicity and starts at $\mathcal{O}(\alpha_s^6)$, leading to a strong dependence on the renormalisation scale and consequently to comparatively large perturbative uncertainties.
Consequently, the two predictions remain compatible within their respective uncertainty bands across the full set of observables considered.
Nevertheless, the observed deviations between the central predictions exhibit a clear and systematic pattern, becoming increasingly pronounced in regions of phase space that are expected to be sensitive to high-energy logarithms.
Since the statistical uncertainties are negligible compared to the scale variations, this behaviour points to genuine perturbative effects associated with high-energy radiation rather than to numerical fluctuations.
These observations therefore provide strong evidence that logarithmic corrections beyond fixed order play an important role in the high-energy regime and should be incorporated through an all-order treatment.

\subsection{Analysis of the Individual Contributions}\label{subsec:lo4j_individual}

Having established the impact of high-energy resummation on the observables considered in this analysis, it is instructive to examine the individual components contributing to the leading-order cross section.
To this end, we decompose the prediction into resummable and non-resummable contributions, thereby quantifying the fraction of the cross section that is described by the all-order formalism implemented in \HEJ.
This study is motivated by the observation that a non-negligible fraction of the total cross section originates from configurations that do not currently admit an all-order description within \HEJ.
Analysing the relative importance of these components for different observables provides further insight into the phase-space regions where the resummation captures the dominant dynamics and where subleading configurations remain numerically relevant.

Particular attention is devoted to configurations in which the Higgs boson is emitted between the two bottom quarks, which constitute the dominant identified subset of the non-resummable component.
This behaviour is closely related to the event selection, which requires both the diphoton system and the $b\bar{b}$ pair to be produced centrally.
Under these conditions, the Higgs boson can either be emitted outside the bottom-quark pair, leading to configurations that admit an all-order description within \HEJ, or between the two bottom quarks, resulting in configurations that are not currently resummed.
Although the latter are formally subleading and contribute only at next-to-next-to-leading logarithmic accuracy, as discussed in appendix~\ref{app:nnll}, they nevertheless account for a sizeable fraction of the cross section and therefore warrant a dedicated investigation.

\begin{figure}[htbp]
    \centering

    \begin{subfigure}[b]{0.49\textwidth}
        \centering
        \includegraphics[width=\textwidth]{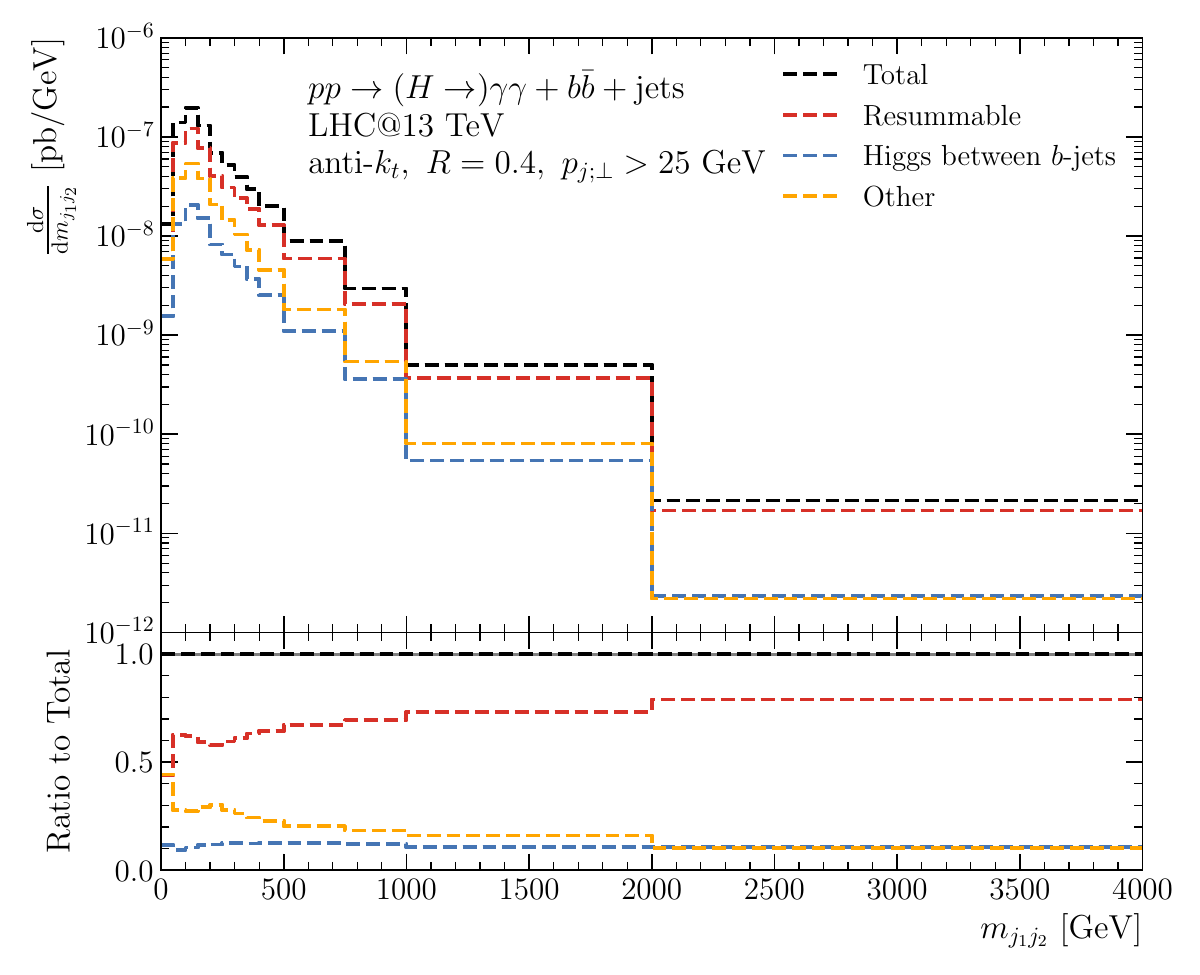}
        \caption{Dijet invariant mass.}
    \label{fig:hh_jjmass_sub}
    \end{subfigure}
    \hfill
    \begin{subfigure}[b]{0.49\textwidth}
        \centering
        \includegraphics[width=\textwidth]{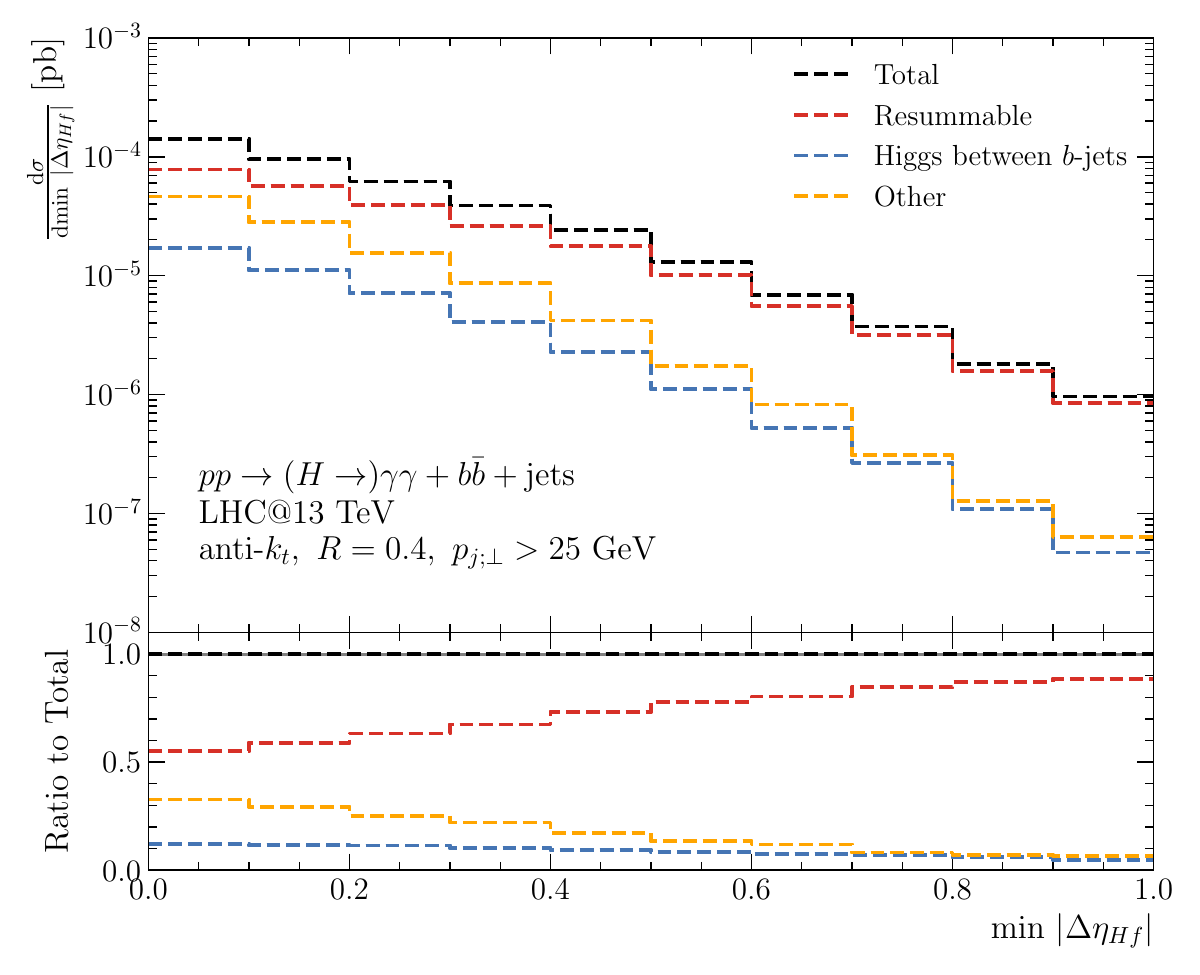}
        \caption{Minimum pseudo-rapidity separation.}
    \label{fig:hh_ggjmindeltaeta_sub}
    \end{subfigure}

\caption{
Individual leading-order contributions to 4-jet inclusive observables sensitive to high-energy effects.
\ref{fig:hh_jjmass_sub}: Invariant mass of the hardest jets.
\ref{fig:hh_ggjmindeltaeta_sub}: Minimum pseudo-rapidity separation between the diphoton system and any other outgoing final-state particle.
The dashed red curve corresponds to the fraction of the cross section that admits an all-order description within the \HEJ\ framework.
The dashed blue and dashed yellow curves instead represent contributions that are only described at fixed order within \HEJ.
In particular, the dashed blue curve denotes the contribution from events in which the Higgs boson is emitted between the two $b$-jets.
Finally, the dashed black curve shows the total leading-order prediction.
}
\label{fig:hh_jj_sub_observables}
\end{figure}

We begin by considering two observables that were already identified in the previous section as particularly sensitive to high-energy corrections, namely the invariant mass of the two hardest jets and the minimum pseudo-rapidity separation between the Higgs boson and any other final-state particle, shown in figures~\ref{fig:hh_jjmass_sub} and~\ref{fig:hh_ggjmindeltaeta_sub}, respectively.
For both observables, the dominant contribution over the entire spectrum originates from configurations that admit an all-order description within \HEJ.
The resummable component increases steadily towards the high-energy region, reaching approximately $80\%$ of the total cross section at large $m_{j_1j_2}$ and up to $90\%$ for large values of $\min |\Delta\eta_{Hf}|$.
Correspondingly, the fraction of the cross section arising from non-resummable configurations decreases as the MRK limit is approached.
These observables therefore provide a particularly clear illustration of the increasing dominance of the logarithmic structures captured by the high-energy resummation in regions characterised by large invariant masses and wide rapidity separations.

\begin{figure}[htbp]
    \centering

    \begin{subfigure}[b]{0.49\textwidth}
        \centering
        \includegraphics[width=\textwidth]{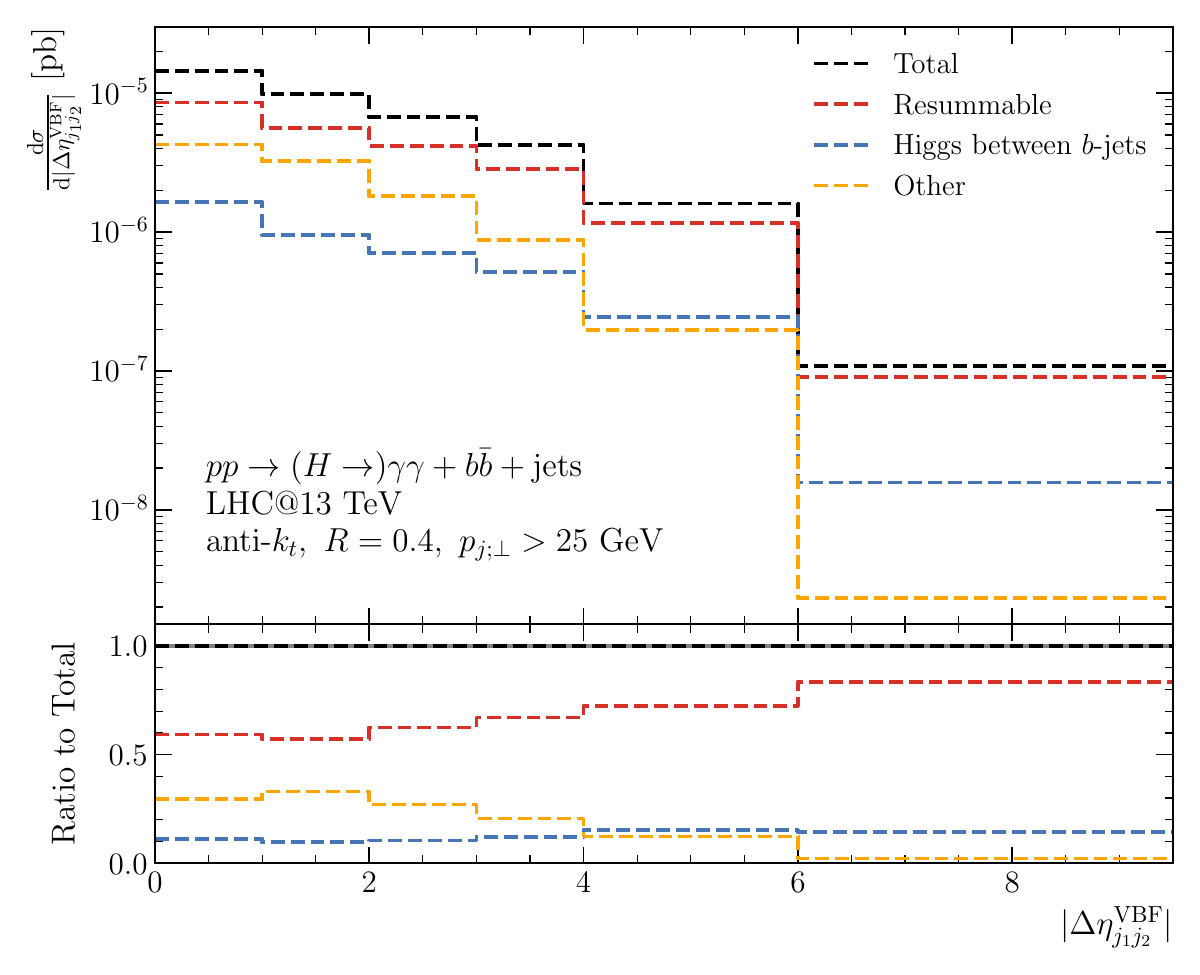}
        \caption{VBF-dijet pseudo-rapidity separation.}
    \label{fig:hh_vbfdeltaeta_sub}
    \end{subfigure}
    \hfill
    \begin{subfigure}[b]{0.49\textwidth}
        \centering
        \includegraphics[width=\textwidth]{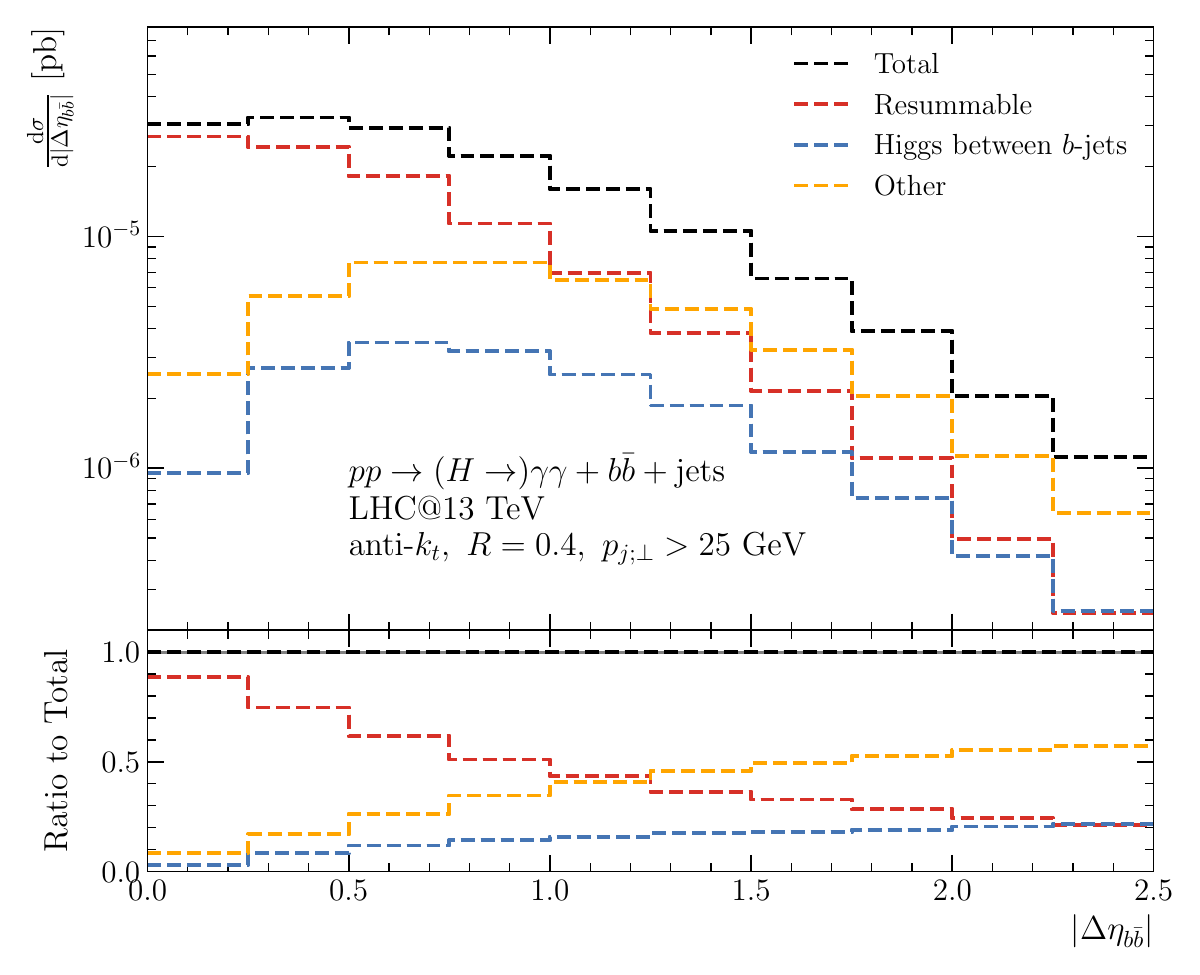}
        \caption{$b$-dijet pseudo-rapidity separation.}
    \label{fig:hh_bbdeltaeta_sub}
    \end{subfigure}

\caption{
Individual leading-order contributions to 4-jet inclusive observables.
\ref{fig:hh_vbfdeltaeta_sub}: Pseudo-rapidity separation between the two VBF jets.
\ref{fig:hh_bbdeltaeta_sub}: Pseudo-rapidity separation between the two b jets.
The curves are as in figure~\ref{fig:hh_jj_sub_observables}.
}
\label{fig:hh_jj_sub_1_observables}
\end{figure}

A similar pattern is observed for the pseudo-rapidity separation between the two VBF-tagged jets, shown in figure~\ref{fig:hh_vbfdeltaeta_sub}.
Also in this case, the resummable component dominates throughout the entire spectrum and exceeds $80\%$ of the total cross section in the last bin.
Two features are particularly noteworthy.
First, the contribution from non-resummable configurations other than those containing a Higgs boson emitted between the two bottom quarks becomes negligible at large rapidity separations.
Consequently, the residual non-resummable component in this region is almost entirely accounted for by Higgs-between-$b\bar{b}$ configurations.
Second, the relative contribution of these events remains approximately constant across the full spectrum, amounting to roughly $10$-$20\%$ of the total cross section.
This behaviour reflects the fact that the VBF dynamics is largely insensitive to the detailed rapidity ordering of the Higgs boson and the central bottom-quark pair, which is also evident from the event selection itself.

A markedly different behaviour is observed for the pseudo-rapidity separation of the bottom-quark pair in figure~\ref{fig:hh_bbdeltaeta_sub}.
In this case, the relative importance of the non-resummable component increases with $|\Delta\eta_{b\bar b}|$, eventually becoming comparable to, and even exceeding, the resummable contribution.
This trend has a simple kinematic origin.
As the rapidity separation between the two bottom quarks increases, configurations in which the Higgs boson is emitted between them become increasingly likely.
At the same time, the relatively loose rapidity requirements imposed on the VBF-tagged jets favour configurations in which an additional jet is also produced within the rapidity interval spanned by the $b$-jets.
Such events belong to the remaining non-resummable component and therefore contribute to the observed enhancement.
The large-$|\Delta\eta_{b\bar b}|$ region thus provides a concrete example of a phase-space domain in which formally subleading configurations acquire substantial numerical importance.

\begin{figure}[htbp]
    \centering

    \begin{subfigure}[b]{0.49\textwidth}
        \centering
        \includegraphics[width=\textwidth]{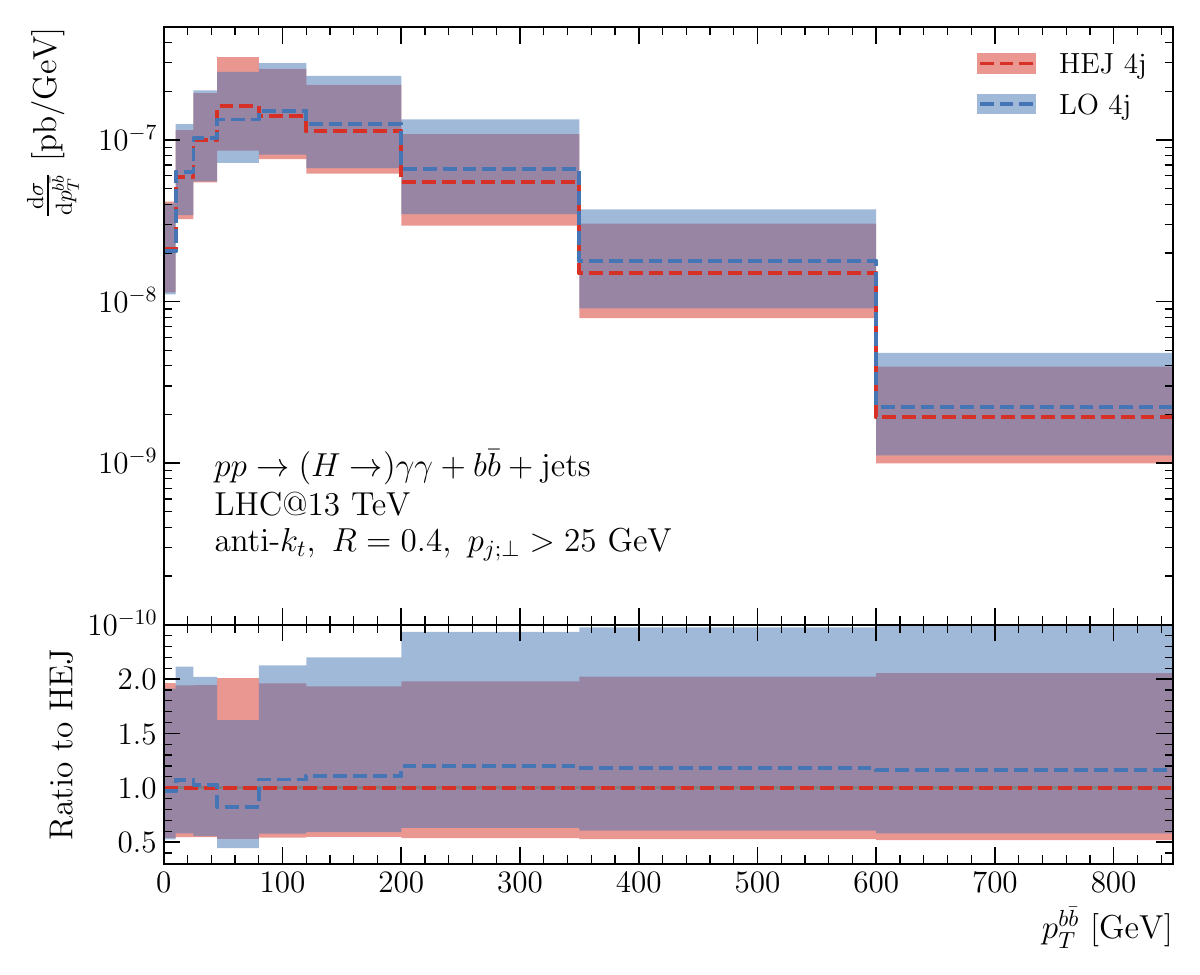}
        \caption{$b$-dijet transverse momentum.}
    \label{fig:hh_bbpt}
    \end{subfigure}
    \hfill
    \begin{subfigure}[b]{0.49\textwidth}
        \centering
        \includegraphics[width=\textwidth]{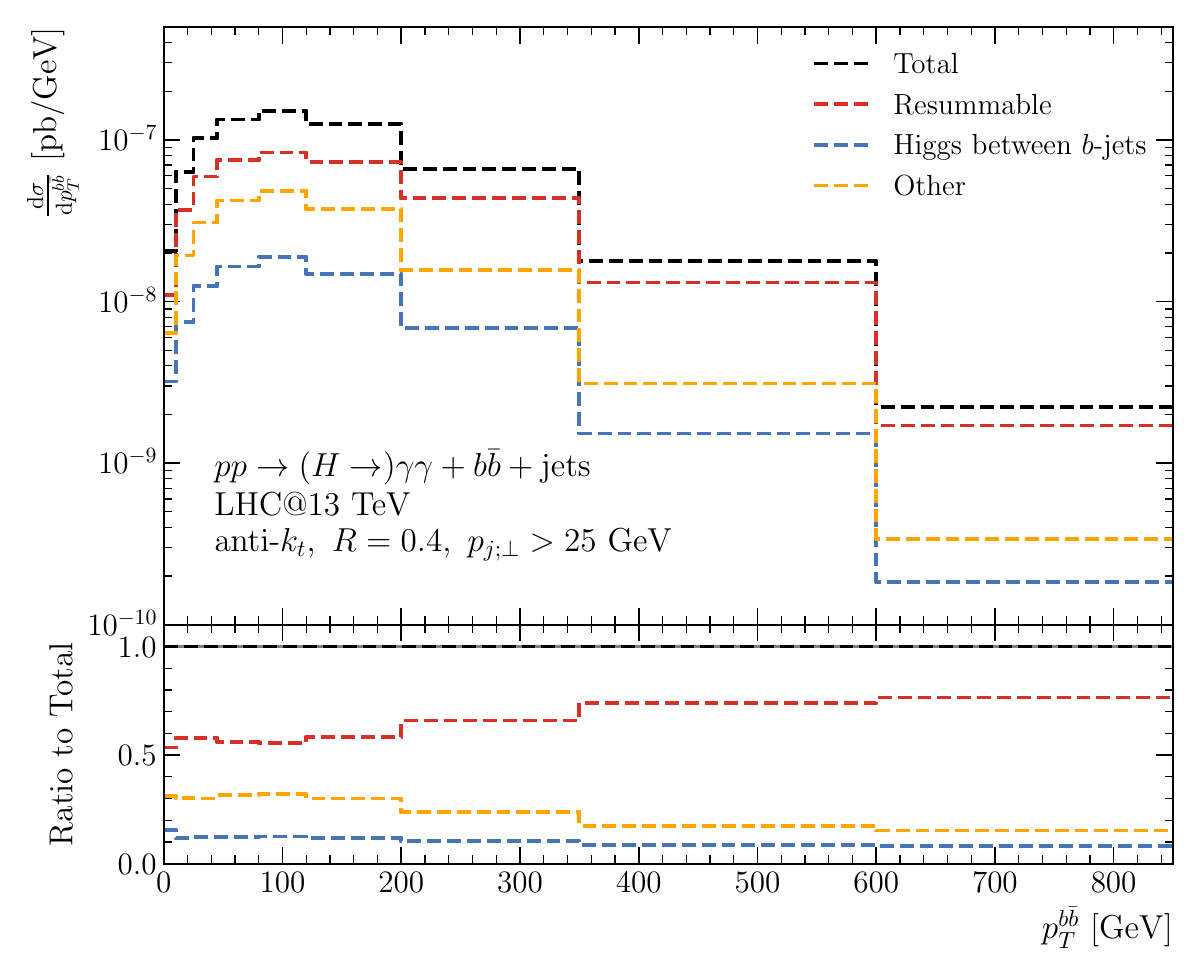}
        \caption{$b$-dijet transverse momentum.}
    \label{fig:hh_bbpt_sub}
    \end{subfigure}

\caption{
On the left panel~\ref{fig:hh_bbpt}, the comparison between \HEJ and LO predictions for the transverse momentum of the $b$-jet candidates.
The panel on the right~\ref{fig:hh_bbpt_sub} instead shows the individual leading-order contributions to the transverse momentum of the $b$-jet candidates.
The curves in the right panel are as in figure~\ref{fig:hh_jj_sub_observables}.
}
\label{fig:hh_jj_sub_2_observables}
\end{figure}

A more detailed picture emerges from figure~\ref{fig:hh_bbpt_sub}, in which the transverse-momentum spectrum of the $b\bar{b}$ system is decomposed into its individual leading-order components.
The fraction of the cross section associated with resummable configurations increases steadily with $p_T^{b\bar{b}}$, reaching approximately $80\%$ in the high-transverse-momentum region.
Nevertheless, a large resummable fraction does not necessarily imply sizeable high-energy corrections.
Indeed, despite the dominance of configurations that admit an all-order description, the corresponding logarithmic contributions remain numerically moderate, resulting in excellent agreement between the \HEJ\ and LO predictions across the entire spectrum, as shown in figure~\ref{fig:hh_bbpt}.

This behaviour provides a further illustration of the perturbative stability of the \HEJ\ framework.
Whenever high-energy logarithms are not numerically significant, the resummation leaves the fixed-order prediction largely unchanged, while retaining at least leading-order accuracy by construction.

\section{Quark Pair Contribution to Higgs Boson plus Jets Production}\label{sec:HJETS}

In this section, we compare predictions for Higgs boson production in association with jets obtained with \HEJ to those from fixed next-to-leading order calculations and to experimental measurements.
The analyses are implemented in \texttt{Rivet}~\cite{Bierlich:2024vqo} and correspond to data collected at the LHC at centre-of-mass energies of 13~TeV~\cite{CMS:2018ctp,CMS:2022wpo} and 8~TeV~\cite{ATLAS:2014yga}.
The \HEJ\ predictions presented in the following extend the previous study~\cite{Andersen:2022zte} through the inclusion of the resummed quark-pair component.
To assess the numerical impact of this contribution, we also perform a dedicated comparison with the \HEJ\ results presented in~\cite{Andersen:2022zte}, as discussed in appendix~\ref{app:qqbar}.
The event generation and resummation setup follows closely that of~\cite{Andersen:2022zte}.
We therefore refer the reader to that work for a detailed description of the setup and technical implementation.

For all the predictions presented in this section, we use the anti-$k_t$ jet-algorithm~\cite{Cacciari:2008gp} implemented in~\texttt{FastJet}~\cite{Cacciari:2011ma} and the NNPDF30@NNLO PDF set~\cite{NNPDF:2014otw} provided by~\texttt{LHAPDF}~\cite{Buckley:2014ana}.
\texttt{Sherpa}~\cite{Sherpa:2019gpd} is used to generate leading-order events for Higgs boson production in association with up to five jets through the matrix-element generators \texttt{Comix}~\cite{Gleisberg:2008fv} and~\texttt{OpenLoops}~\cite{Buccioni:2019sur}.
Exact top-quark mass dependence is included where available (i.e. for $n=1,2$), while for higher jet multiplicities the amplitudes are evaluated in the infinite top-mass approximation.
High-energy resummation is subsequently applied using the \HEJ~2 framework, described in detail in~\cite{Andersen:2018tnm,Andersen:2023kuj}.
In this approach, the fixed-order events serve as input to which the all-order corrections, both real and virtual, are added for each Born-level phase-space configuration.

Finite quark-mass effects are incorporated within the \HEJ\ resummation following the procedure described in~\cite{Andersen:2022zte}.
In particular, finite top-quark mass corrections are evaluated using \texttt{QCDLoop}~\cite{Carrazza:2016gav}, which is interfaced to the \HEJ\ framework.

Concerning the fixed next-to-leading order 1-jet and 2-jet computation, we employ \texttt{Sherpa} and \texttt{OpenLoops} in the infinite top-quark mass limit and without resummation.
We use these results for comparison with \HEJ\ and the experimental data.
To improve the accuracy of the \HEJ\ total cross section to next-to-leading order, we can multiply the \HEJ\ predictions by a flat factor of $\sigma_{\rm NLO}/\sigma_{\HEJ}$.
Specifically, we multiply the \HEJ\ results for the inclusive 1-jet (or~2-jet) distributions by the ratio of the inclusive 1-jet (resp.~2-jet) cross-section at NLO divided by the inclusive 1-jet (resp.~2-jet) cross-section of \HEJ, namely
\begin{equation}\label{eq:HEJ@NLO}
\left( \frac{{\rm d}\sigma}{{\rm d}\mathcal{O}} \right)_{\HEJ {\rm NLO}n{\rm J}} = \frac{\sigma_{\rm NLO n{\rm J}}}{\sigma_{\HEJ n{\rm J}}} \left( \frac{{\rm d}\sigma}{{\rm d}\mathcal{O}} \right)_{\HEJ},
\end{equation}
where $\sigma_{{\rm NLO}n{\rm J}}$ and $\sigma_{\HEJ n {\rm J}}$ with $n = 1, 2$ denote the inclusive NLO and \HEJ\ $n$-jet cross section, respectively.
This multiplication changes the normalisation of distributions, and reduces the scale variation.
The contributions to the \HEJ\ cross section from the exclusive three or more jet component are matched only at Born level.
However, we still rescale them using the ratio in~\eqref{eq:HEJ@NLO}, as they enter the inclusive one- and two-jet observables.

Throughout this section, we employ the central renormalisation and factorisation scale choice discussed in section~\ref{sec:HEJ_HH_bkg} and also used in~\cite{Andersen:2022zte}, see equation~\eqref{eq:scale_choice}.
For the special case of 1-jet events, both scales are always taken to be equal to $m_{\rm H}$, as $m_{12}$ is set to zero.
To provide an estimate of the theory uncertainty due to the scale dependence, a conventional 7-point scale variation with a factor of two is employed.
An alternative scale choice, $\mu_{\rm F} = \mu_{\rm R} = H_{\rm T}/2$, was investigated in~\cite{Andersen:2022zte}.
No significant differences were observed with respect to the central scale choice employed here.

\subsection{Predictions at 8 TeV}\label{sec:8TeV}

In this section, we present the updated \HEJ results for the inclusive and differential cross sections for Higgs boson production in association with jets at centre-of-mass energy $\sqrt{\hat{s}} = 8~\rm{TeV}$.
The predictions are based on the ATLAS analysis~\cite{ATLAS:2014yga} implemented in~\texttt{Rivet}.
The most relevant cuts are summarised in table~\ref{table:8TeV-baseline}, see~\cite{ATLAS:2014yga} for the comprehensive list.
In the experimental study, the Higgs boson candidate is reconstructed from the diphoton decay channel, while jets are clustered using the anti-$k_{t}$ algorithm with a radius parameter of $R = 0.4$.

We first compare the resummed \HEJ\ predictions with those obtained at NLO QCD through \texttt{Sherpa} in the gluon-gluon-fusion (ggF) channel, where jets consists of light quarks and gluons.
Then, we proceed by including the non-ggF contribution from electroweak vector-boson-fusion (VBF), associated production with a vector boson (VH) and associated production with a top quark pair ($t\bar{t}H$).
The overall non-ggF contribution is labelled together as ``HX'', and the value of this component is taken from the experimental paper.
For a meaningful comparison to data, we added it to both \HEJ and NLO QCD predictions, where possible.
This is indicated with ``+HX'' in the legend.
\begin{table}[htbp]
\begin{center}
\begin{tabular}{|l|c|}
 \hline
Description & Baseline cuts  \\
 \hline
Photon transverse momentum & $p_T(\gamma) > 25 \text{ GeV}$  \\
Diphoton invariant mass & $105 \text{ GeV} < m_{\gamma\gamma} < 160 \text{ GeV} $ \\
Pseudo-rapidity of the photons & $|\eta_{\gamma} |<2.37$ \\ & excluding $1.37< |\eta_\gamma| < 1.56$\\
Ratio of harder photon $p_T$ to diphoton invariant mass & $p_T(\gamma_1)/m_{\gamma \gamma} > 0.35 $  \\
Ratio of softer photon $p_T$ to diphoton invariant mass & $p_T(\gamma_2)/m_{\gamma \gamma} > 0.25 $  \\
Photon isolation cut & $\text{Iso}^\gamma_\text{gen} < 14 \text{ GeV}$ \\
\hline
Jet transverse momentum & $p_T(j) > 30 \text{ GeV}$  \\
Jet rapidity & $|y_j| <4.4$  \\
\hline
\end{tabular}
\caption{
Baseline cuts of the 8 TeV analysis, following the ATLAS analysis of~\cite{ATLAS:2014yga}.
$\text{Iso}^\gamma_\text{gen}$ denotes the sum of transverse energies of stable particles in a cone of radius $\Delta R$ = 0.4 around each photon.}
\label{table:8TeV-baseline}
\end{center}
\end{table}
The results presented in the following are separated into 1-jet and 2-jet observables, defined as events containing at least one jet and at least two jets, respectively.
Although the resummation of quark-pair configurations formally contributes only to final states with two or more jets, it can still affect inclusive 1-jet observables.
Therefore, both categories provide sensitivity to its impact.

Overall, the inclusion of the quark-pair resummation produces only moderate changes in the \HEJ\ predictions for the observables considered.
As discussed in appendix~\ref{app:qqbar}, the resulting modifications are typically at the level of a few percent, $\mathcal{O}(1$-$10\%)$, and remain well within the scale-uncertainty bands of the \HEJ\ predictions presented in~\cite{Andersen:2022zte}.
This behaviour is consistent with the formally subleading nature of these contributions and confirms that the phenomenological conclusions of the previous study remain largely unchanged.
At the same time, the improved description further illustrates the perturbative stability of the \HEJ\ framework, where successive logarithmic corrections provide increasingly refined predictions without significantly altering the overall phenomenological picture.
\begin{figure}[htbp]
    \centering

    \begin{subfigure}[b]{0.49\textwidth}
        \centering
        \includegraphics[width=\textwidth]{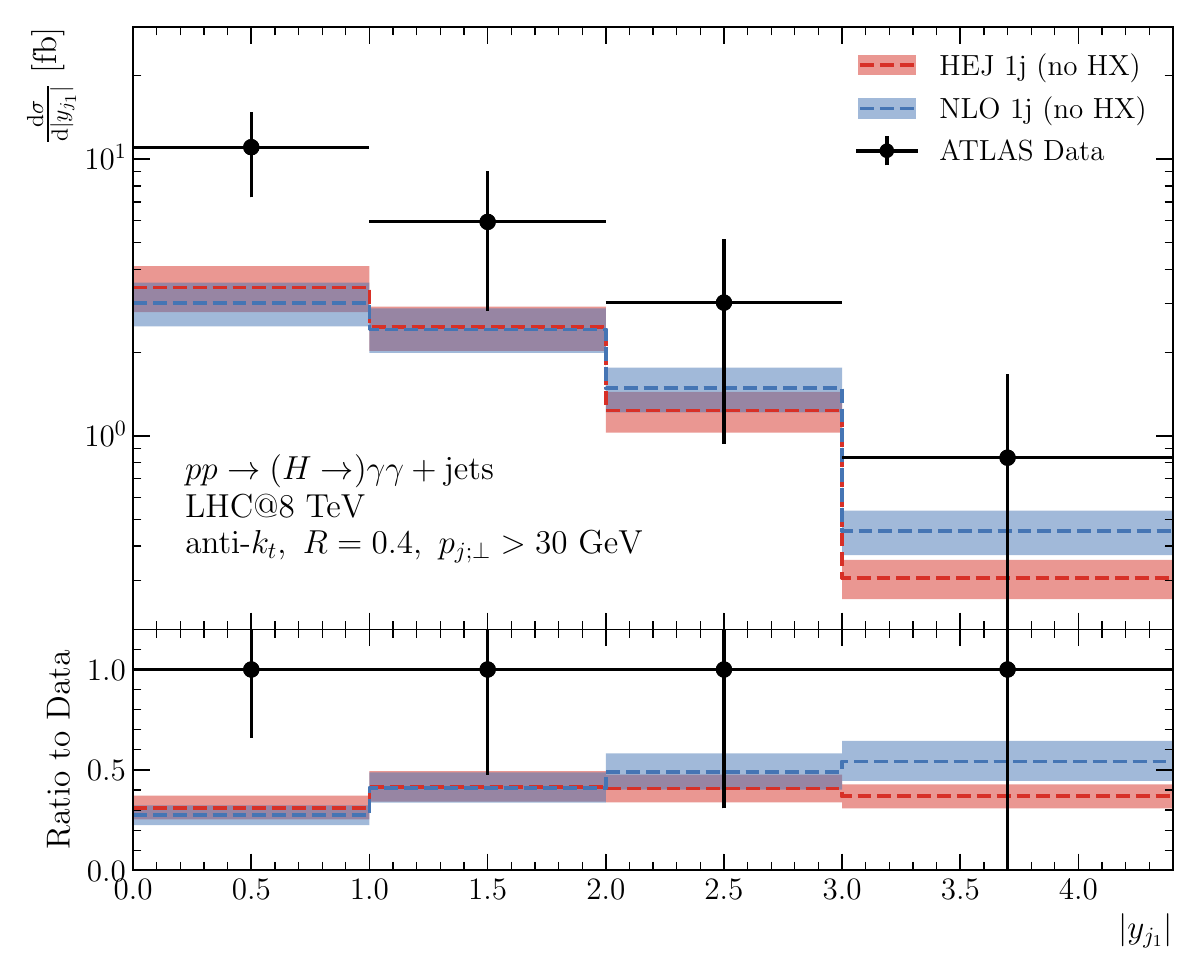}
        \caption{Leading jet rapidity.}
    \label{fig:8TeV_yj1_hejqqx_nlo_1j}
    \end{subfigure}
    \hfill
    \begin{subfigure}[b]{0.49\textwidth}
        \centering
        \includegraphics[width=\textwidth]{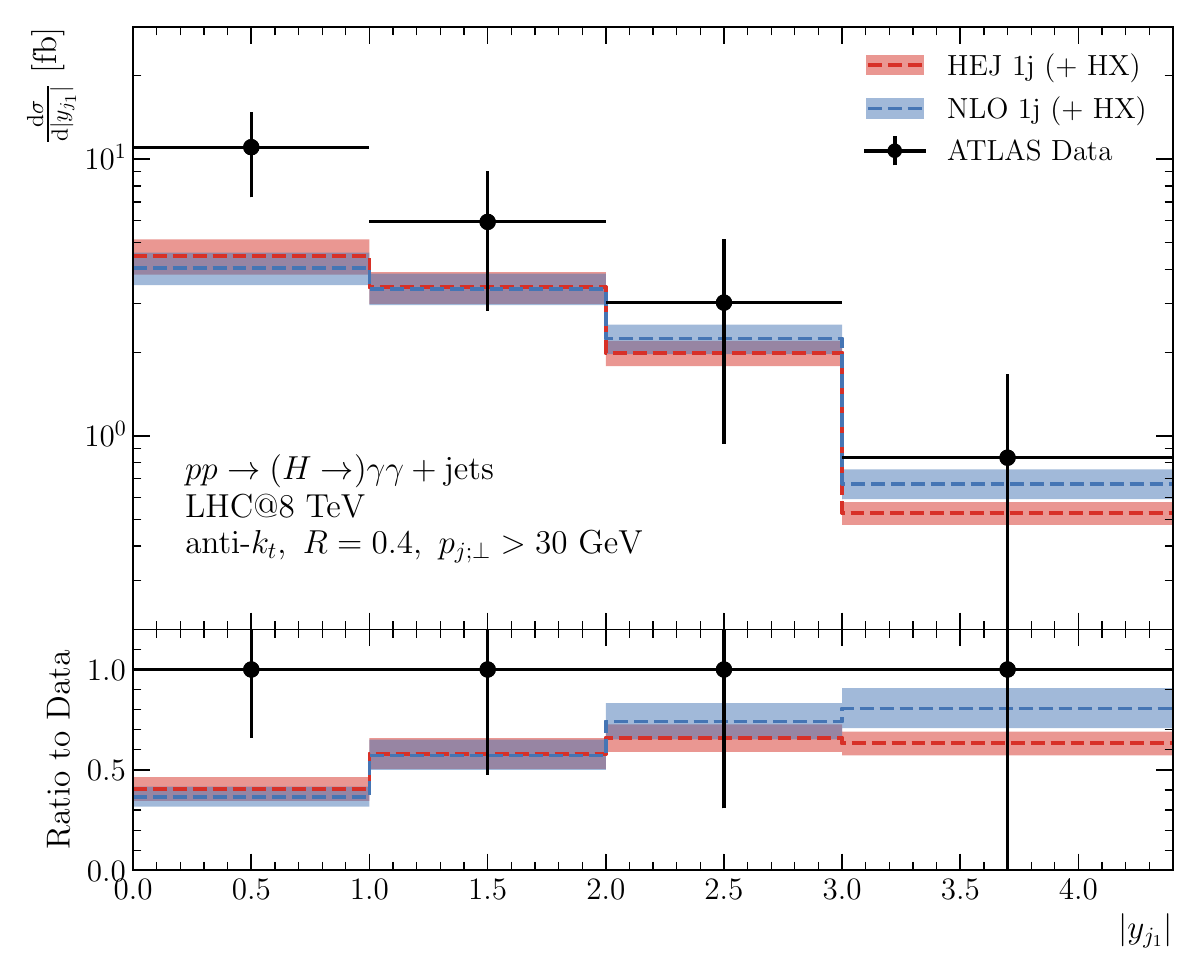}
        \caption{Leading jet rapidity.}
    \label{fig:8TeV_yj1_hejqqx_nlo_1j_with_ewk}
    \end{subfigure}

    \vspace{1em}

     \begin{subfigure}[b]{0.49\textwidth}
        \centering
        \includegraphics[width=\textwidth]{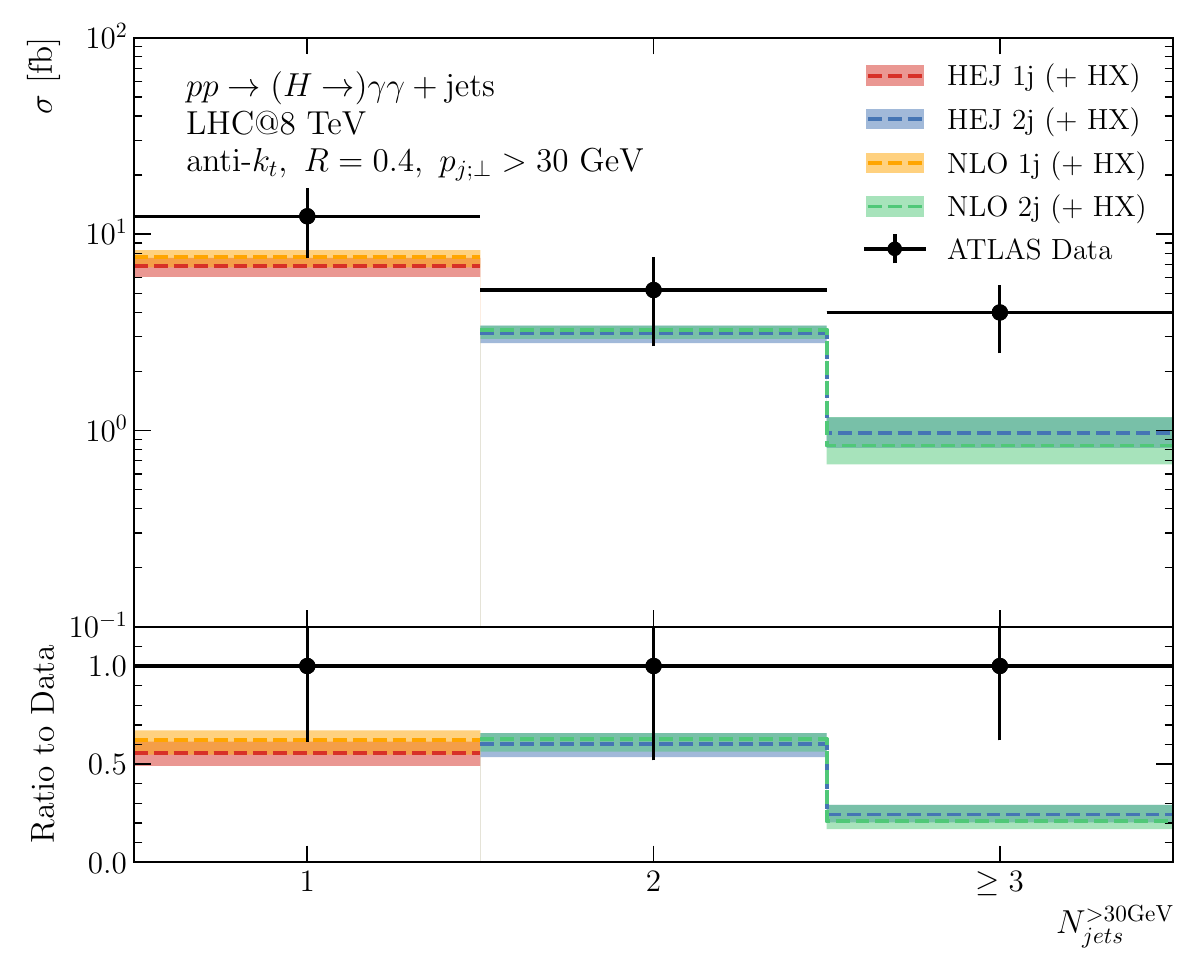}
        \caption{Number of jets.}
    \label{fig:8TeV_Njet_pt30_hejqqx_nlo_with_ewk}
    \end{subfigure}

\caption{
Comparison between \HEJ, NLO and ATLAS data for 1-jet inclusive observables:~\ref{fig:8TeV_yj1_hejqqx_nlo_1j} and~\ref{fig:8TeV_yj1_hejqqx_nlo_1j_with_ewk}.
Comparison between \HEJ, NLO and ATLAS data for the jet multiplicity observable:~\ref{fig:8TeV_Njet_pt30_hejqqx_nlo_with_ewk}.
The first two bins are each exclusive in the jet multiplicity ($N_{jets} = 1, 2$), whereas the last bin is inclusive ($N_{jets} \geq 3$).
The 1-jet \HEJ\ predictions are rescaled by the inclusive cross section ratio $\sigma_{\rm NLO1J}/\sigma_{\HEJ{\rm1J}}$, while the \HEJ\ predictions of the 2 and 3-jet bins of~\ref{fig:8TeV_Njet_pt30_hejqqx_nlo_with_ewk} are rescaled by $\sigma_{{\rm NLO2J}}/\sigma_{\HEJ {\rm2J}}$.
\ref{fig:8TeV_yj1_hejqqx_nlo_1j}: Leading jet rapiditidy in comparison to experimental data without the ``HX'' component.
\ref{fig:8TeV_yj1_hejqqx_nlo_1j_with_ewk}: Leading jet rapiditidy in comparison to experimental data with the ``HX'' component.
\ref{fig:8TeV_Njet_pt30_hejqqx_nlo_with_ewk}: Number of jets in the 1-, 2- and 3-jet bins.
The ``HX'' component and the experimental data are extracted from~\cite{ATLAS:2014yga}.
}
\label{fig:8TeV_hejqqx_vs_nlo_1j_yj1}
\end{figure}
\begin{figure}[htbp]
    \centering

    \begin{subfigure}[b]{0.49\textwidth}
        \centering
        \includegraphics[width=\textwidth]{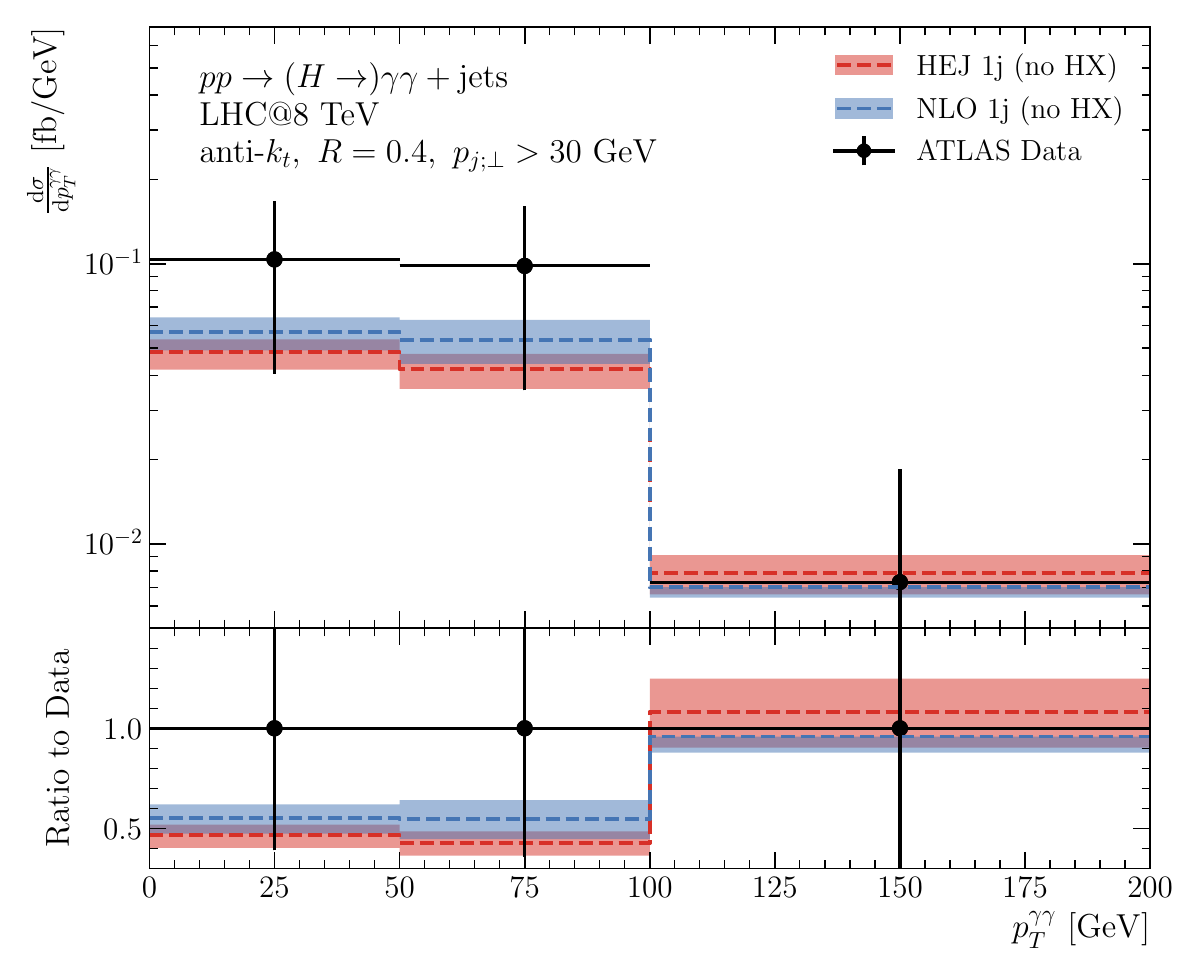}
        \caption{$p_T$ of the diphoton system, $N_{\rm jets} = 1$.}
    \label{fig:8TeV_pTtgg_Njet1_hejqqx_nlo_1j}
    \end{subfigure}
    \hfill
    \begin{subfigure}[b]{0.49\textwidth}
        \centering
        \includegraphics[width=\textwidth]{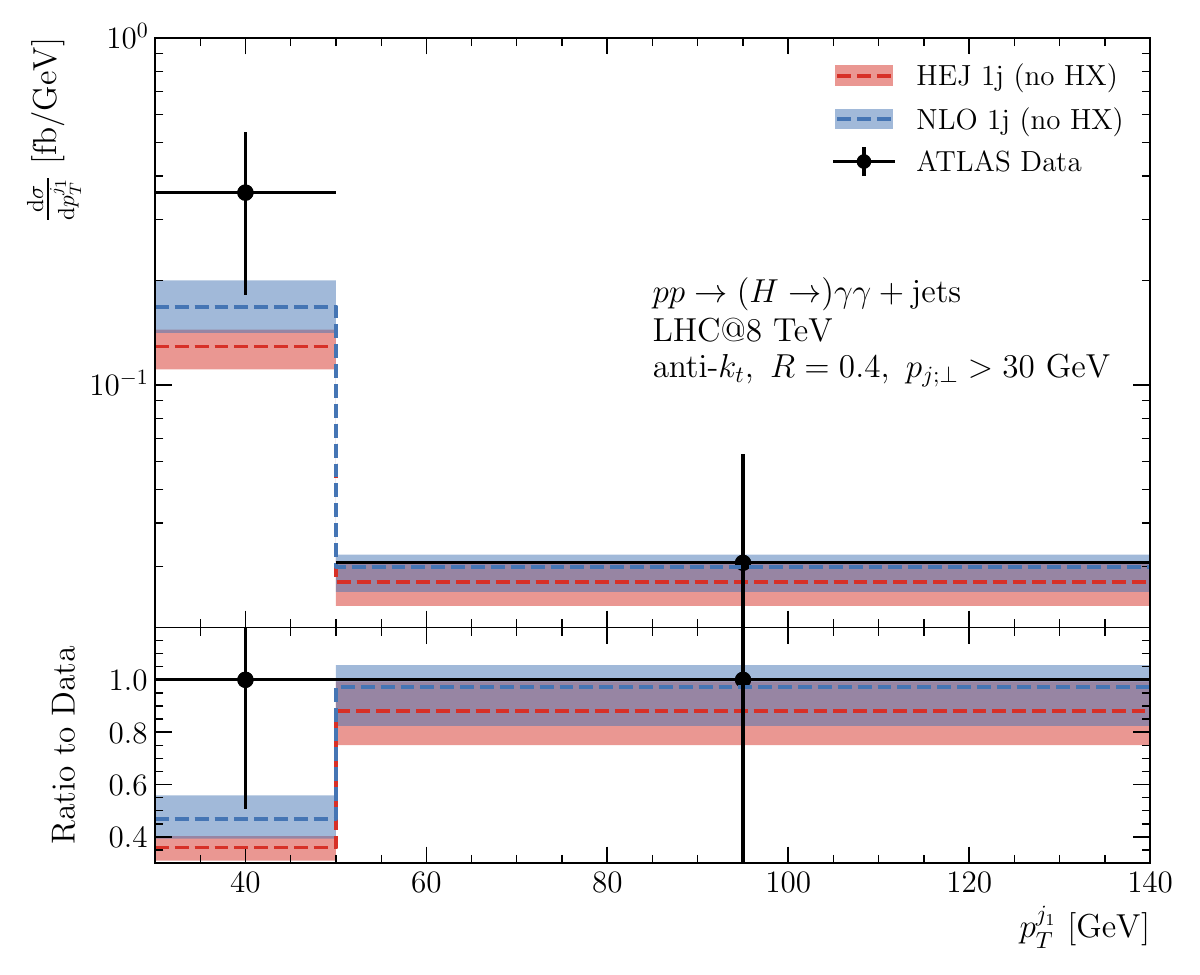}
        \caption{Leading jet $p_T$, $N_{\rm jets} = 1$.}
    \label{fig:8TeV_pTj1_Njet1_hejqqx_nlo_1j}
    \end{subfigure}

\caption{
Comparison between \HEJ, NLO and ATLAS data for 1-jet inclusive observables.
The 1-jet \HEJ\ predictions are rescaled by the inclusive cross section ratio $\sigma_{\rm NLO1J}/\sigma_{\HEJ{\rm1J}}$.
\ref{fig:8TeV_pTtgg_Njet1_hejqqx_nlo_1j}: Higgs boson transverse momentum reconstructed from the diphoton system in the 1-jet bin.
\ref{fig:8TeV_pTj1_Njet1_hejqqx_nlo_1j}: Leading jet transverse momentum in the 1-jet bin.
The ``HX'' component was not available in~\cite{ATLAS:2014yga}.
}
\label{fig:8TeV_hejqqx_vs_nlo_1j_pTggpTj1}
\end{figure}
\begin{figure}[htbp]
    \centering

    \begin{subfigure}[b]{0.49\textwidth}
        \centering
        \includegraphics[width=\textwidth]{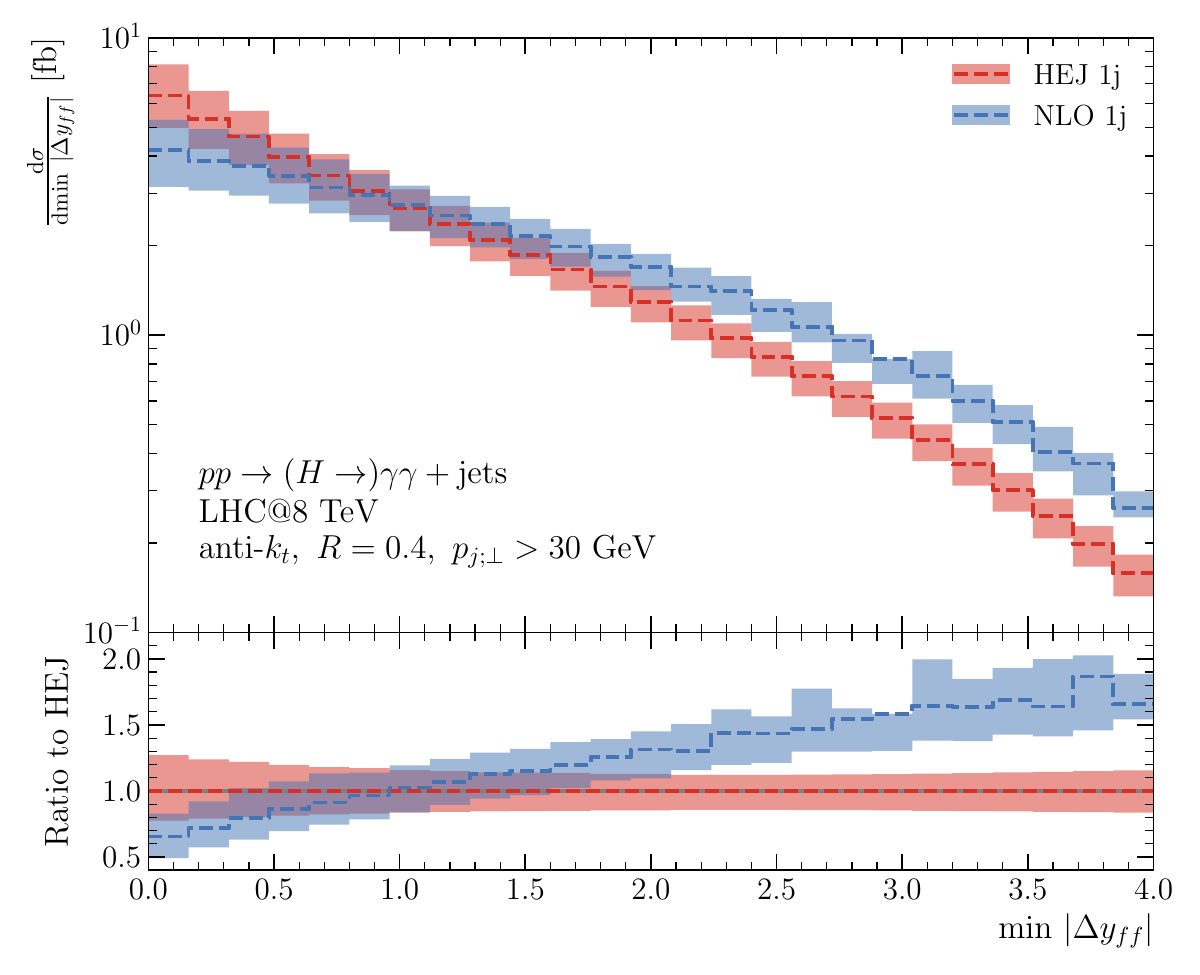}
        \caption{Minimum rapidity separation.}
    \label{fig:8TeV_minDyff_hejqqx_nlo_1j}
    \end{subfigure}
    \hfill
    \begin{subfigure}[b]{0.49\textwidth}
        \centering
        \includegraphics[width=\textwidth]{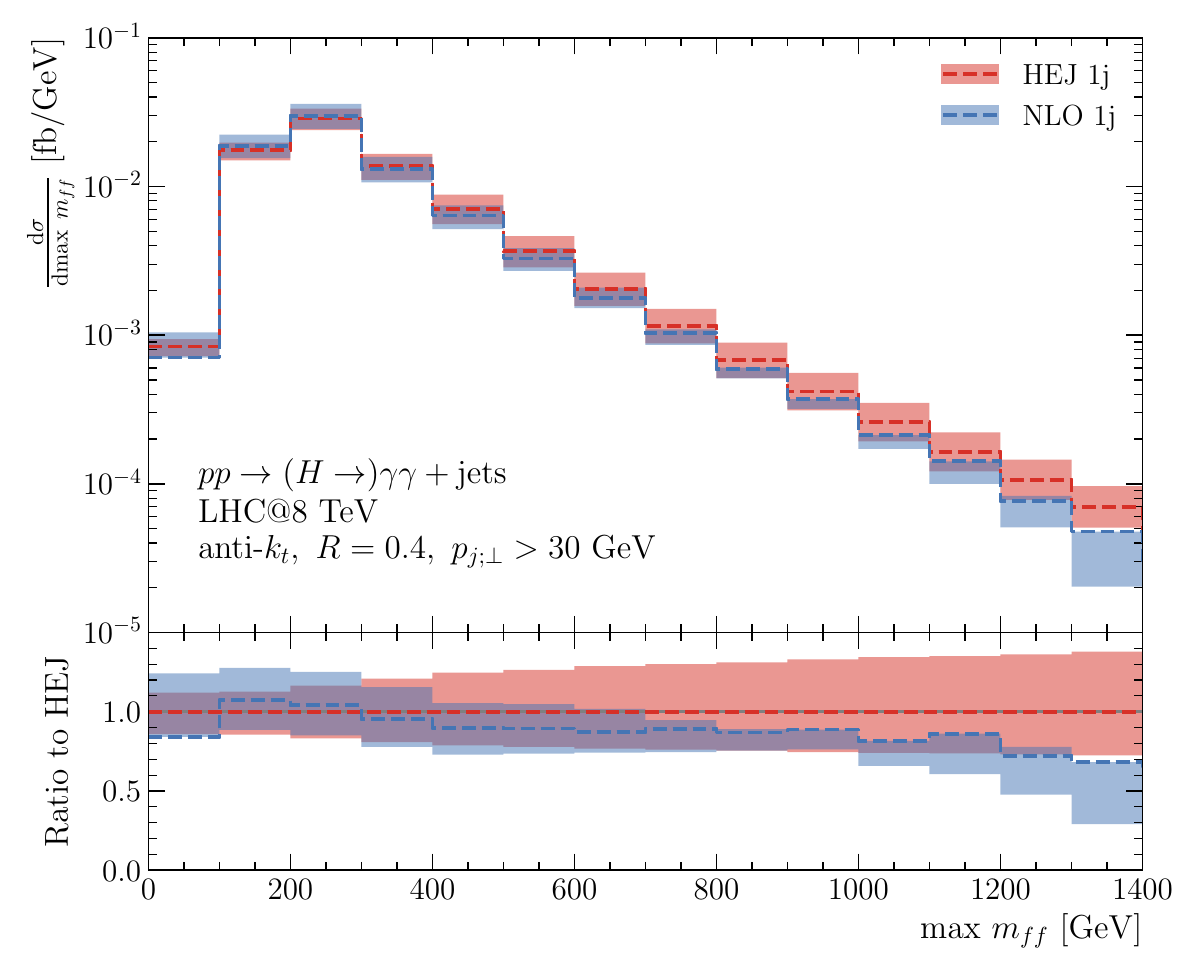}
        \caption{Maximum invariant mass.}
    \label{fig:8TeV_maxmff_long_hejqqx_nlo_1j}
    \end{subfigure}

\caption{
High-energy sensitive 1-jet distributions.
\ref{fig:8TeV_minDyff_hejqqx_nlo_1j}: Minimum rapidity separation between any two outgoing particles (Higgs boson or jets).
\ref{fig:8TeV_maxmff_long_hejqqx_nlo_1j}: Maximum invariant mass between any two outgoing particles (Higgs boson or jets).
\HEJ predictions are rescaled by the inclusive cross section ratio $\sigma_{{\rm NLO1J}}/\sigma_{\HEJ 1{\rm J}}$.
}
\label{fig:8TeV_hejqqx_vs_nlo_1j_minmax}
\end{figure}
\begin{figure}[htbp]
    \centering

    \begin{subfigure}[b]{0.49\textwidth}
        \centering
        \includegraphics[width=\textwidth]{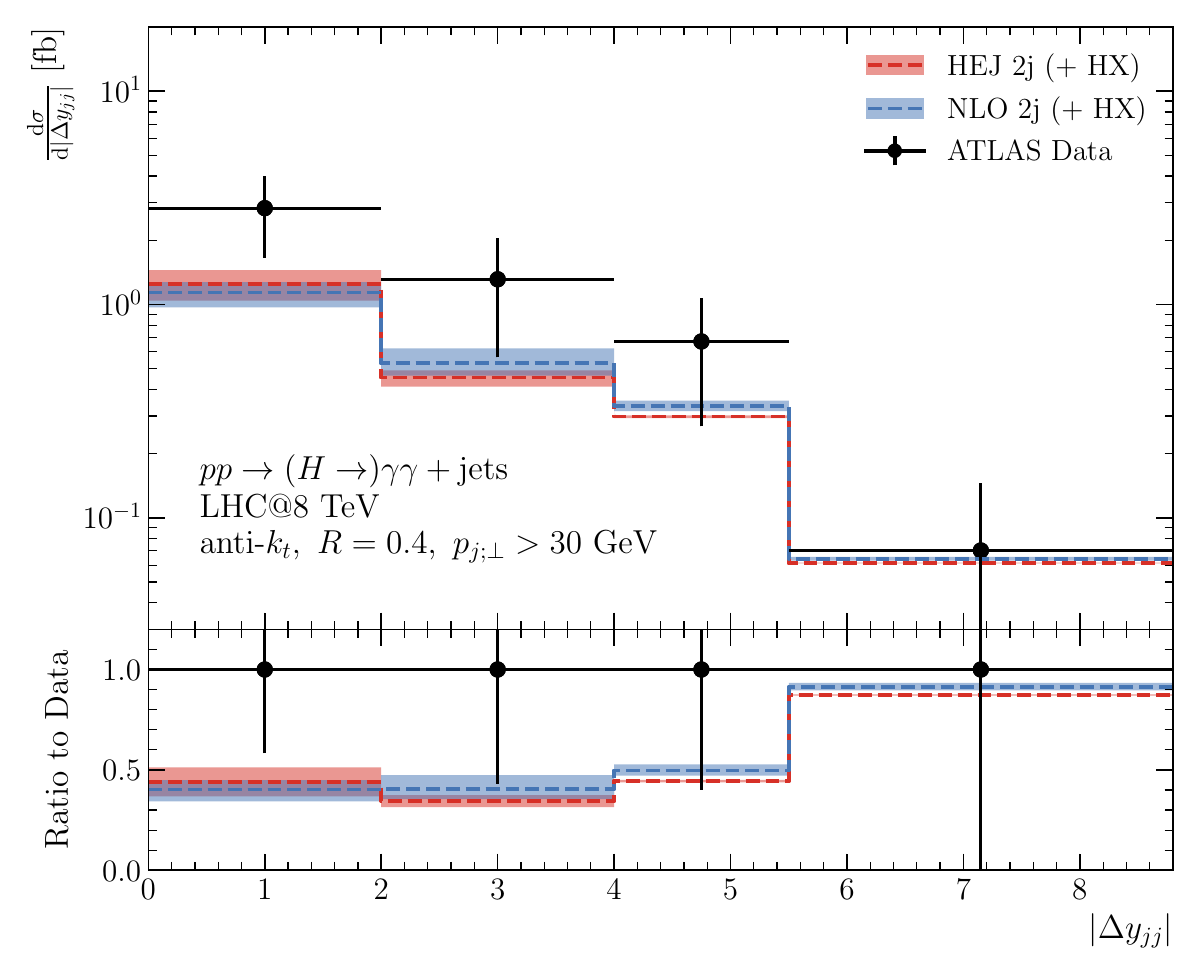}
        \caption{Dijet rapidity separation.}
    \label{fig:8TeV_dyjj_hejqqx_nlo_2j}
    \end{subfigure}
    \hfill
    \begin{subfigure}[b]{0.49\textwidth}
        \centering
        \includegraphics[width=\textwidth]{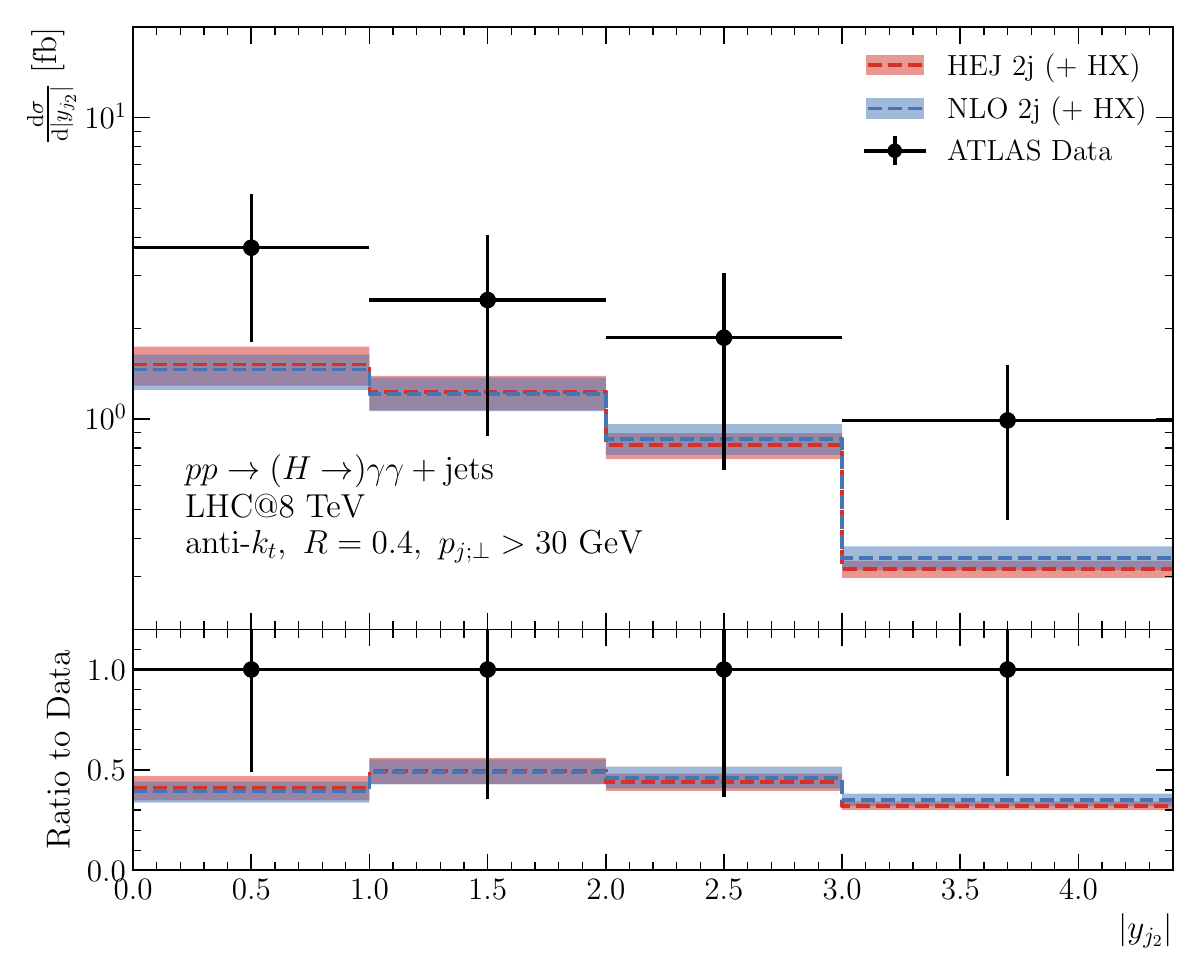}
        \caption{Subleading jet rapidity.}
    \label{fig:8TeV_yj2_hejqqx_nlo_2j}
    \end{subfigure}

    \vspace{1em}

    \begin{subfigure}[b]{0.49\textwidth}
        \centering
        \includegraphics[width=\textwidth]{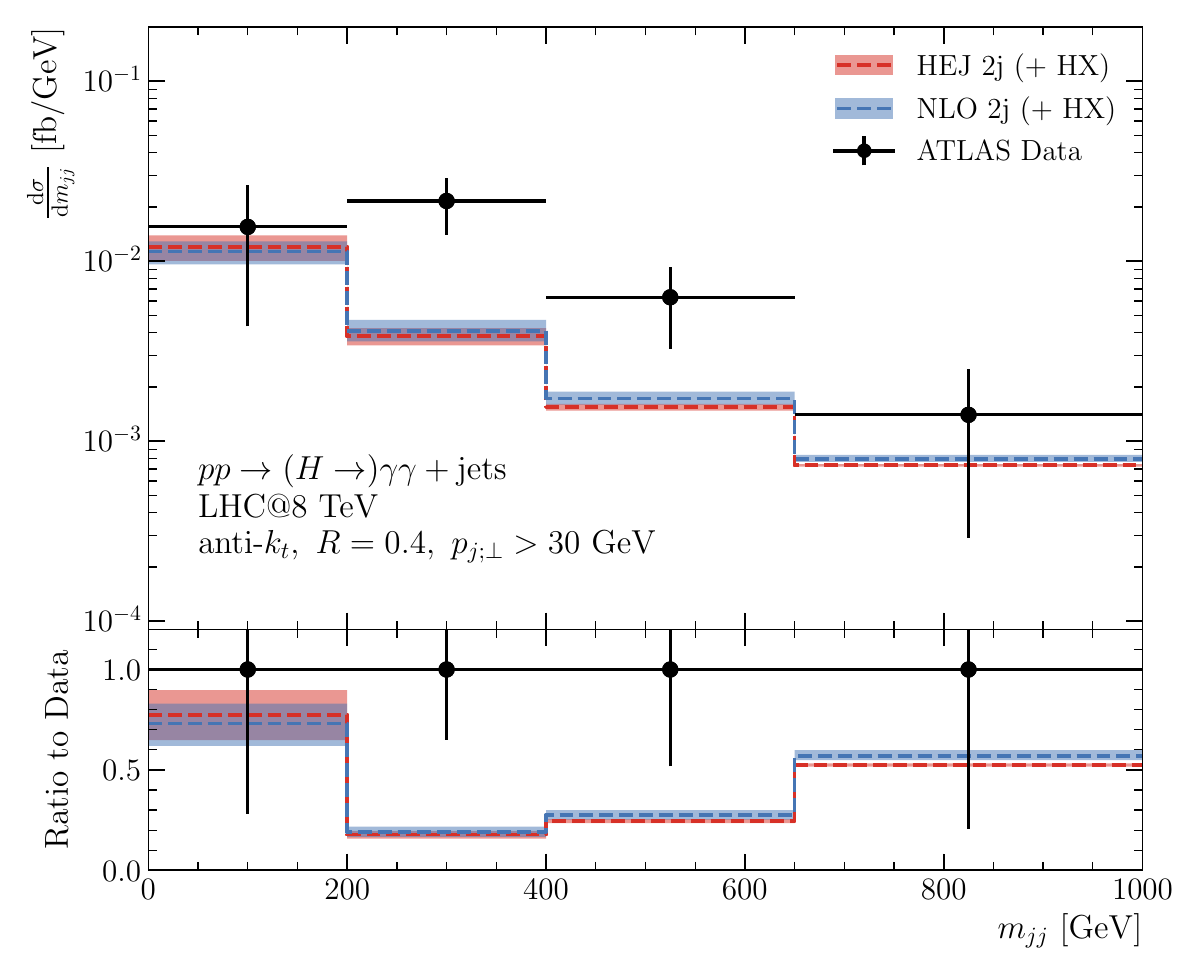}
        \caption{Dijet invariant mass.}
    \label{fig:8TeV_mjj_hejqqx_nlo_2j}
    \end{subfigure}
    \hfill
    \begin{subfigure}[b]{0.49\textwidth}
        \centering
        \includegraphics[width=\textwidth]{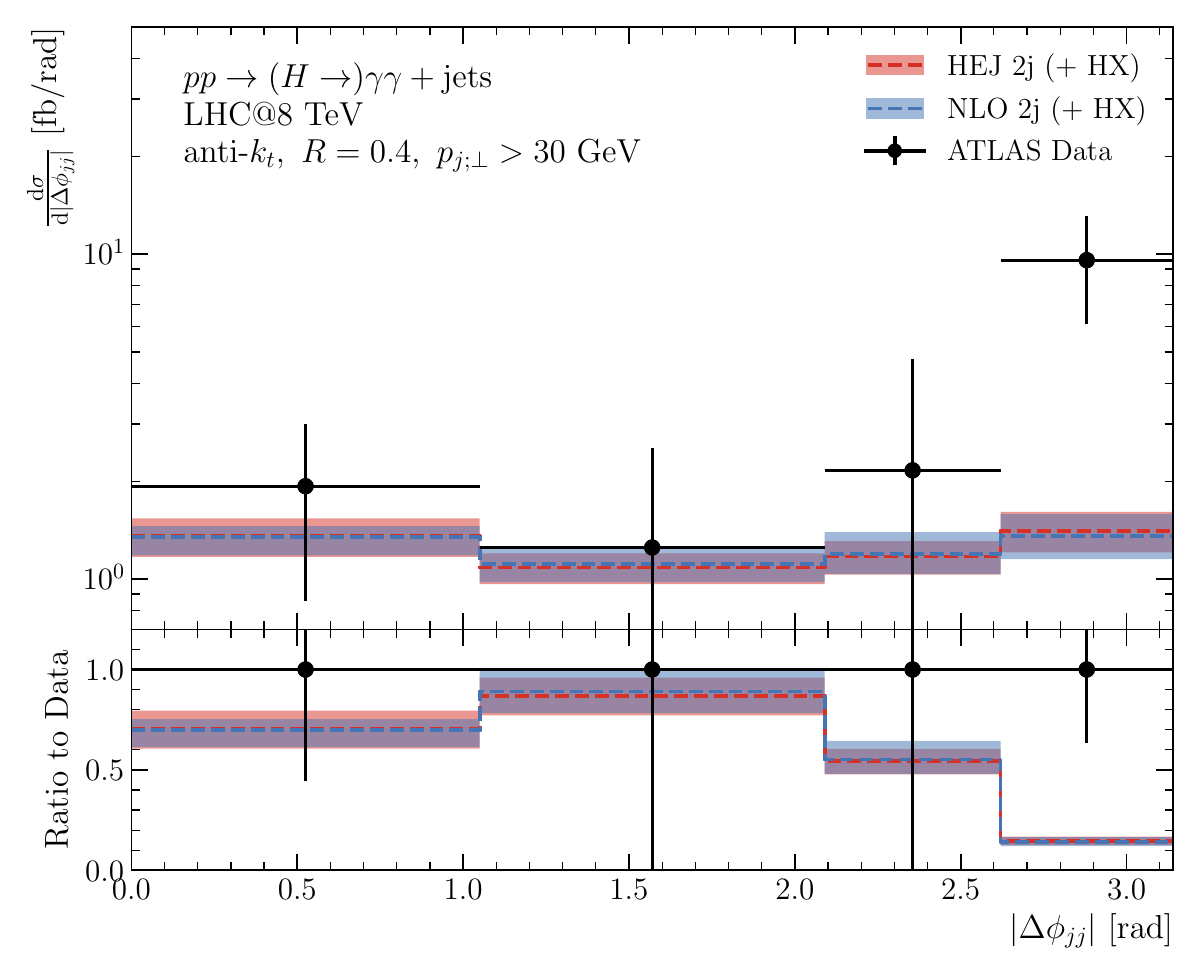}
        \caption{$\Delta \phi$ between the leading 2 jets.}
    \label{fig:8TeV_dphijj_long_hejqqx_nlo_2j}
    \end{subfigure}

\caption{
Comparison between \HEJ, NLO and ATLAS data for 2-jet inclusive observables.
The 2-jet \HEJ\ predictions are rescaled by the inclusive cross section ratio $\sigma_{\rm NLO2J}/\sigma_{\HEJ{\rm2J}}$.
\ref{fig:8TeV_dyjj_hejqqx_nlo_2j}: Dijet rapidity separation.
\ref{fig:8TeV_yj2_hejqqx_nlo_2j}: Subleading jet rapidity.
\ref{fig:8TeV_mjj_hejqqx_nlo_2j}: Dijet invariant mass.
\ref{fig:8TeV_dphijj_long_hejqqx_nlo_2j}: Azimuthal angle difference between the leading 2 jets.
The ``HX'' component is extracted from~\cite{ATLAS:2014yga}.
}
\label{fig:8TeV_hejqqx_vs_nlo_2j_dyjjyj2mjjdphijj}
\end{figure}
\begin{figure}[htbp]
    \centering

    \begin{subfigure}[b]{0.49\textwidth}
        \centering
        \includegraphics[width=\textwidth]{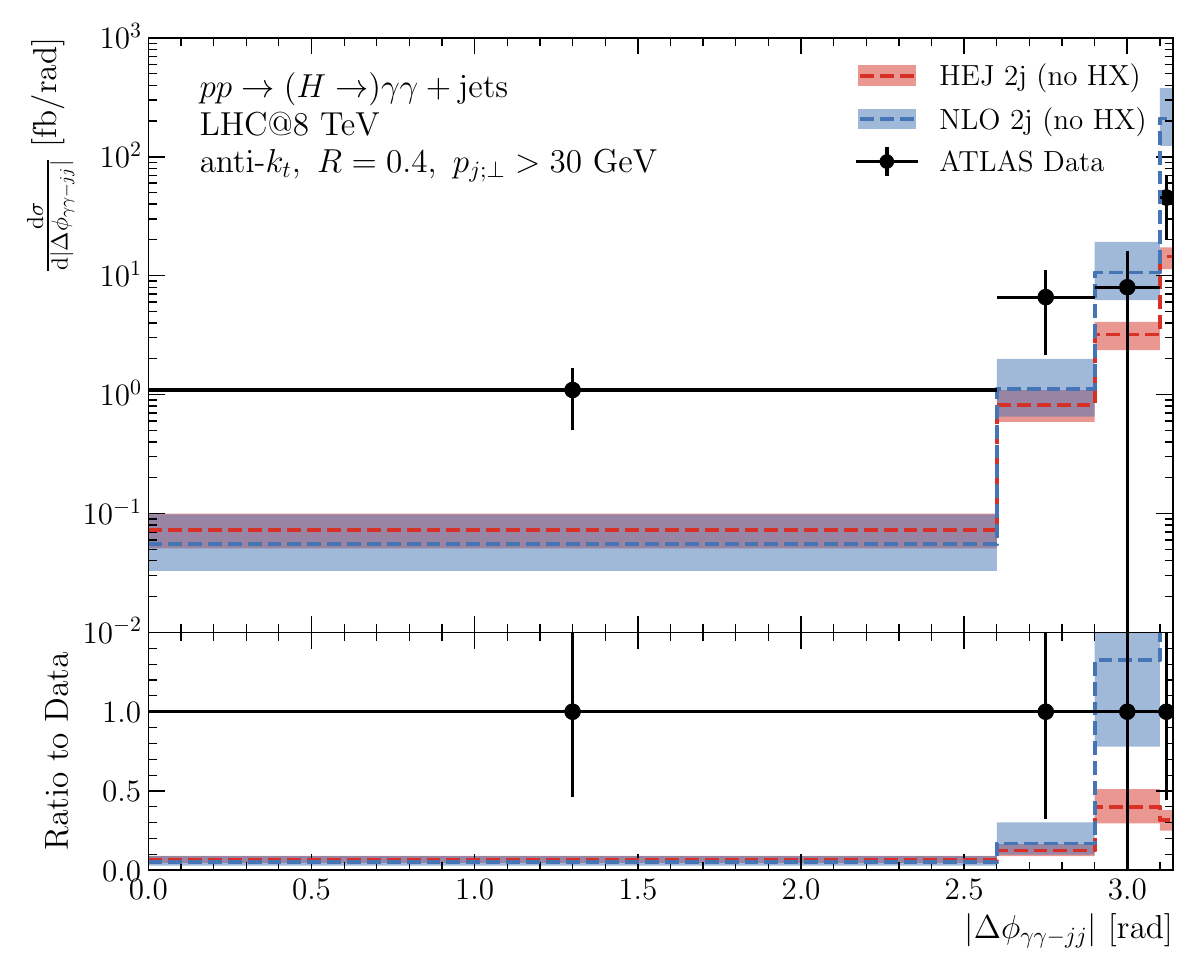}
        \caption{$\Delta \phi$ between dijet and diphoton systems.}
    \label{fig:8TeV_dphi_ggjj_hejqqx_nlo_2j}
    \end{subfigure}
    \hfill
    \begin{subfigure}[b]{0.49\textwidth}
        \centering
        \includegraphics[width=\textwidth]{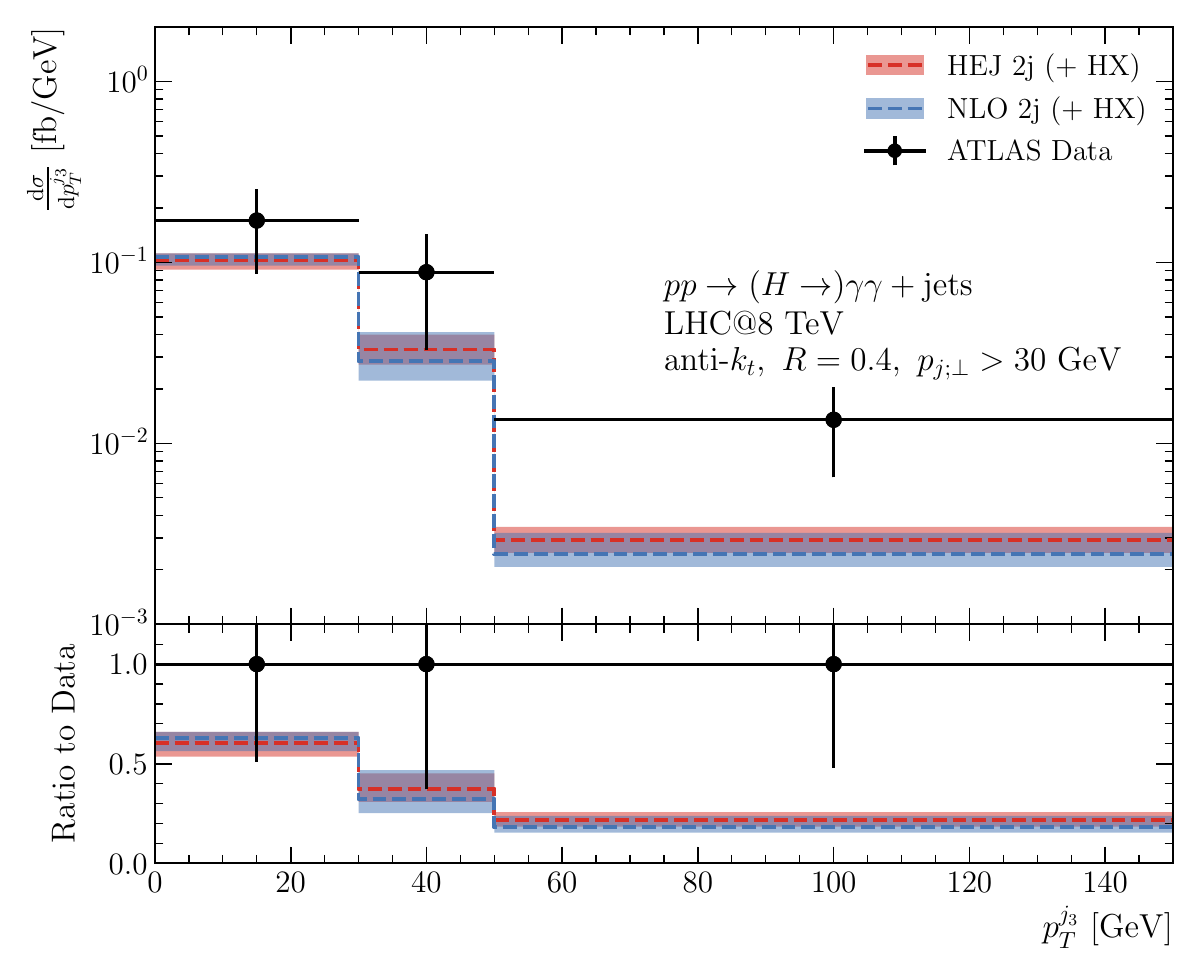}
        \caption{Third-leading jet transverse momentum.}
    \label{fig:8TeV_pTj3_hejqqx_nlo_2j}
    \end{subfigure}

    \vspace{1em}

    \begin{subfigure}[b]{0.49\textwidth}
        \centering
        \includegraphics[width=\textwidth]{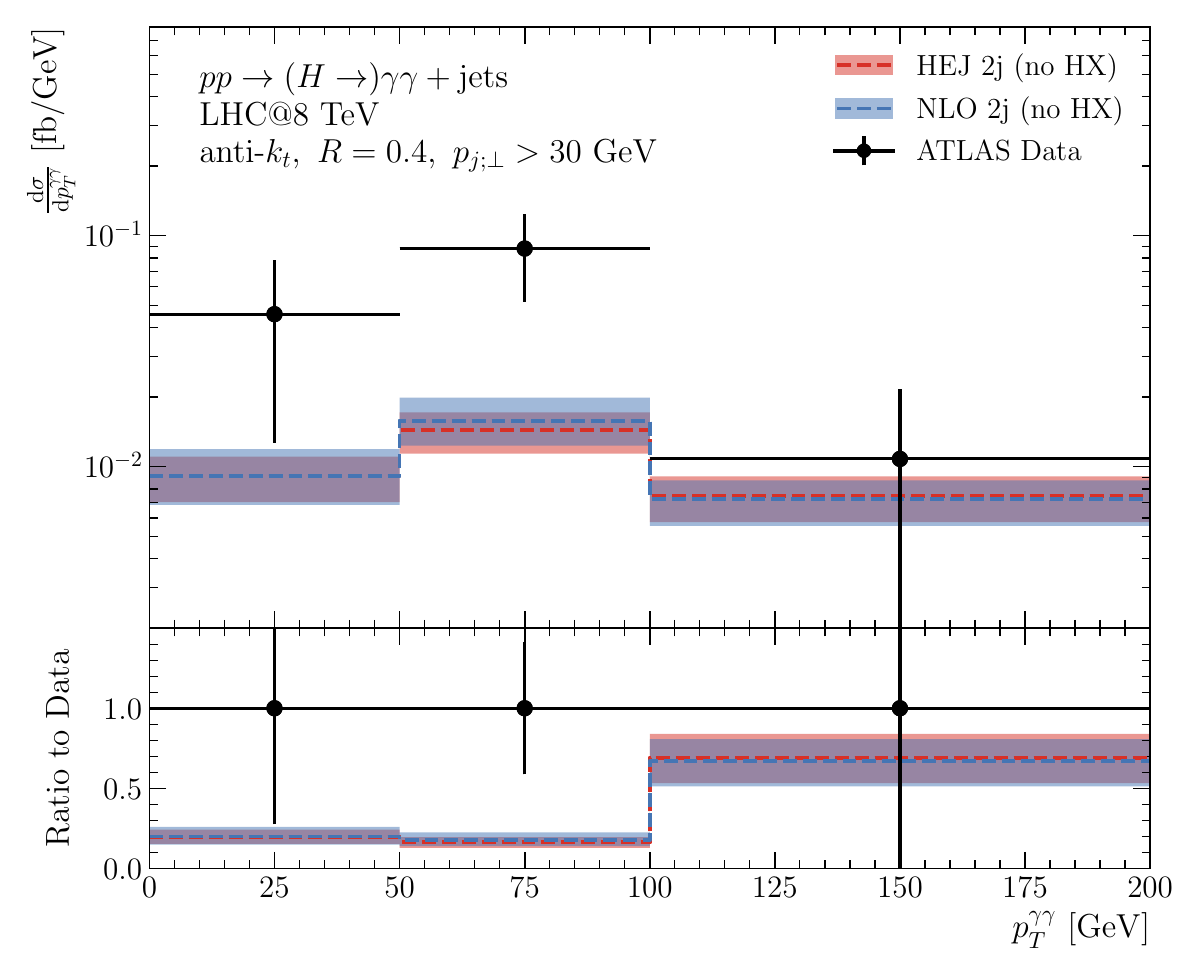}
        \caption{$p_{T}$ of the diphoton system, $N_{{\rm jets}}=2$.}
    \label{fig:8TeV_pTtgg_hejqqx_nlo_2j}
    \end{subfigure}
    \hfill
    \begin{subfigure}[b]{0.49\textwidth}
        \centering
        \includegraphics[width=\textwidth]{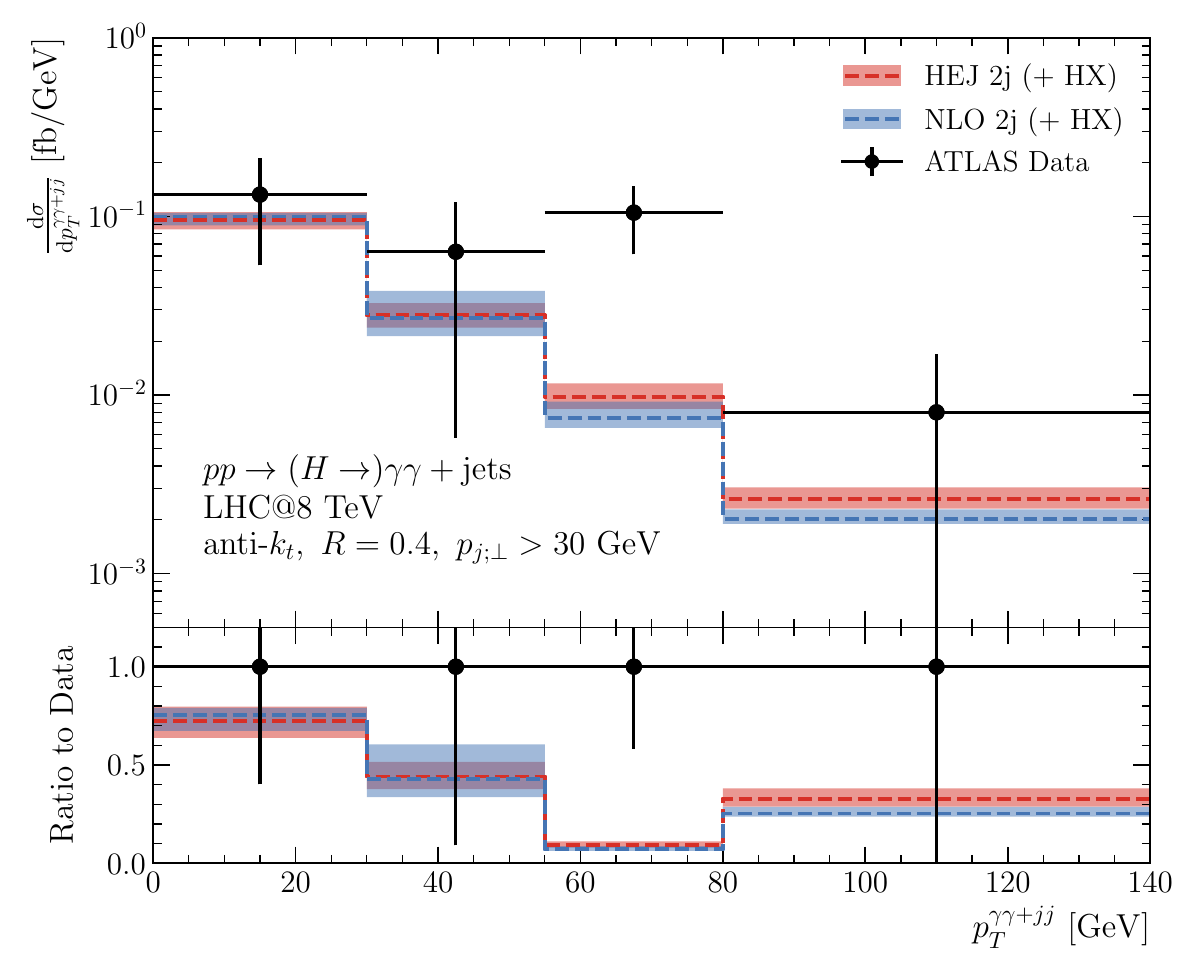}
        \caption{$p_{T}$ of the diphoton-dijet system.}
    \label{fig:8TeV_pTggjj_hejqqx_nlo_2j}
    \end{subfigure}

\caption{
Comparison between \HEJ, NLO and ATLAS data for 2-jet inclusive observables:~\ref{fig:8TeV_dphi_ggjj_hejqqx_nlo_2j},~\ref{fig:8TeV_pTtgg_hejqqx_nlo_2j} and~\ref{fig:8TeV_pTggjj_hejqqx_nlo_2j}.
Comparison between \HEJ, NLO and ATLAS data for the 3-jet inclusive observable:~\ref{fig:8TeV_pTj3_hejqqx_nlo_2j}.
All 2- and 3-jet \HEJ\ predictions are rescaled by the inclusive cross section ratio $\sigma_{\rm NLO2J}/\sigma_{\HEJ{\rm2J}}$.
\ref{fig:8TeV_dphi_ggjj_hejqqx_nlo_2j}: Azimuthal angle difference between dijet and diphoton systems.
\ref{fig:8TeV_pTj3_hejqqx_nlo_2j}: Transverse momentum of the third-leading jet.
\ref{fig:8TeV_pTtgg_hejqqx_nlo_2j}: Higgs boson transverse momentum reconstructed from the diphoton system in the $\geq$ 2-jet bin.
\ref{fig:8TeV_pTggjj_hejqqx_nlo_2j}: Higgs boson transverse momentum reconstructed from the diphoton system plus the dijet system: $(p_{H} + p_{j_1} + p_{j_2})_{\perp}$.
The ``HX'' component is extracted from~\cite{ATLAS:2014yga}.
The latter was not available for~\ref{fig:8TeV_dphi_ggjj_hejqqx_nlo_2j} and~\ref{fig:8TeV_pTtgg_hejqqx_nlo_2j}.
}
\label{fig:8TeV_hejqqx_vs_nlo_2j_dphipTj3pTtgg_pTggjj}
\end{figure}
This consistent behaviour is corroborated by the NLO reweighting factors, $\sigma_{{\rm NLO}n{\rm J}}/\sigma_{\HEJ n{\rm J}}$, defined in equation~\eqref{eq:HEJ@NLO} and reported in table~\ref{table:8TeV_reweighting-factors} for the present 8~TeV ATLAS analysis~\cite{ATLAS:2014yga}.
The inclusive 1-jet and 2-jet cross sections are obtained from the rapidity distributions of the leading and subleading jets, shown in figures~\ref{fig:8TeV_yj1_hejqqx_nlo_1j} and~\ref{fig:8TeV_yj2_hejqqx_nlo_2j}, respectively.

For inclusive 1-jet observables, the next-to-leading logarithmic (NLL) reweighting factors -- labelled UNO and QQX in table~\ref{table:8TeV_reweighting-factors} -- are found to be very close to the leading-logarithmic (LL) result, denoted by FKL, indicating that subleading corrections have only a marginal numerical impact in this case.
Here, FKL refers to the pure LL contribution, UNO includes in addition the first class of NLL corrections arising from unordered emissions, while QQX further incorporates configurations containing the emission of a quark pair.
A detailed discussion of this classification is given in section~\ref{subsec:AMPCLASS}.
In contrast, for 2-jet observables the NLL factors exhibit a more pronounced deviation from the LL result, particularly between the FKL and QQX configurations.
This pattern suggests that quark-pair resummation effects, while still moderate, become increasingly relevant in more exclusive final states, providing a controlled but non-negligible correction to single Higgs production in association with jets.
\begin{table}[hbtp]
\begin{center}
\begin{tabular}{lccc}
\hline
Analysis  & \multicolumn{3}{l}{8 TeV} \\
\hline
Scale     & $\mu_{\rm F},\mu_{\rm R}$  & $(\mu_{\rm F},\mu_{\rm R})/2$  &  $2(\mu_{\rm F},\mu_{\rm R})$   \\
\hline
1j (FKL) factor & 1.85 & 1.51 & 2.13  \\
1j (UNO) factor & 1.86 & 1.53 & 2.14  \\
1j (QQX) factor & 1.87 & 1.54 & 2.15  \\
\hline
2j (FKL) factor & 1.90 & 1.41 & 2.30 \\
2j (UNO) factor & 1.96 & 1.46 & 2.36 \\
2j (QQX) factor & 2.00 & 1.50 & 2.41 \\
\hline
\end{tabular}
\caption{
NLO reweighting factors for the 8~TeV ATLAS analysis~\cite{ATLAS:2014yga} with $\mu_{\rm F} = \mu_{\rm R} = \max(m_{12},m_{\rm H})$.
The labels FKL, UNO and QQX denote the LL contribution, the LL contribution supplemented by NLL corrections from unordered emissions, and the latter further supplemented by NLL corrections from quark-pair configurations, respectively.
}
\label{table:8TeV_reweighting-factors}
\end{center}
\end{table}

\subsection{Predictions at 13 TeV}\label{sec:13TeV}

In this section, we present updated predictions for a CMS analysis~\cite{CMS:2018ctp,CMS:2022wpo} at a centre-of-mass energy of $\sqrt{\hat{s}} = 13~\mathrm{TeV}$.
In addition, we include high-energy sensitive distributions to highlight the differences between \HEJ\ and fixed-order predictions at NLO.

The experimental study considers observables for Higgs boson production both inclusively and in association with one jet, with the Higgs decaying into the di-photon channel.
The baseline selection criteria for photon and jet identification are listed in table~\ref{table:13TeV-baseline}, while the jet pseudo-rapidity cuts are observable-dependent and provided in table~\ref{table:13TeV-baseline-observables}.
Jets are reconstructed using the anti-$k_{t}$ jet algorithm with a radius parameter of $R = 0.4$.
\begin{table}[hbtp]
\begin{center}
\begin{tabular}{ |l|c| }
 \hline
Description & Baseline cuts  \\
 \hline
 Leading photon transverse momentum & $p_T(\gamma_1) > 30 \text{ GeV}$  \\
 Subleading photon transverse momentum & $p_T(\gamma_2) > 18 \text{ GeV}$  \\
 Diphoton invariant mass & $m_{\gamma\gamma} > 90 \text{ GeV} $ \\
 Pseudo-rapidity of the photons & $|\eta_{\gamma} |<2.5$ \\ & excluding $1.4442< |\eta_\gamma| < 1.566$ \\
 Ratio of harder photon $p_T$ to diphoton invariant mass & $p_T(\gamma_1)/m_{\gamma \gamma} > \frac{1}{3} $  \\
 Ratio of softer photon $p_T$ to diphoton invariant mass & $p_T(\gamma_2)/m_{\gamma \gamma} > \frac{1}{4} $  \\
 Photon isolation cut & $\text{Iso}^\gamma_\text{gen} < 10 \text{ GeV}$ \\
 Jet transverse momentum & $p_T(j) > 30 \text{ GeV}$  \\
 \hline
\end{tabular}
\caption{
Baseline photon and jet cuts of the 13 TeV analysis, following the CMS analysis of~\cite{CMS:2018ctp,CMS:2022wpo}.
$\text{Iso}^\gamma_\text{gen}$ denotes the sum of transverse energies of stable particles in a cone of radius $\Delta R$ = 0.3 around each photon.
}
\label{table:13TeV-baseline}
\end{center}
\end{table}
\begin{table}[hbtp]
\begin{center}
\begin{tabular}{ |l|l| }
 \hline
 Observable & Pseudo-rapidity jet cut \\
 \hline
 Number of jets $N_{jets}$, figure~\ref{fig:13TeV_Njet_pt30_hejqqx_nlo_with_ewk} & $|\eta_j| <2.5$ (all jets)  \\
 $p_{T}^{j_1}$, figures~\ref{fig:13TeV_pTj1_hejqqx_nlo_1j}-\ref{fig:13TeV_pTj1_hejqqx_nlo_1j_with_ewk} & $|\eta_{j_1}| <2.5$ and $|\eta_j| <4.7$ (other jets)  \\
 $|y_{j_1}|$, figure~\ref{fig:13TeV_yj1_hejqqx_nlo_1j} & $|\eta_{j_1}| <2.5$ and $|\eta_j| <4.7$ (other jets)  \\
 $\min |\Delta y_{ff}|$, figure~\ref{fig:13TeV_minDyff_hejqqx_nlo_1j} & $|\eta_j| <4.7$ (all jets)  \\
 $\max m_{ff}$, figure~\ref{fig:13TeV_maxmff_hejqqx_nlo_1j} & $|\eta_j| <4.7$ (all jets)  \\
\hline
\end{tabular}
\caption{
Pseudo-rapidity jet cuts used for the 13 TeV analysis observables presented in this section, following the CMS analysis of~\cite{CMS:2018ctp,CMS:2022wpo}.
It is worth noting that $j_1$ denotes the hardest jet among the \textit{central} jets in a given event.
Consequently, for the $p_{T}^{j_1}$ and $|y_{j_1}|$ distributions, $j_1$ does not necessarily correspond to the overall hardest jet in the event.
}
\label{table:13TeV-baseline-observables}
\end{center}
\end{table}
Analogously to the 8~TeV analysis discussed in section~\ref{sec:8TeV}, both \HEJ\ and fixed-order predictions describe $pp \to H (\to \gamma \gamma) + n$-jet production exclusively via gluon fusion (ggF).
Contributions beyond ggF are collectively denoted as ``HX'' and are extracted from the experimental publications wherever available.

\begin{figure}[htbp]
    \centering

    \begin{subfigure}[b]{0.49\textwidth}
        \centering
        \includegraphics[width=\textwidth]{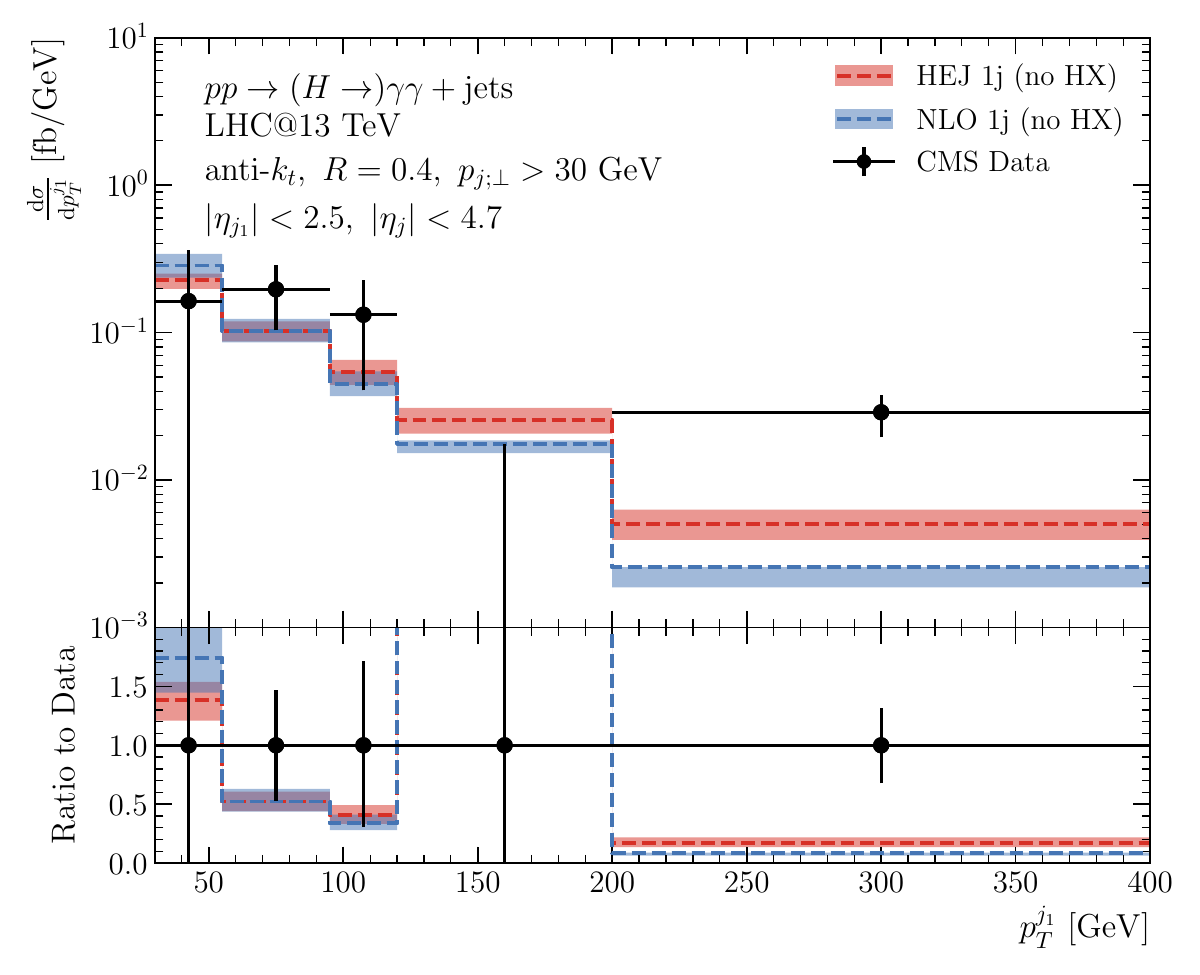}
        \caption{Leading jet transverse momentum.}
    \label{fig:13TeV_pTj1_hejqqx_nlo_1j}
    \end{subfigure}
    \hfill
    \begin{subfigure}[b]{0.49\textwidth}
        \centering
        \includegraphics[width=\textwidth]{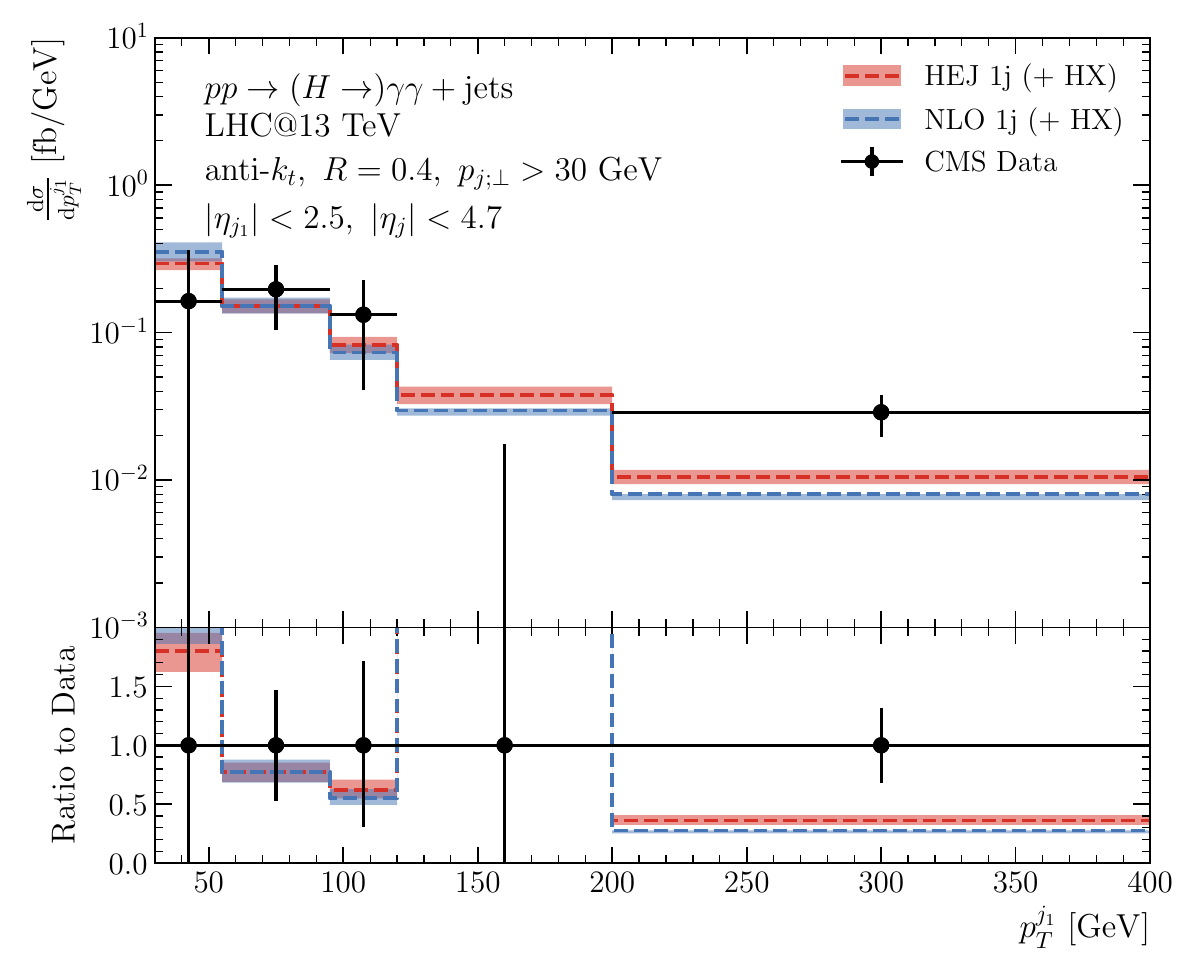}
        \caption{Leading jet transverse momentum.}
    \label{fig:13TeV_pTj1_hejqqx_nlo_1j_with_ewk}
    \end{subfigure}

    \vspace{1em}

    \begin{subfigure}[b]{0.49\textwidth}
        \centering
        \includegraphics[width=\textwidth]{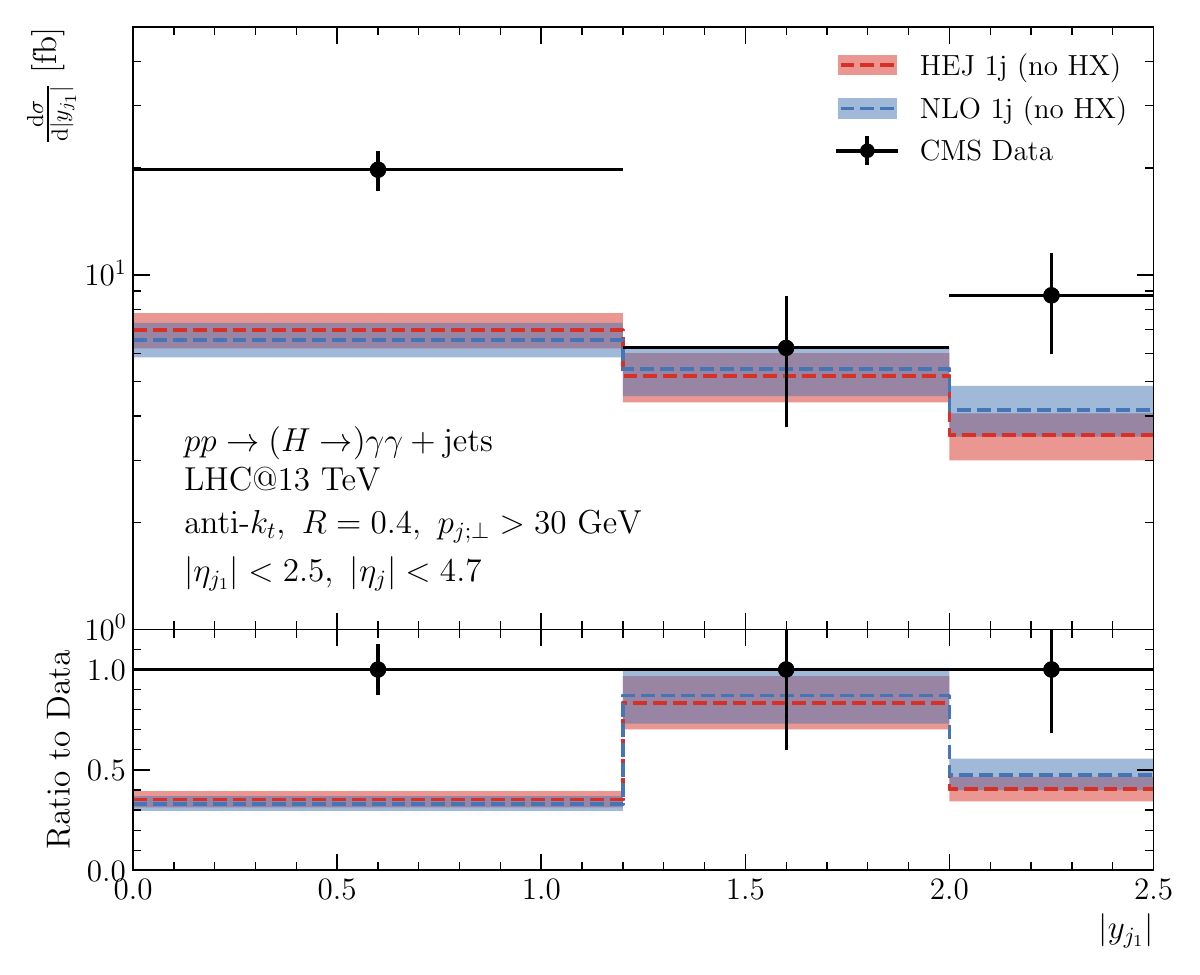}
        \caption{Leading central jet rapidity.}
    \label{fig:13TeV_yj1_hejqqx_nlo_1j}
    \end{subfigure}
    \hfill
    \begin{subfigure}[b]{0.49\textwidth}
        \centering
        \includegraphics[width=\textwidth]{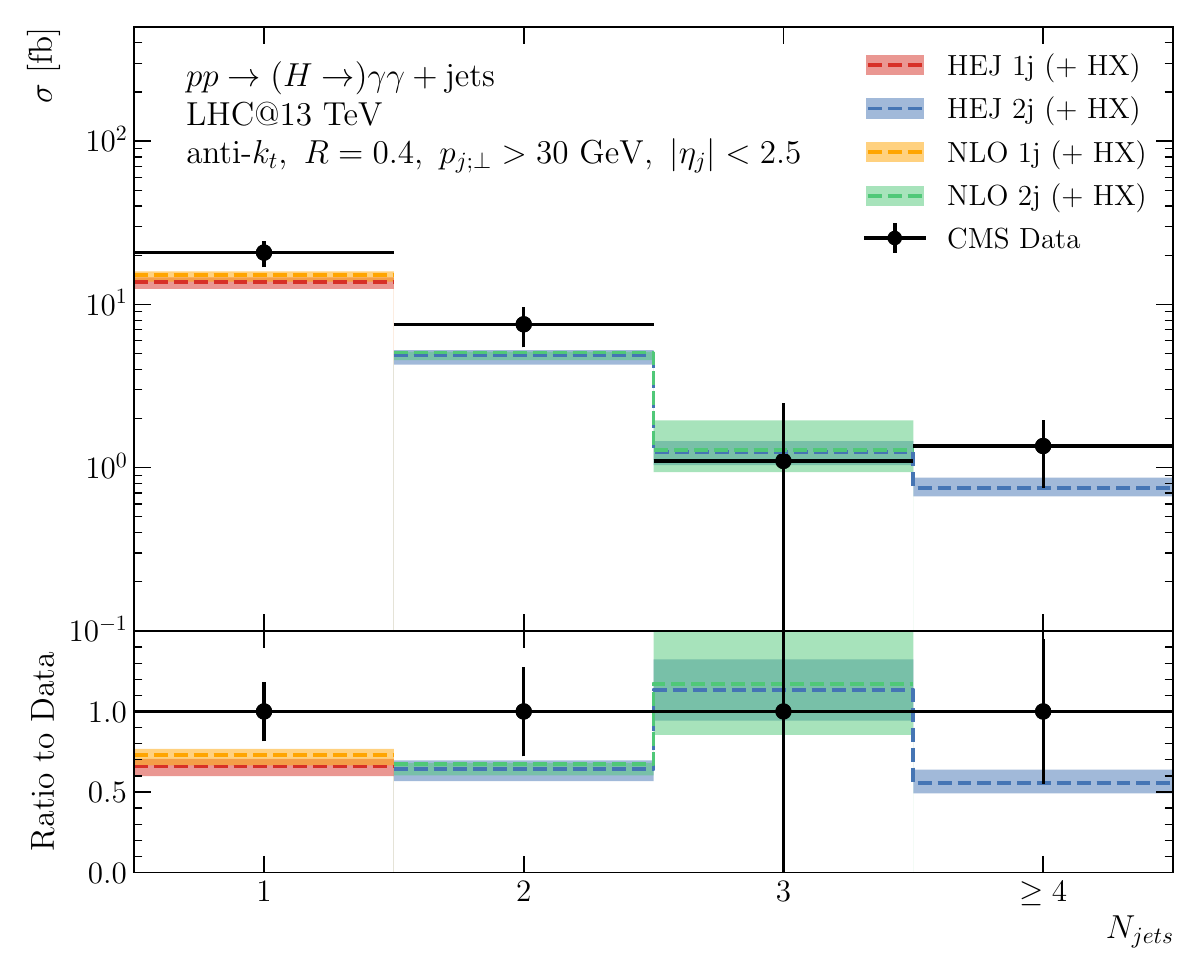}
        \caption{Number of jets.}
    \label{fig:13TeV_Njet_pt30_hejqqx_nlo_with_ewk}
    \end{subfigure}

\caption{
Comparison between \HEJ, NLO and CMS data~\cite{CMS:2018ctp} for 1-jet inclusive observables:~\ref{fig:13TeV_pTj1_hejqqx_nlo_1j} and~\ref{fig:13TeV_pTj1_hejqqx_nlo_1j_with_ewk}.
Comparison between \HEJ, NLO and CMS data~\cite{CMS:2022wpo} for the jet multiplicity observable, shown in figure~\ref{fig:13TeV_Njet_pt30_hejqqx_nlo_with_ewk}, and for the remaining 1-jet observable, namely the rapidity of leading central jet, in figure~\ref{fig:13TeV_yj1_hejqqx_nlo_1j}.
The first three bins in panel~\ref{fig:13TeV_Njet_pt30_hejqqx_nlo_with_ewk} are each exclusive in the jet multiplicity ($N_{jets} = 1,2,3$), whereas the last bin is inclusive ($N_{jets} \geq 4$).
The 1-jet \HEJ\ predictions are rescaled by the inclusive cross section ratio $\sigma_{\rm NLO1J}/\sigma_{\HEJ{\rm1J}}$, while the \HEJ predictions of the 2-, 3- and 4-jet bins of~\ref{fig:13TeV_Njet_pt30_hejqqx_nlo_with_ewk} are rescaled by $\sigma_{{\rm NLO2J}}/\sigma_{\HEJ {\rm2J}}$.
\ref{fig:13TeV_pTj1_hejqqx_nlo_1j}: Leading jet transverse momentum in comparison to experimental data without the ``HX'' component.
\ref{fig:13TeV_pTj1_hejqqx_nlo_1j_with_ewk}: Leading jet transverse momentum in comparison to experimental data with the ``HX'' component.
\ref{fig:13TeV_yj1_hejqqx_nlo_1j}: Rapidity of the leading central jet in comparison to experimental data without the ``HX'' component.
\ref{fig:13TeV_Njet_pt30_hejqqx_nlo_with_ewk}: Number of jets in the 1-, 2-, 3- and 4-jet bins.
The ``HX'' component is extracted from those CMS publications.
}
\label{fig:13TeV_hejqqx_vs_nlo_1j_pTj1}
\end{figure}

In figures~\ref{fig:13TeV_pTj1_hejqqx_nlo_1j} and~\ref{fig:13TeV_pTj1_hejqqx_nlo_1j_with_ewk}, we present the transverse momentum distribution of the leading jet $j_1$.
This jet is defined as the hardest central jet ($ \vert \eta_{j_1} \vert < 2.5$), and is therefore not necessarily the hardest jet in the event.
We compare resummed and fixed-order predictions -- with and without the HX component -- to data from the CMS analysis~\cite{CMS:2018ctp}, rather than~\cite{CMS:2022wpo}, as the former covers a wider kinematic range.

The impact of the quark-pair resummation at 13~TeV closely mirrors the behaviour observed at 8~TeV.
Across all considered observables, the resulting corrections remain at the level of a few percent and lie well within the scale-uncertainty bands of the previous \HEJ\ predictions~\cite{Andersen:2022zte}.
The consistency of this pattern across different collider energies provides further evidence for the perturbative stability of the \HEJ\ framework and the controlled numerical impact of higher-logarithmic contributions.
\begin{figure}[htbp]
    \centering

    \begin{subfigure}[b]{0.49\textwidth}
        \centering
        \includegraphics[width=\textwidth]{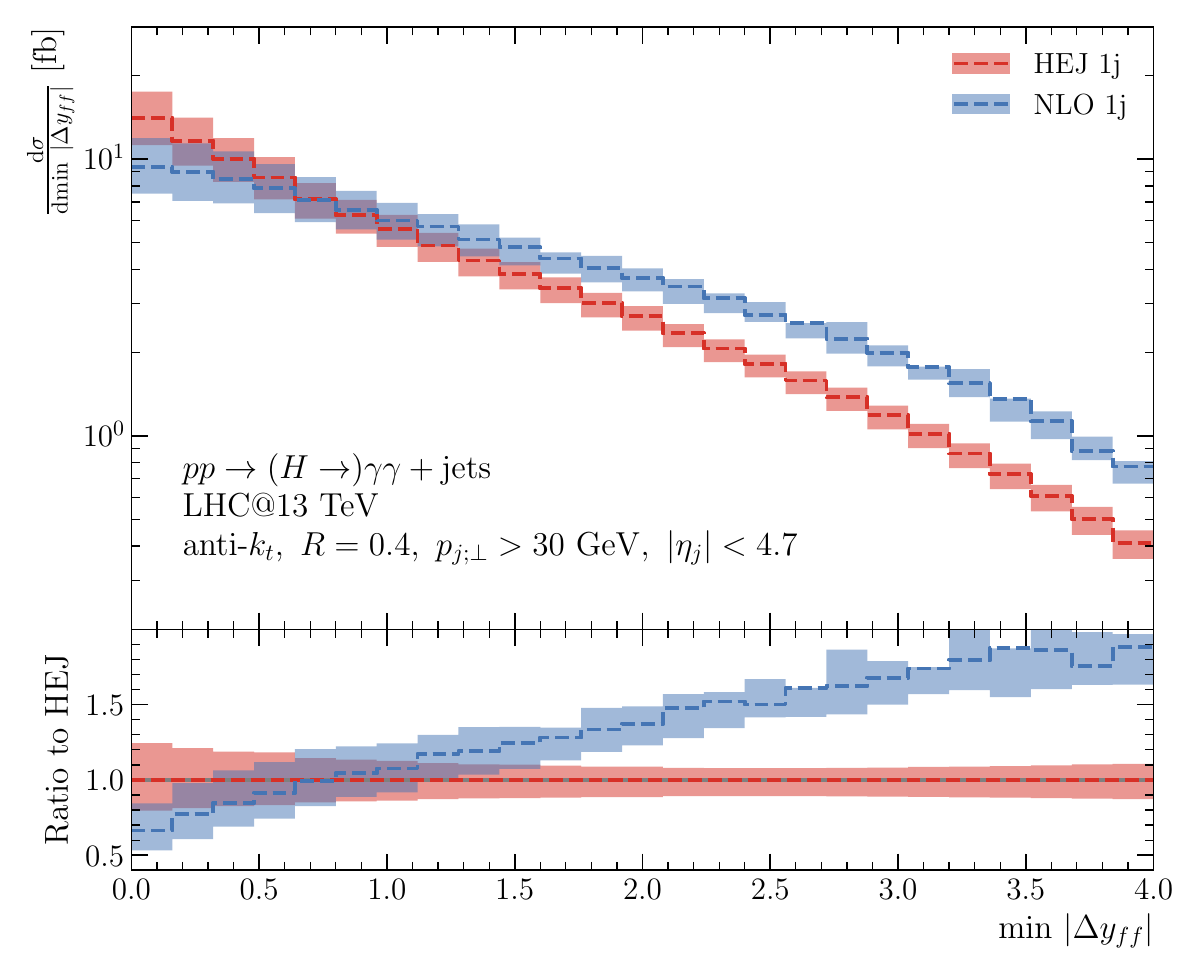}
        \caption{Minimum rapidity separation.}
    \label{fig:13TeV_minDyff_hejqqx_nlo_1j}
    \end{subfigure}
    \hfill
    \begin{subfigure}[b]{0.49\textwidth}
        \centering
        \includegraphics[width=\textwidth]{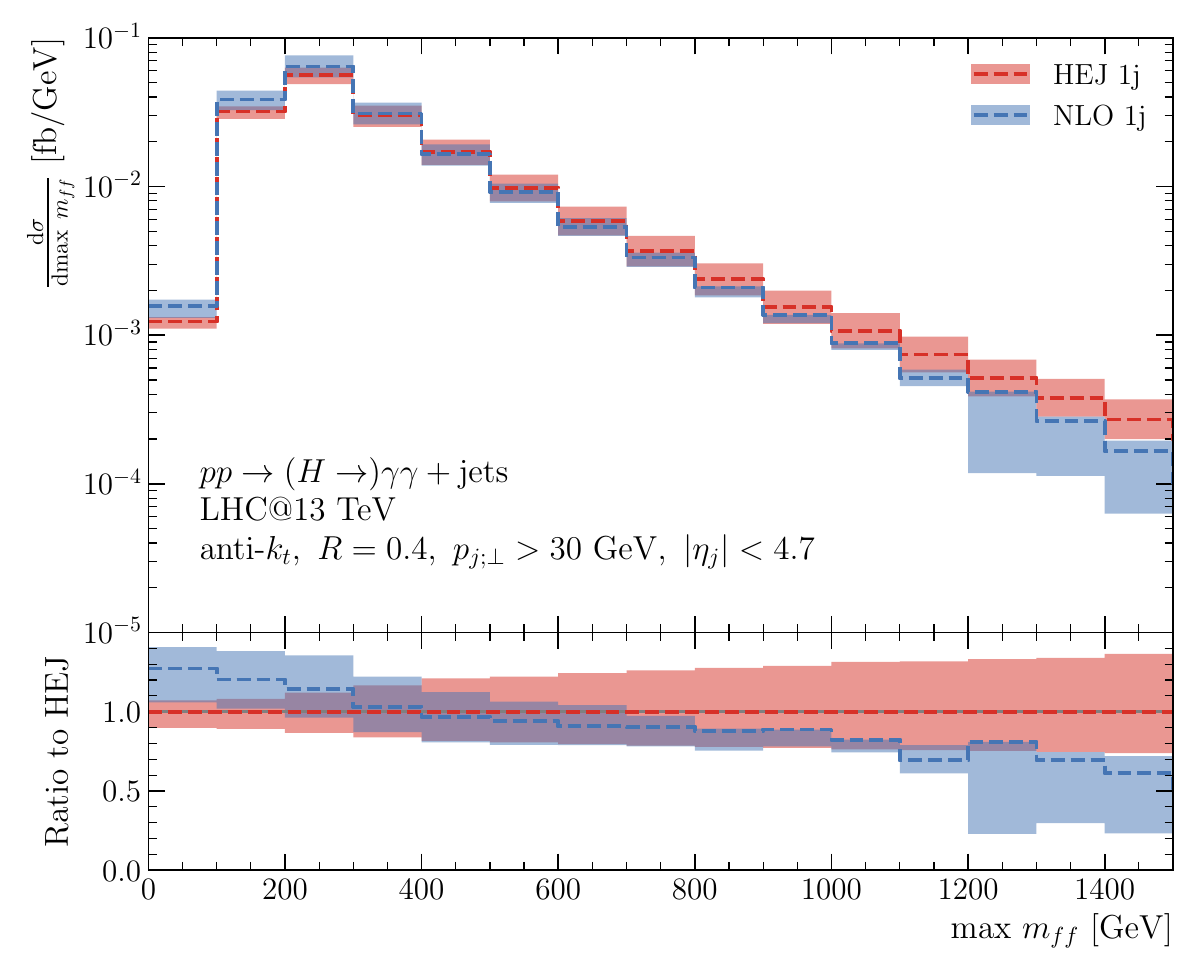}
        \caption{Maximum invariant mass.}
    \label{fig:13TeV_maxmff_hejqqx_nlo_1j}
    \end{subfigure}

\caption{
High-energy sensitive 1-jet distributions.
\ref{fig:13TeV_minDyff_hejqqx_nlo_1j}: Minimum rapidity separation between any two outgoing particles (Higgs boson or jets).
\ref{fig:13TeV_maxmff_hejqqx_nlo_1j}: Maximum invariant mass between any two outgoing particles (Higgs boson or jets).
\HEJ predictions are rescaled by the inclusive cross section ratio $\sigma_{{\rm NLO1J}}/\sigma_{\HEJ 1{\rm J}}$.}
\label{fig:13TeV_hejqqx_vs_nlo_1j_minmax}
\end{figure}
This is supported by the NLO reweighting factors, $\sigma_{\mathrm{NLO}n\mathrm{J}}/\sigma_{\HEJ n\mathrm{J}}$, reported in table~\ref{table:13TeV_reweighting-factors} for the 13~TeV CMS analysis~\cite{CMS:2018ctp,CMS:2022wpo}.

The inclusive 1-jet cross sections are extracted from the leading jet rapidity distribution, see figure~\ref{fig:13TeV_yj1_hejqqx_nlo_1j}.
On the other hand, the inclusive 2-jet cross sections are obtained from the appropriate $N_{jets} \geq 2$ bins in figure~\ref{fig:13TeV_Njet_pt30_hejqqx_nlo_with_ewk}.
Therefore, the sum of 2- and 3-jet bins for the fixed-order prediction and the sum of 2-, 3- and 4-jet bins for \HEJ\ are used.

For inclusive 1-jet observables, the NLL reweighting factors (UNO and QQX) are close to the LL result (FKL), indicating a negligible impact of subleading corrections.
Their numerical value is reported in table~\ref{table:13TeV_reweighting-factors}.
The notation is unchanged with respect to the previous $8~\mathrm{TeV}$ analysis, and we refer the reader to table~\ref{table:8TeV_reweighting-factors} for the definition of the FKL, UNO and QQX contributions.
In contrast, for 2-jet observables the NLL factors deviate more noticeably from LL, particularly between FKL and QQX.
This indicates that quark-pair resummation effects, while still moderate, become more relevant in more exclusive final states.
\begin{table}[htbp]
\begin{center}
\begin{tabular}{lccc}
\hline
Analysis  & \multicolumn{3}{l}{13 TeV} \\
\hline
Scale     & $\mu_{\rm F},\mu_{\rm R}$  & $(\mu_{\rm F},\mu_{\rm R})/2$  &  $2(\mu_{\rm F},\mu_{\rm R})$   \\
\hline
1j (FKL) factor & 1.55 & 1.28 & 1.81  \\
1j (UNO) factor & 1.58 & 1.30 & 1.83  \\
1j (QQX) factor & 1.60 & 1.31 & 1.85  \\
\hline
2j (FKL) factor & 1.58 & 1.14 & 1.95 \\
2j (UNO) factor & 1.63 & 1.19 & 2.01 \\
2j (QQX) factor & 1.69 & 1.23 & 2.07 \\
\hline
\end{tabular}
\caption{
NLO reweighting factors for the 13~TeV~ATLAS analysis~\cite{CMS:2018ctp,CMS:2022wpo} with $\mu_{\rm F} = \mu_{\rm R} = \max(m_{12},m_{\rm H})$.
The definitions of the FKL, UNO and QQX labels can be found in table~\ref{table:8TeV_reweighting-factors}.
}
\label{table:13TeV_reweighting-factors}
\end{center}
\end{table}

\section{Conclusions}
\label{sec:conclusions}

We have presented the first study of gluon-fusion Higgs boson production in association with a bottom-quark pair and two additional hard jets at high energies as a background to vector-boson-fusion Higgs-boson pair production in the $b\bar{b}\gamma\gamma$ final state within \HEJ.
The largest differences between \HEJ\ and LO consistently arise for observables probing large invariant masses and wide rapidity separations, precisely where high-energy logarithmic corrections are expected to become relevant.
This correlation strongly suggests that the observed changes originate from the all-order logarithmic contributions captured by \HEJ, rather than from generic higher-order effects.
The present results therefore indicate that high-energy logarithmic corrections constitute a relevant component of the perturbative description of gluon-fusion Higgs production in VBF-like topologies, clearly demonstrating that VBF selections are more effective than predicted by fixed-order calculations.
The inclusion of high-energy QCD logarithms is thus particularly important in view of the role played by this process as a major background to VBF Higgs-boson pair production, one of the key channels for probing both the Higgs self-interactions and the electroweak couplings of the Higgs boson.

These results have been made possible by extending the \HEJ\ framework for Higgs boson production in association with jets to include the all-order resummation of quark-pair emissions.
The classification of the relevant partonic configurations has been revisited and the logarithmic hierarchy of the corresponding amplitudes has been analysed in detail.
The numerical impact of the newly resummed quark-pair configurations has been assessed through dedicated phenomenological studies at both 8 and 13~TeV.
Across all observables considered, the inclusion of the quark-pair resummation leads to corrections at the level of a few percent, typically within the scale-uncertainty bands of previous \HEJ\ predictions.
The phenomenological picture established in earlier studies therefore remains unchanged, while the improved logarithmic accuracy further demonstrates the perturbative stability of the \HEJ\ framework and the controlled nature of successive logarithmic corrections.

More generally, the results presented in this work provide further evidence that the \HEJ\ framework offers a robust and systematically improvable description of Higgs-boson production in the high-energy limit of QCD.

All the \HEJ predictions of the current study have been obtained using the public release of \HEJ 2.4 at \href{http://hej.hepforge.org}{http://hej.hepforge.org}.

\acknowledgments

The authors would like to express their gratitude to both current and former members of the \HEJ collaboration.
We are particularly grateful to J.~R.~Andersen, J.~M.~Smillie, A.~Papaefstathiou, J.~Paltrinieri, S.~Jaskiewicz and M.~Rosca for valuable comments and insightful discussions throughout this work.
Andreas Maier acknowledges financial support from the Spanish Ministry of Science and Innovation (MICINN) through the Spanish State Research Agency, under Severo Ochoa Centres of Excellence Programme 2025-2029 (CEX2024001442-S).
This work was supported in part by the Spanish Ministry of Science and Innovation (PID2020-112965GB-I00,PID2023-146142NB-I00), and by the Departament de Recerca i Universities from Generalitat de Catalunya to the Grup de Recerca 00649 (Codi: 2021 SGR 00649).
This project has received funding from the European Union’s Horizon 2020 research and innovation programme under grant agreement No 824093.
IFAE is partially funded by the CERCA program of the Generalitat de Catalunya. The authors acknowledge support from the COMETA COST Action CA22130.

\clearpage

\appendix
\section{Impact of Higher Jet Multiplicities}\label{app:hsixjets}

In section~\ref{sec:HEJ_HH_bkg}, we presented predictions from \HEJ\ and at LO for $pp \to H(\to \gamma\gamma) + b\bar{b} + {\geq}\,2$ jets at $\sqrt{\hat{s}} = 13~\mathrm{TeV}$, where all results correspond to 4-jet inclusive observables.
To obtain \HEJ\ predictions for such inclusive quantities, exclusive samples must be merged.
The primary \HEJ\ results include only the 4-jet and 5-jet exclusive contributions, which are subsequently combined to yield 4-jet inclusive predictions.
This raises the question of how significant the impact of higher jet multiplicities may be.

The most direct approach would be to generate LO input events for $H + n$ jets with $n \geq 6$ and pass them to \HEJ\ to perform the resummation.
However, this method proves to be computationally prohibitive.
We therefore adopt an alternative strategy to estimate the higher-jet-multiplicity component amenable to \HEJ\ resummation.
Computing only the resummable contributions from exclusive $H+4j$ and $H+5j$ production, we observe a flat ratio between the two across all distributions.
This ratio is then applied to the exclusive $H+5j$ result to derive an estimate of the exclusive, resummed \HEJ\ contribution for $H+6j$.

\begin{figure}[htbp]
    \centering

    \begin{subfigure}[b]{0.49\textwidth}
        \centering
        \includegraphics[width=\textwidth]{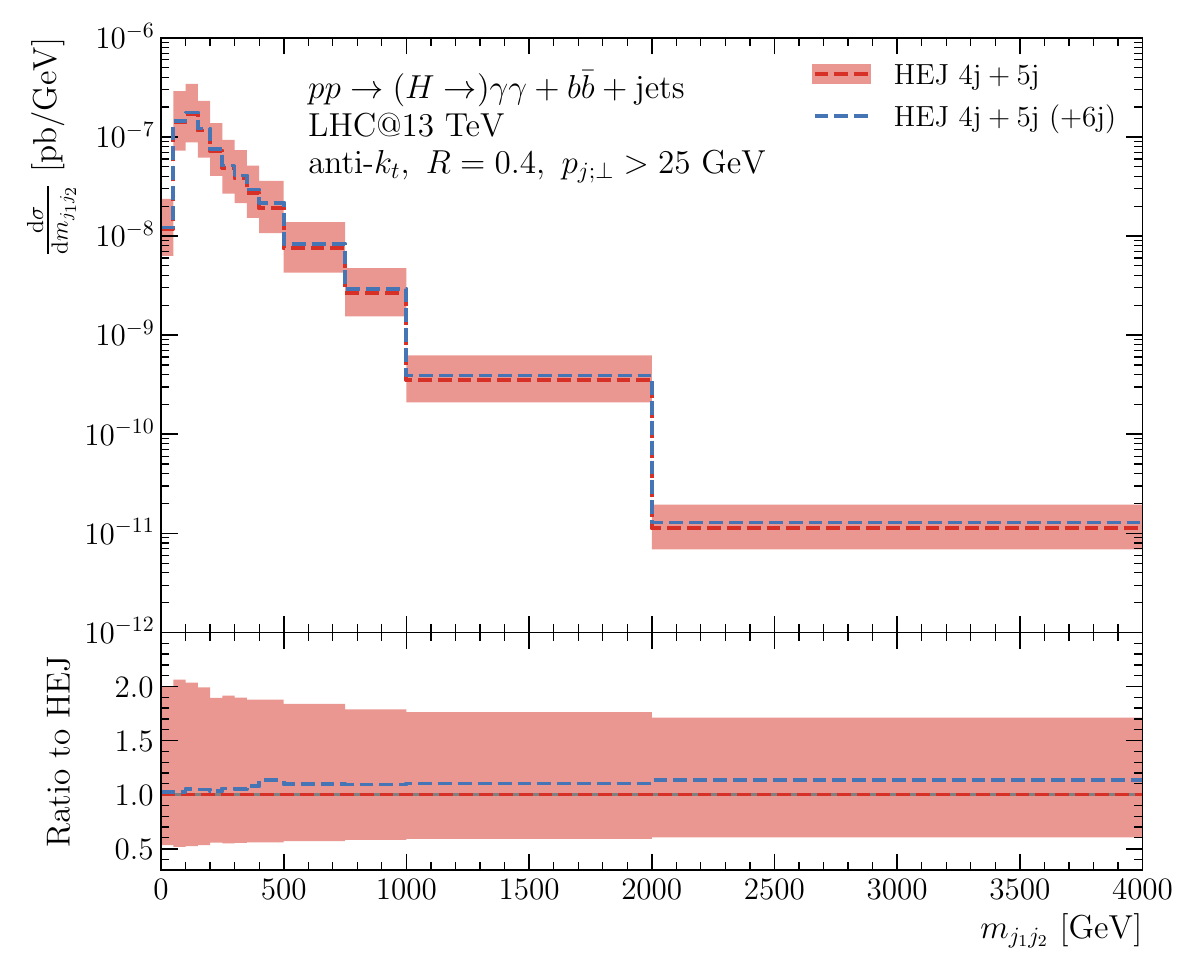}
        \caption{Dijet invariant mass.}
    \label{fig:hh_jjmass_h6j}
    \end{subfigure}
    \hfill
    \begin{subfigure}[b]{0.49\textwidth}
        \centering
        \includegraphics[width=\textwidth]{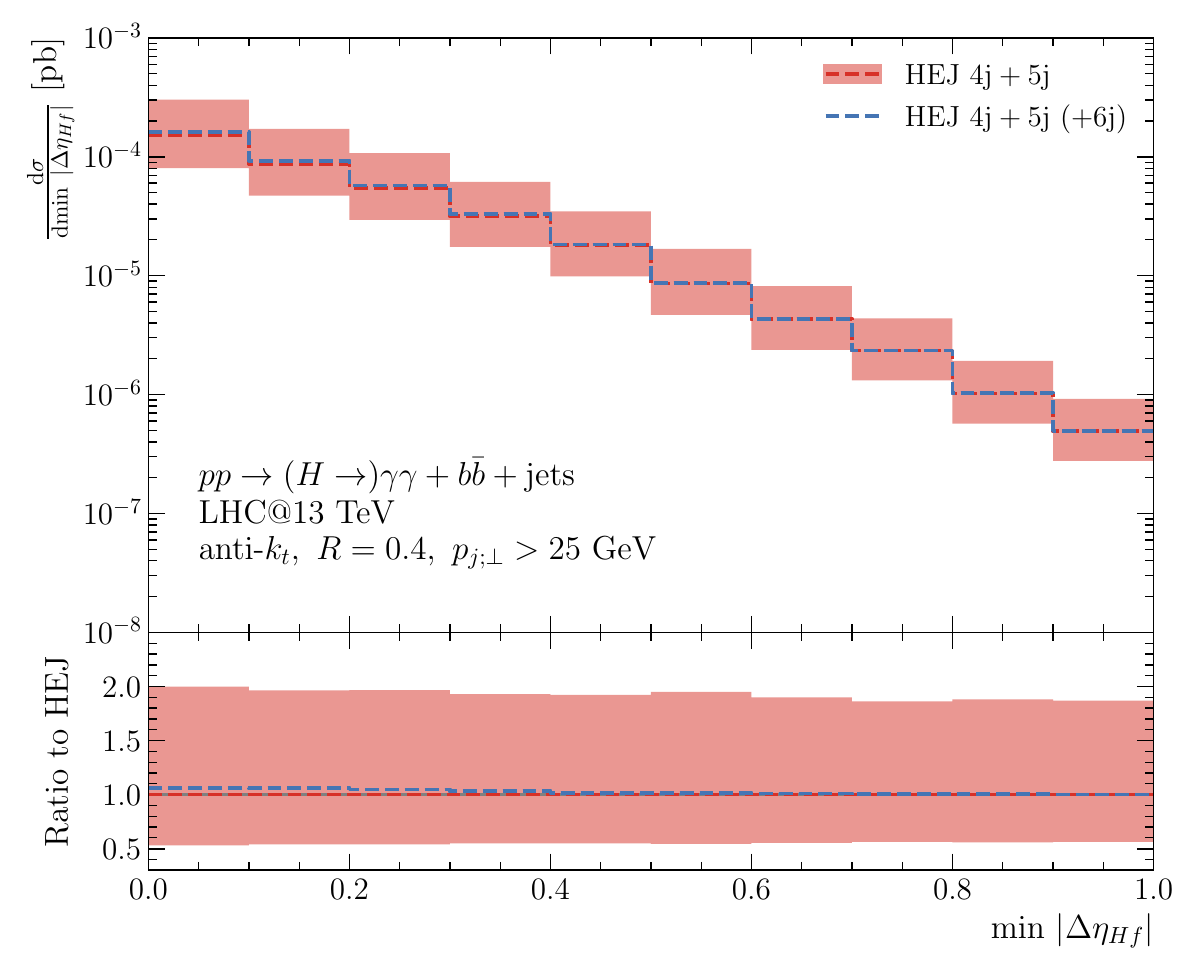}
        \caption{Minimum pseudo-rapidity separation.}
    \label{fig:hh_mindeltaeta_h6j}
    \end{subfigure}

    \vspace{1em}

    \begin{subfigure}[b]{0.49\textwidth}
        \centering
        \includegraphics[width=\textwidth]{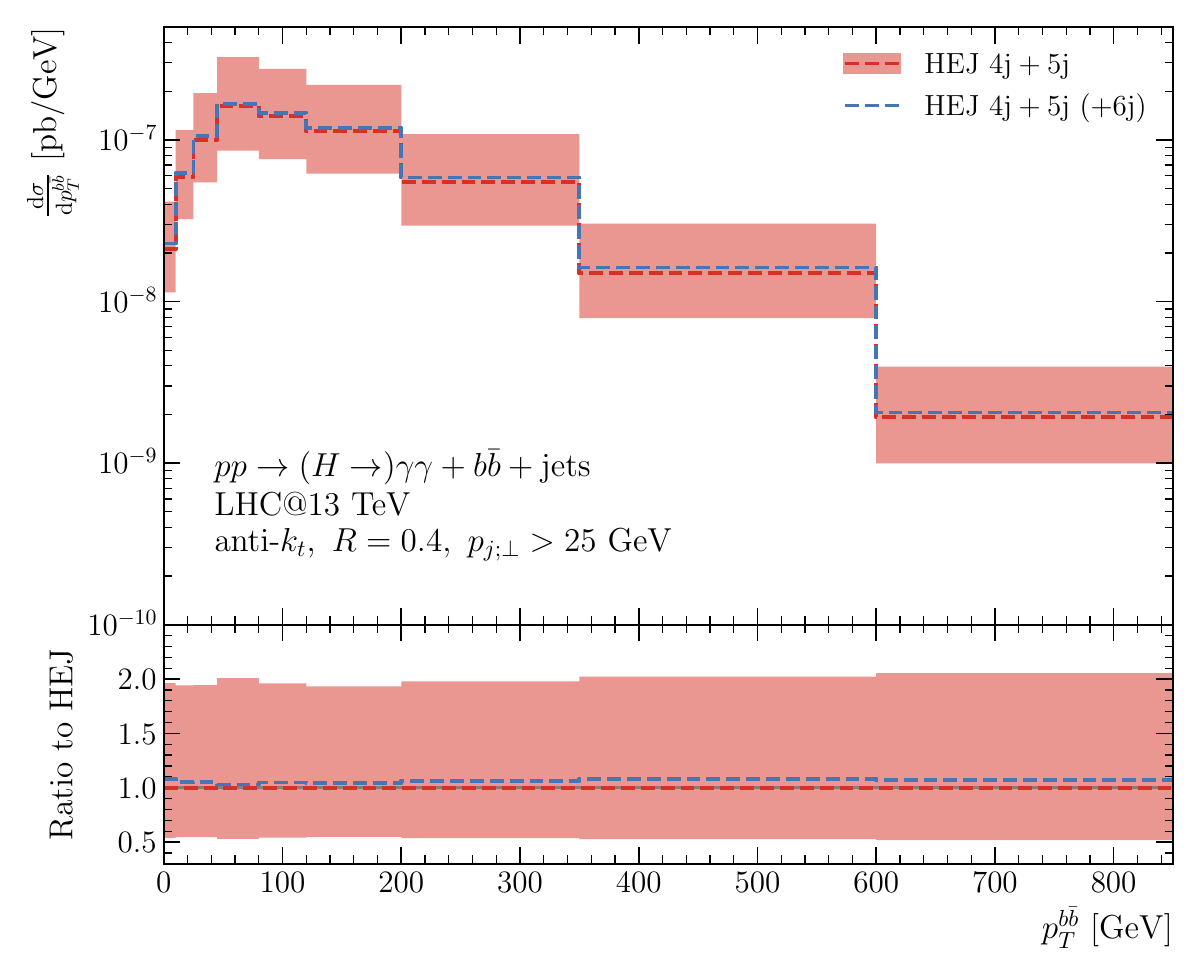}
        \caption{$b$-dijet transverse momentum.}
    \label{fig:hh_bbpt_h6j}
    \end{subfigure}
    \hfill
    \begin{subfigure}[b]{0.49\textwidth}
        \centering
        \includegraphics[width=\textwidth]{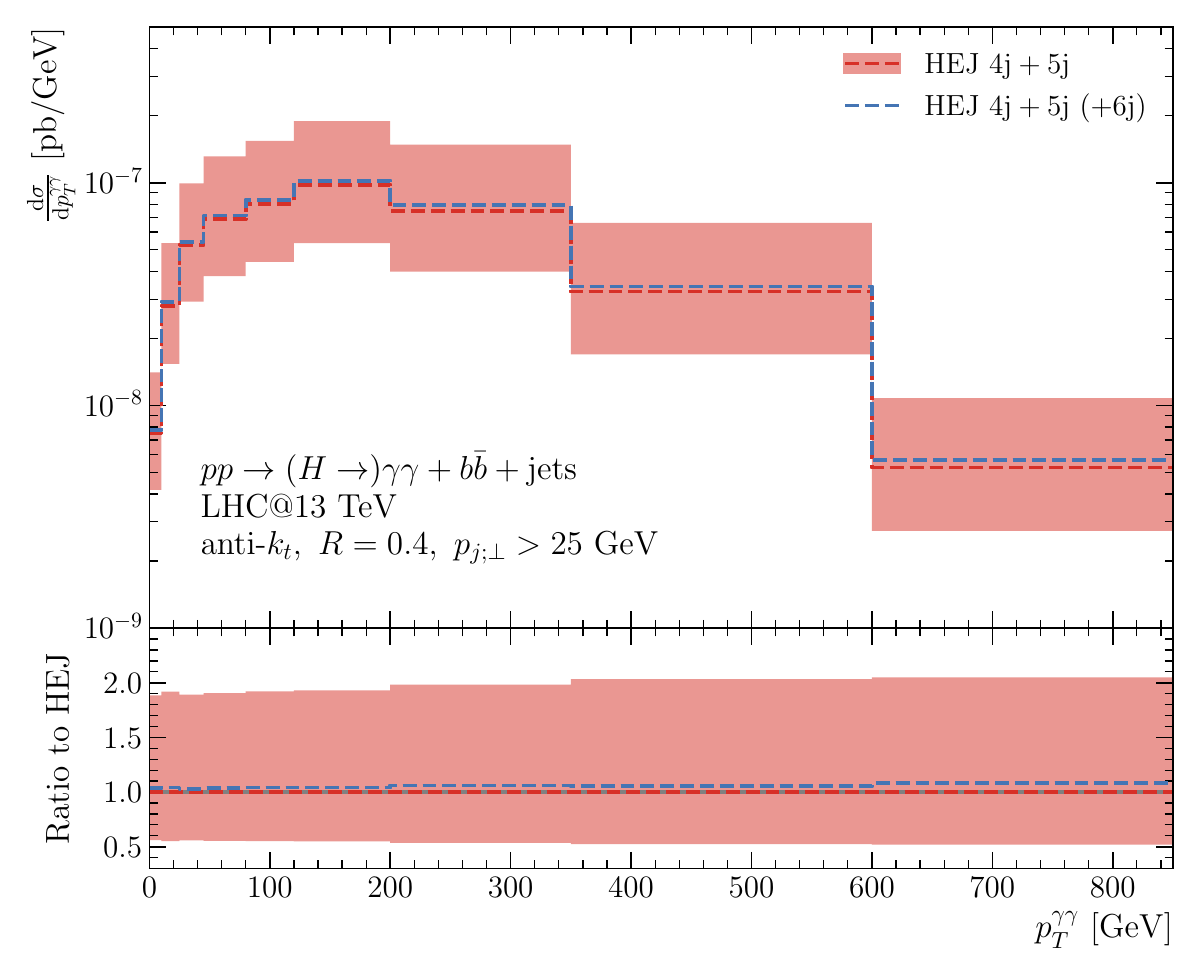}
        \caption{Diphoton-system transverse momentum.}
    \label{fig:hh_ggpt_h6j}
    \end{subfigure}

\caption{
\HEJ predictions for 4-jet inclusive observables.
\ref{fig:hh_jjmass_h6j}: Invariant mass of the two hardest jets.
\ref{fig:hh_mindeltaeta_h6j}: Minimum pseudo-rapidity separation between the reconstructed Higgs boson, identified through the diphoton system, and any other outgoing final-state particle.
\ref{fig:hh_bbpt_h6j}: Transverse momentum of the $b\bar{b}$ system.
\ref{fig:hh_ggpt_h6j}: Transverse momentum of the diphoton system.
The dashed red curve corresponds to the nominal \HEJ prediction presented in the main text~(see section~\ref{sec:HEJ_HH_bkg}), including the exclusive 4-jet and 5-jet contributions.
The dashed blue curve instead shows the full \HEJ result after including an estimate for the exclusive 6-jet contribution as well.
The impact of the latter remains well within the scale-uncertainty band represented by the red shaded region.
}
\label{fig:hh_h6j_observables}
\end{figure}

The 6-jet contribution is part of the missing higher-order corrections, whose overall size is conventionally estimated through scale variation. In order to verify that the error due to the missing 6-jet component is indeed covered by this uncertainty estimate, a comparison between \HEJ\ predictions with and without the estimated 6-jet component is shown in figure~\ref{fig:hh_jjmass_h6j} for the invariant mass of the two hardest jets, and in figure~\ref{fig:hh_mindeltaeta_h6j} for the minimum pseudo-rapidity separation between the Higgs boson and any outgoing final-state particle.
In both cases, the 6-jet contribution falls well within the scale-uncertainty bands shown in red.

This behaviour is consistently observed across all distributions presented in this work, including the transverse momentum of the $b\bar{b}$ system (figure~\ref{fig:hh_bbpt_h6j}) and the transverse momentum of the Higgs boson (figure~\ref{fig:hh_ggpt_h6j}).
Since the higher-jet-multiplicity contributions are everywhere subleading with respect to the scale uncertainties, we conclude that their omission from the main results is well justified, and that the 4-jet and 5-jet exclusive samples provide a reliable and sufficient description of the 4-jet inclusive observables considered here.

\section{Impact of the Bottom-Quark Mass}\label{app:bmass}

In general, the calculation of the cross section for Higgs boson together with $b$-jets production can be performed within two different flavour schemes.
In the four-flavour scheme (4FS), bottom quarks are treated as massive particles and no bottom Parton Distribution Functions (PDFs) are introduced.
As a consequence, the bottom quarks are generated entirely in the final state from light-quark and gluon initial states already at leading order.
The calculation therefore requires 4FS PDFs together with a consistent definition of the strong coupling $\alpha_s$, avoiding artificially large logarithmic contributions at higher perturbative orders.

In contrast, the five-flavour scheme (5FS) treats bottom quarks as massless partons within the proton.
In this framework, potentially large logarithms associated with collinear $g\to b\bar b$ splittings are resummed into the bottom PDFs through the Dokshitzer-Gribov-Lipatov-Altarelli-Parisi (DGLAP) evolution equation.
The 5FS calculation is therefore particularly suitable for inclusive observables characterised by scales significantly larger than the bottom-quark mass.

\begin{figure}[htbp]
    \centering

    \begin{subfigure}[c]{0.49\textwidth}
        \centering
        \includegraphics[width=\textwidth]{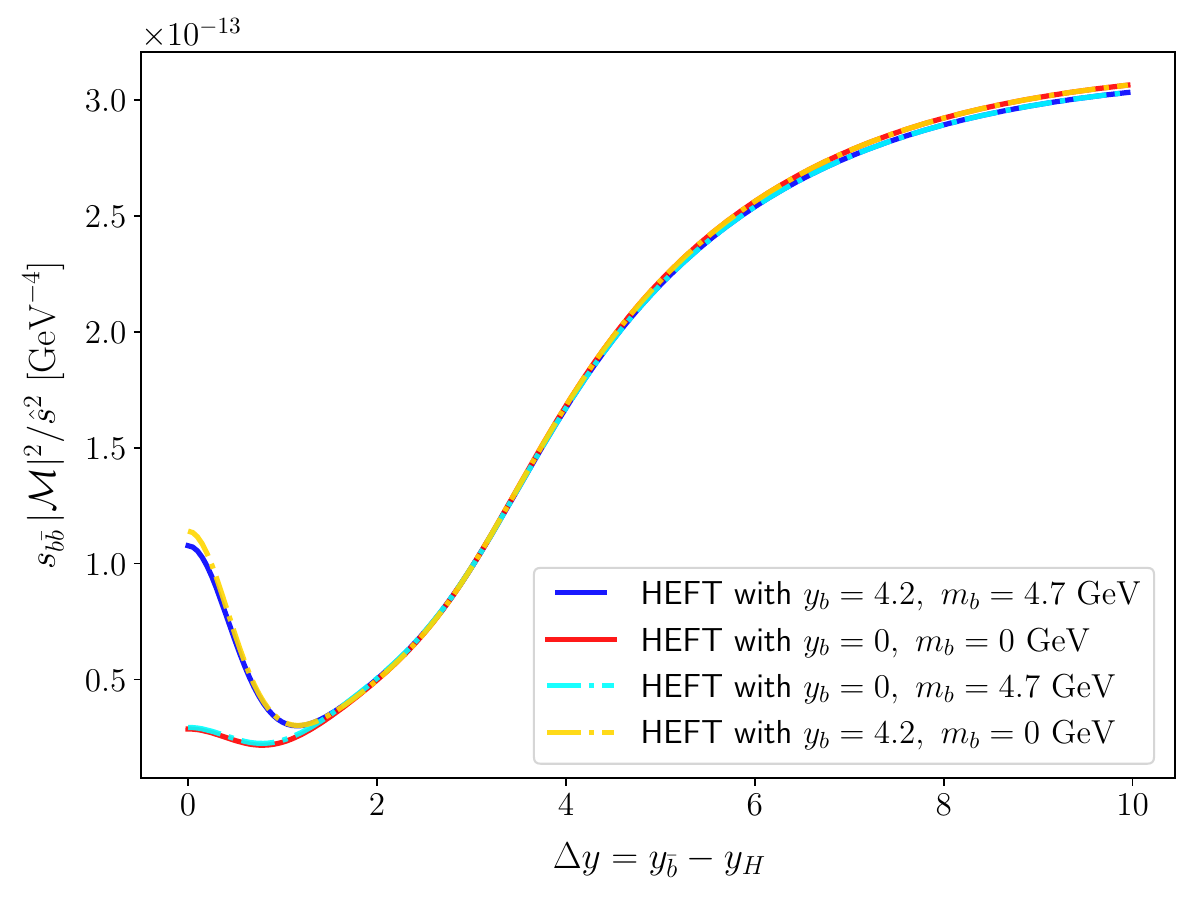}
        \caption{$s_{b \bar{b}} \vert \mathcal{M} \vert^2/\hat{s}^2$ for $ gg \to H b \bar{b}$.}
    \label{fig:impact_mb_a}
    \end{subfigure}
    \hfill
    \begin{subfigure}[c]{0.49\textwidth}
        \centering
        \includegraphics[width=\textwidth]{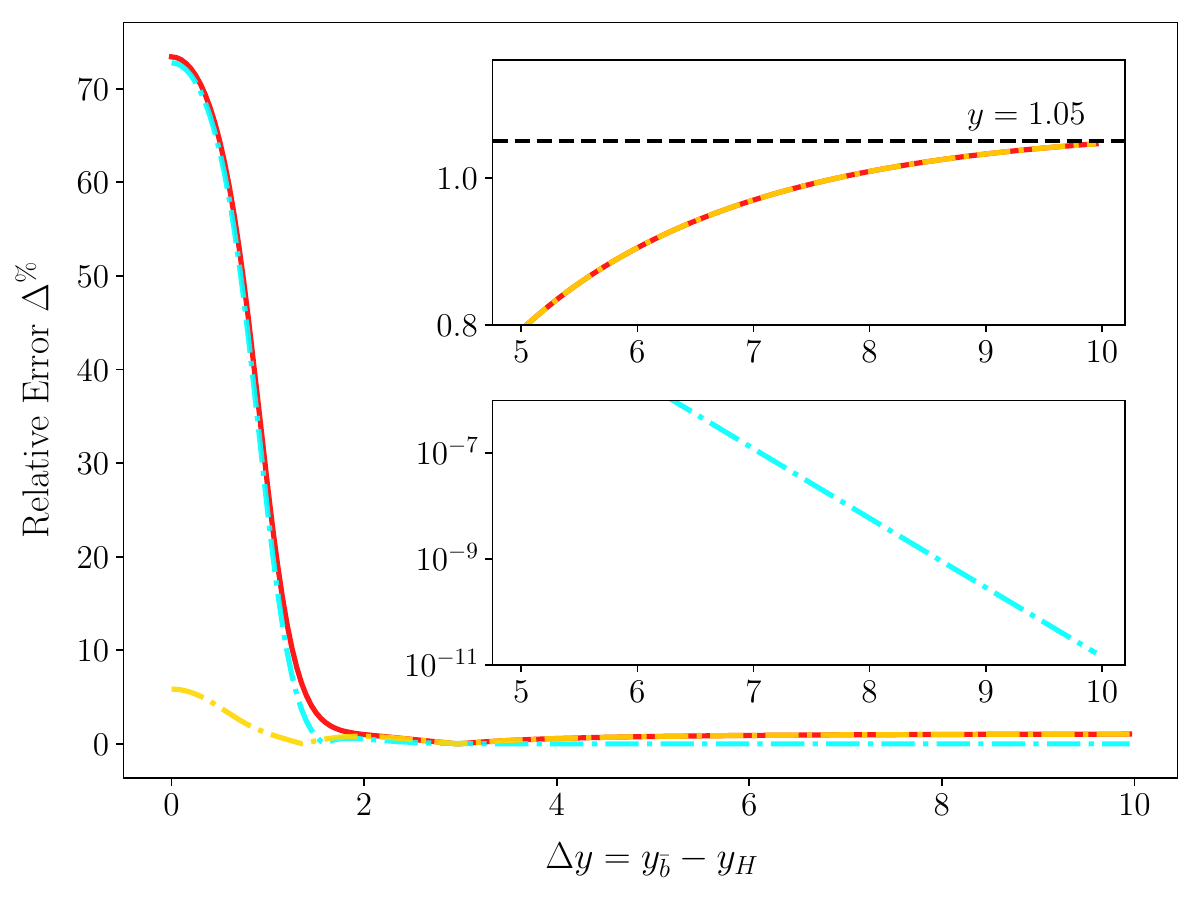}
        \caption{Relative error~$(\%)$.}
    \label{fig:impact_mb_b}
    \end{subfigure}

    \vspace{1em}

    \begin{subfigure}[c]{0.49\textwidth}
        \centering
        \includegraphics[width=\textwidth]{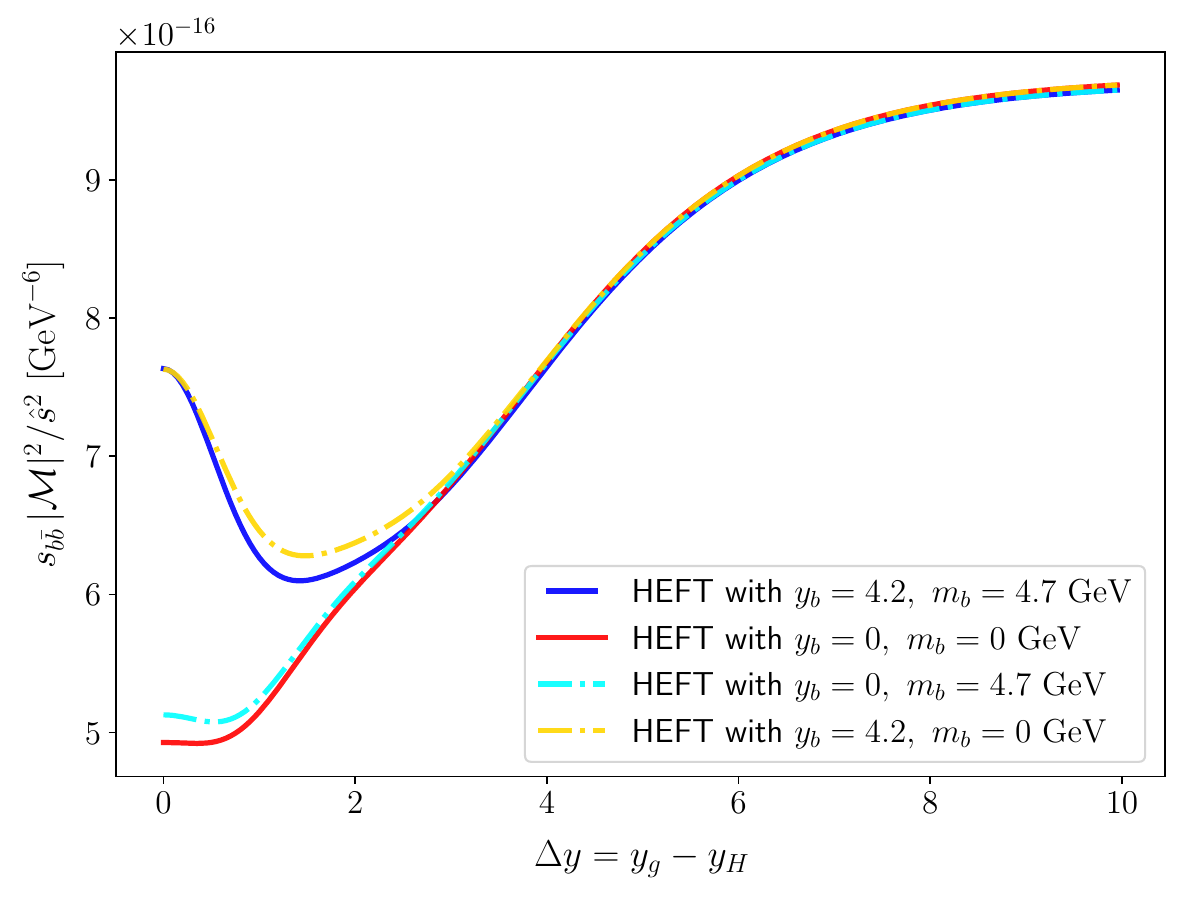}
        \caption{$s_{b \bar{b}} \vert \mathcal{M} \vert^2/\hat{s}^2$ for $ gg \to H b \bar{b} g$.}
    \label{fig:impact_mb_c}
    \end{subfigure}
    \hfill
    \begin{subfigure}[c]{0.49\textwidth}
        \centering
        \includegraphics[width=\textwidth]{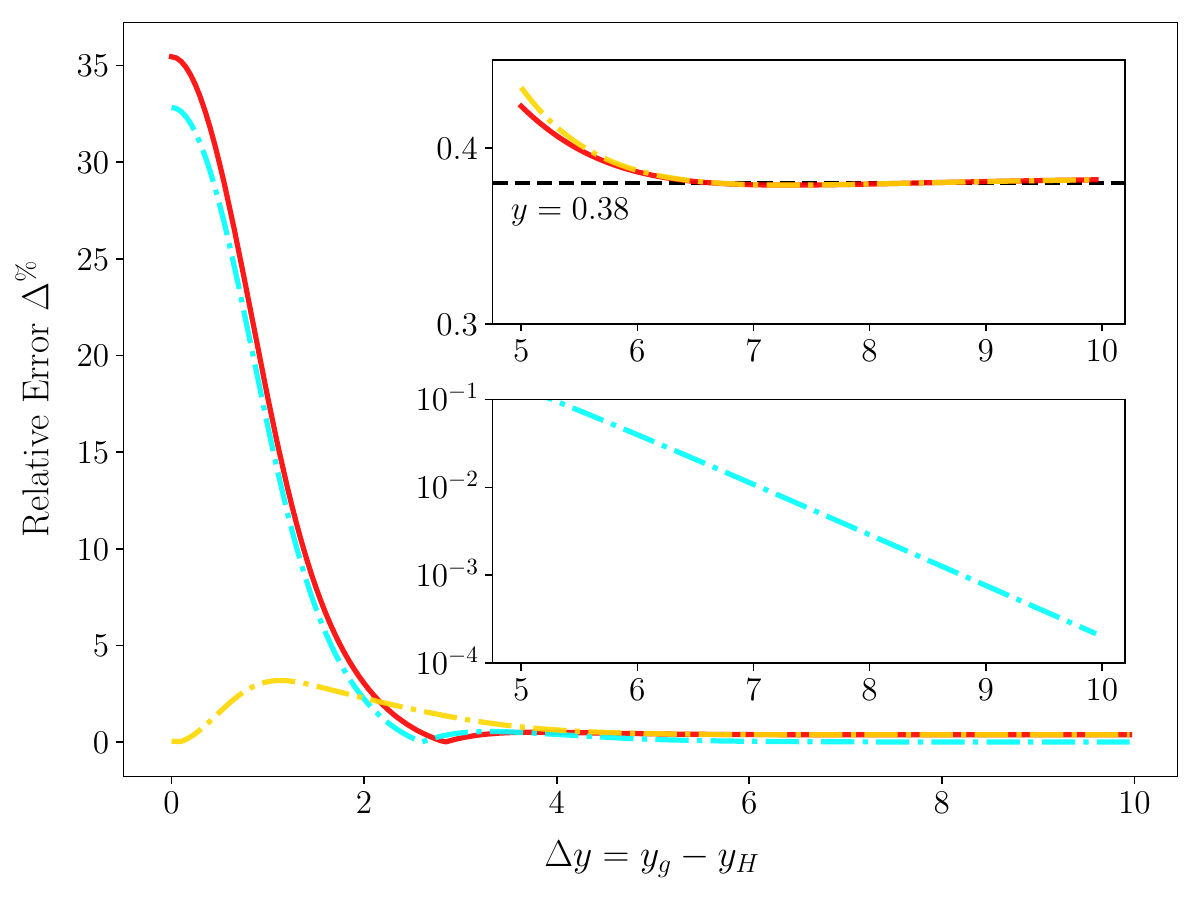}
        \caption{Relative error~$(\%)$.}
    \label{fig:impact_mb_d}
    \end{subfigure}

    \vspace{1em}

    \begin{subfigure}[c]{0.49\textwidth}
        \centering
        \includegraphics[width=\textwidth]{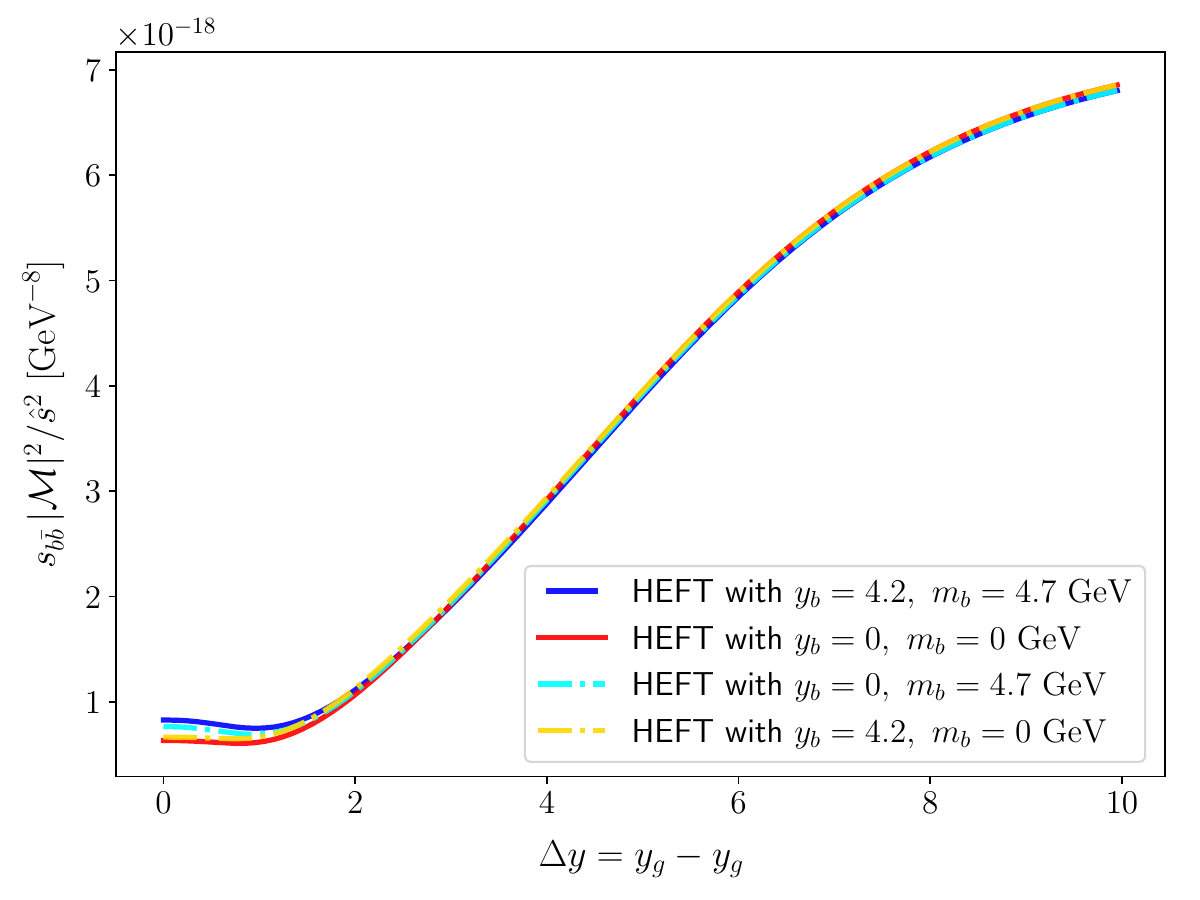}
        \caption{$s_{b \bar{b}} \vert \mathcal{M} \vert^2/\hat{s}^2$ for $ gg \to g b \bar{b} H g$.}
    \label{fig:impact_mb_e}
    \end{subfigure}
    \hfill
    \begin{subfigure}[c]{0.49\textwidth}
        \centering
        \includegraphics[width=\textwidth]{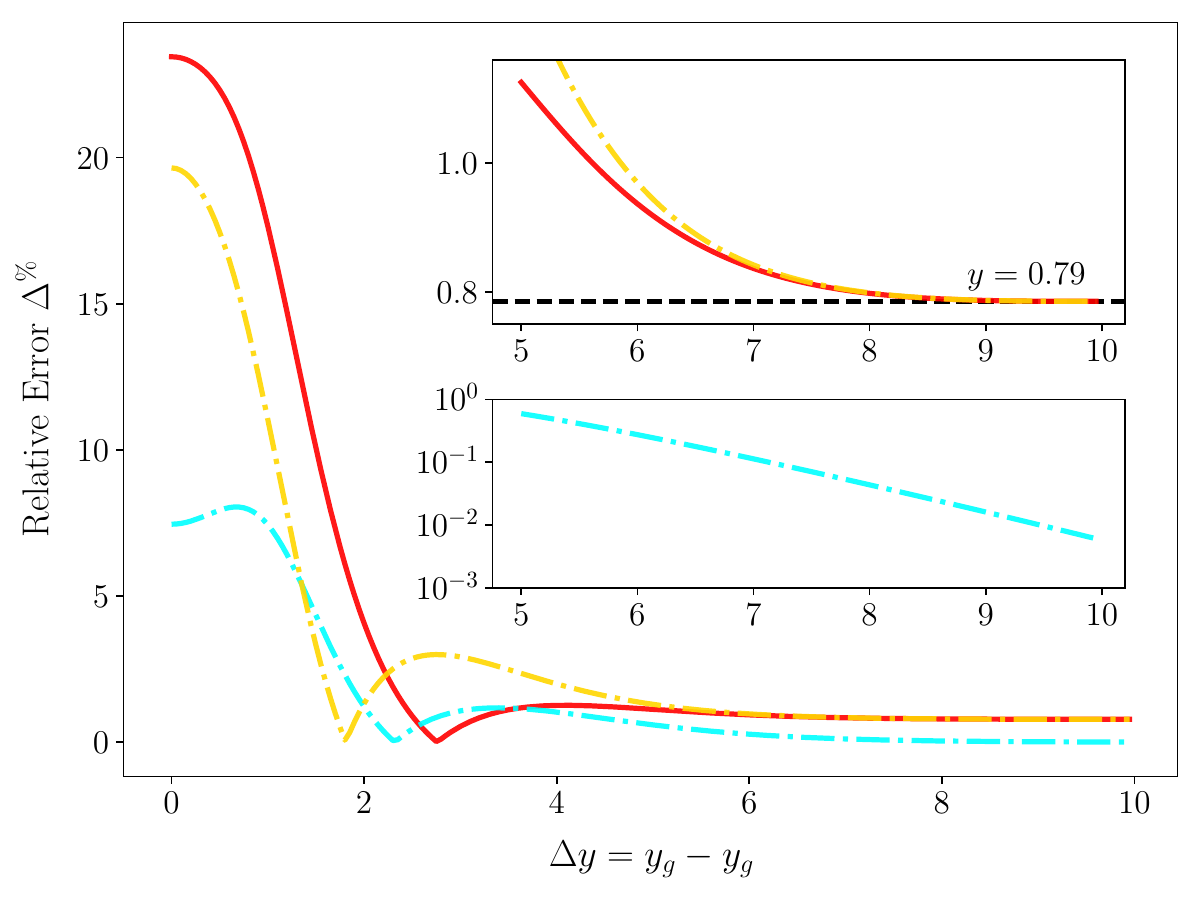}
        \caption{Relative error~$(\%)$.}
    \label{fig:impact_mb_f}
    \end{subfigure}

\caption{
On the left panels, the behaviour of $\lvert \mathcal{M} \rvert^2$ as a function of the rapidity separation is shown for three different processes at leading order in the HEFT model and in the infinite top-quark mass limit.
The solid blue curve corresponds to the reference prediction, retaining both $m_b$ and $y_b$.
The solid red curve shows the result obtained by neglecting both effects.
The dot-dashed cyan and yellow curves represent the cases in which only the bottom-quark mass or only the bottom-Yukawa coupling is retained, respectively.
The right panels display the relative difference of the various approximations with respect to the reference prediction.
}
\label{fig:impact_mb}
\end{figure}
In the present work, we adopt the 5FS.
To validate this approximation, we compare the LO HEFT matrix element squared in the infinite top-quark mass limit under four different assumptions: (i) retaining the full dependence on the bottom-quark mass and Yukawa coupling, (ii) neglecting both contributions, (iii) retaining only finite bottom-quark mass effects while omitting all $Hb\bar{b}$ interactions, and (iv) retaining only the bottom-Yukawa coupling while taking the bottom quark to be massless.
The results shown in figure~\ref{fig:impact_mb} have been obtained using {\tt MadGraph5\_aMC@NLO}~\cite{Alwall:2014hca}.
The amplitudes are computed in a one-dimensional phase-space as a function of the rapidity separation between all pairs of particles.
The momentum configuration used for figure~\ref{fig:impact_mb} is reported in table~\ref{tab:explorer_impact_mb}.

Although the two hybrid setups are not physically consistent -- since a massless bottom quark would also imply a vanishing Yukawa coupling -- they provide a useful diagnostic tool.
By considering them separately, we can disentangle the purely kinematic effects associated with a finite bottom-quark mass from those originating from the $Hb\bar{b}$ interaction.
This distinction is particularly relevant for assessing whether the high-energy behaviour in the MRK limit is dominated by the QCD component of the amplitude or receives sizeable contributions from the bottom-Higgs Yukawa interaction.
We stress that, within the HEFT framework employed here, the effective Higgs-gluon coupling is induced exclusively by the top-quark Yukawa coupling.
Consequently, the bottom-Yukawa dependence enters only through the explicit tree-level $Hb\bar{b}$ interaction, making it possible to isolate and study its contribution independently of the effective $ggH$ vertex.

As the MRK limit is approached, corresponding to increasing rapidity separations $\Delta y$ among all pairs of final-state particles, the relative difference between the massive and massless scenarios stabilises at approximately the $1\%$ level, as shown in the right panels of figure~\ref{fig:impact_mb}.
It is worth noting that, in this region of phase space, the tree-level bottom-Higgs contributions become entirely negligible, independently of whether finite bottom-quark mass effects are retained.

In contrast, for small rapidity separations both the bottom-quark mass and the bottom-Yukawa coupling can have a sizeable numerical impact.
The most pronounced effect arises from the presence of the Yukawa interaction: the comparison between the vanishing- and non-vanishing-Yukawa cases shows that the tree-level $Hb\bar{b}$ contribution plays an important role in this region.
Finite bottom-quark mass effects are comparatively smaller, but still induce differences at the percent level among the various approximations considered.

It is important, however, to distinguish between these two kinematic regimes.
While large rapidity separations correspond to the approach to the MRK limit and therefore probe the universal asymptotic behaviour of the amplitudes, the small-$\Delta y$ region is not associated with any particular kinematic limit.
Consequently, the conclusions drawn at small rapidity separations may depend on the specific momentum configuration adopted in the analysis.
The behaviour observed in the MRK region, on the other hand, is largely insensitive to the details of the chosen kinematics and reflects the genuine asymptotic high-energy dynamics of the process.
This indicates that, for the purposes of the present analysis and at the level of accuracy considered here, bottom-quark mass and Yukawa effects are entirely subleading.
Consequently, all results presented in this work for $H b\bar{b} + \geq 2$ jets have been obtained within the 5FS.
\begin{table}[H]
\begin{center}
\begin{tabular}{ |c||l| }
\hline
Process & Momenta configuration \\
\hline
$gg \to Hb\bar{b}$ &
\(
\begin{cases}
      y_H = -\Delta, y_b = \frac{\Delta}{2} \text{ and } y_{\bar{b}} = \Delta &\\
      \phi_b = \frac{11}{9}\pi \text{ and } \phi_{\bar{b}} = \frac{\pi}{7} &\\
     p_{b\perp} = 40 \, \text{GeV} \text{ and } p_{\bar{b}\perp} = 40 \, \text{GeV} &
    \end{cases}
\)  \\
$gg \to Hb\bar{b}g$ &
\(
\begin{cases}
      y_H = -\Delta, y_b=-\frac{\Delta}{3}, y_{\bar{b}}=\frac{\Delta}{3} \text{ and } y_g = \Delta &\\
      \phi_b = \frac{3}{4}\pi, \phi_{\bar{b}} = \frac{6}{11}\pi \text{ and } \phi_g = \frac{11}{9}\phi & \\
     p_{b\perp} = 40 \, \text{GeV}, p_{\bar{b}\perp} = 40 \, \text{GeV} \text{ and } p_g = 40 \, \text{GeV} &
    \end{cases}
\)  \\
$gg \to gb\bar{b}Hg$ &
\(
\begin{cases}
      y_g = -\Delta, y_b=-\frac{\Delta}{2}, y_{\bar{b}} = 0, y_{H}=\frac{\Delta}{2} \text{ and } y_g = \Delta &\\
      \phi_b = \frac{6}{11}\pi, \phi_{\bar{b}} = \frac{3}{4}\pi, \phi_{H} = \frac{8}{5}\pi \text{ and } \phi_g = \frac{11}{9}\phi & \\
     p_{b\perp} = 40 \, \text{GeV}, p_{\bar{b}\perp} = 40 \, \text{GeV},p_{H\perp} = 40 \, \text{GeV} \text{ and } p_g = 40 \, \text{GeV} &
    \end{cases}
\)  \\
\hline
\end{tabular}
\end{center}
\caption{The momentum configuration used in figure~\ref{fig:impact_mb}.}
\label{tab:explorer_impact_mb}
\end{table}

\section{Next-to-next-to Logarithmic Corrections}\label{app:nnll}

As shown in section~\ref{subsec:lo4j_individual}, the dominant non-resummable contribution to Higgs boson production in association with a quark pair as a background to VBF di-Higgs production arises from configurations in which the Higgs boson is emitted centrally between the two bottom quarks.
Although formally subleading, these configurations provide a sizeable contribution to the total cross section.
This behaviour is largely driven by the analysis cuts, which require both the Higgs boson and the $b\bar{b}$ system to be produced centrally, thereby enhancing the relative importance of such configurations.

In the following, we outline the main ingredients underlying the suppression of configurations in which the Higgs boson is emitted between the quark pair.
Following the same arguments already presented in section~\ref{sec:HEJFORHIGGS}, relaxing the strong rapidity ordering of equation~\eqref{eq:MRKLIMIT} introduces an additional suppression of the scattering amplitude.
In the MRK limit, this suppression is exponential in the rapidity separation between the particles whose ordering is relaxed, reflecting the replacement of a $t$-channel gluon exchange by a $t$-channel quark exchange.
Consequently, their relative contribution vanishes asymptotically in the MRK limit.

After integration over phase space, the same behaviour manifests itself as a logarithmic suppression.
Since the high-energy logarithms resummed by \HEJ\ originate from the large rapidity intervals associated with real and virtual corrections, through factors of the form $\log(s_{ij}/p_{\perp}^{2})$, configurations with an additional exponential suppression in rapidity contribute only at subleading logarithmic accuracy.
As a result, the Higgs-between-quark-pair configurations discussed here enter formally at NNLL accuracy.

To gain further insight into the case of a Higgs boson emitted between a quark pair, let us consider the production of a Higgs boson in association with four partons, two of which form a quark-antiquark pair.
Here we consider light quarks, which do not couple directly to the Higgs boson since their Yukawa coupling vanishes.
This approximation is equally valid for bottom quarks in the limit $y_b = 0$.
Requiring the Higgs boson and the quark pair to be emitted centrally leaves only two possible configurations: (a) the Higgs boson is emitted adjacent to the quark pair, or (b) between the two quarks.

For the first configuration $f_a f_b \to f_a \cdots q\bar{q} \cdots H \cdots f_b$, Regge theory predicts that the scattering amplitude scales as
\begin{align}
\includegraphics[valign=c,width=0.3\textwidth]{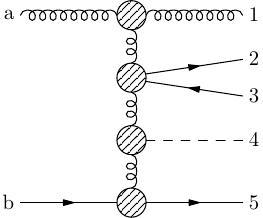} \quad \longrightarrow \mathcal{M}_1 \sim s_{12} \sqrt{s_{23}} s_{34} s_{45}
\label{eq:amplitudesplit_4j_CENQQX}
\end{align}
On the other hand, in the second configuration $f_a f_b \to f_a \cdots q H \bar{q} \cdots f_b$, we effectively relax the strong rapidity ordering, corresponding to the replacement of a $t$-channel gluon exchange between the Higgs boson and the quark pair by a $t$-channel quark exchange.
This leads to the following scaling in the Regge limit
\begin{align}
\includegraphics[valign=c,width=0.3\textwidth]{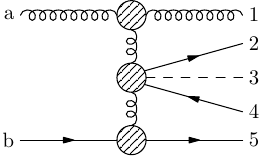}  \quad \longrightarrow \mathcal{M}_2 \sim s_{12} \sqrt{s_{23}} \sqrt{s_{34}} s_{45}
\label{eq:amplitudesplit_4j_CENQHQX}
\end{align}
To better understand the relative scaling of $\mathcal{M}_1$ and $\mathcal{M}_2$ in terms of the centre-of-mass energy squared $\hat{s} \equiv s_{ab}$ and the invariant mass of the quark pair ($s_{23}$ for the first configuration and $s_{24}$ for the second configuration), it is useful to parametrise the four-momentum of each outgoing particle in terms of its transverse momentum $p_{\perp}$, rapidity $y$, and azimuthal angle $\phi$ as
\begin{equation}
p^{\mu} := p = \vert p_{\perp} \vert (\cosh y, \cos \phi, \sin \phi, \sinh y).
\end{equation}
It is straightforward to observe that the invariant mass $s_{ij}$ between any two outgoing final states $i$ and $j$ in the relativistic limit, or equivalently in the massless limit, can be written as
\begin{equation}
s_{ij} \equiv (p_i + p_j)^2 \approx 2 p_i \cdot p_j = 2 \vert p_{i\perp} \vert \vert p_{j\perp}\vert (\cosh \Delta y_{ji} - \cos \Delta \phi_{ji}).
\end{equation}
For large rapidity separations, $\Delta y_{ji} \equiv |y_j - y_i| \gg 1$, the invariant mass grows exponentially, independently of the values of the transverse momenta and azimuthal-angle separation $\Delta\phi_{ji} \equiv |\phi_j-\phi_i|$.
Therefore, in the MRK limit, the invariant mass $s_{ij}$ can be further simplified as
\begin{equation}
s_{ij} \approx \vert p_{i\perp} \vert \vert p_{j\perp}\vert e^{\Delta y_{ji}}.
\end{equation}
Therefore, the matrix element for the first configuration goes as
\begin{equation}
\mathcal{M}_1 \sim e^{\Delta y_{21}} e^{\Delta y_{32}/2} e^{\Delta y_{43}} e^{\Delta y_{54}} = e^{\Delta y_{51}} e^{-\Delta y_{32}/2} \sim \frac{\hat{s}}{\sqrt{s_{23}}},
\label{eq:regge_cenqqx_exp}
\end{equation}
where in the first step we explicitly dropped the transverse momenta, as they only change the overall normalisation.
In addition, in the last step we have exploited the fact that $\hat{s} \equiv s_{ab} \to s_{15}$ in the MRK limit.

Similarly, the matrix element for the second configuration scales as
\begin{equation}
\mathcal{M}_2 \sim e^{\Delta y_{21}} e^{\Delta y_{32}/2} e^{\Delta y_{43}/2} e^{\Delta y_{54}} = e^{\Delta y_{51}} e^{-\Delta y_{42}/2} \sim \frac{\hat{s}}{\sqrt{s_{24}}}.
\label{eq:regge_cenqhqx_exp}
\end{equation}
Using the MRK scaling of the invariants, the denominator can be further rewritten as
\begin{equation}
\sqrt{s_{24}} \sim \sqrt{s_{23}} \sqrt{s_{34}} \sim \sqrt{s_{23}}\, e^{\Delta y_{43}/2},
\end{equation}
so that
\begin{equation}
\mathcal{M}_2 \sim \frac{\hat{s}}{\sqrt{s_{23}}}\, e^{-\Delta y_{43}/2}.
\label{eq:suppression_factor}
\end{equation}

Compared with equation~\eqref{eq:regge_cenqqx_exp}, the second configuration therefore carries an additional exponential suppression in the rapidity interval separating the Higgs boson from the neighbouring quark.
As this rapidity interval increases, the contribution of $\mathcal{M}_2$ becomes progressively smaller relative to the leading configuration, explaining why these events contribute only at subleading logarithmic accuracy.

\begin{figure}[H]
    \centering

    \begin{subfigure}[c]{0.49\textwidth}
        \centering
        \includegraphics[width=\textwidth]{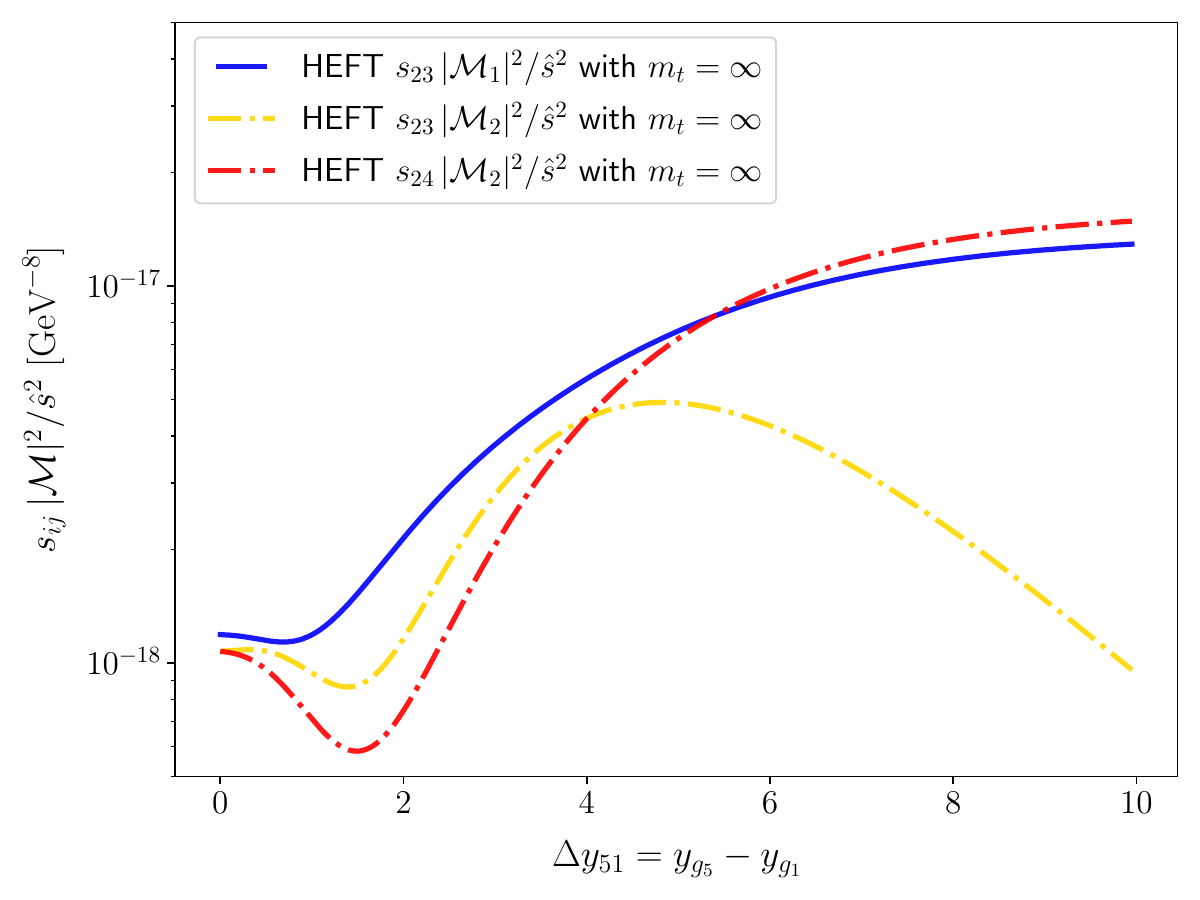}
        \caption{MRK scaling for $ gg \to g u H \bar{u} g$.}
    \label{fig:sub_qhqx_a}
    \end{subfigure}
    \hfill
    \begin{subfigure}[c]{0.49\textwidth}
        \centering
        \includegraphics[width=\textwidth]{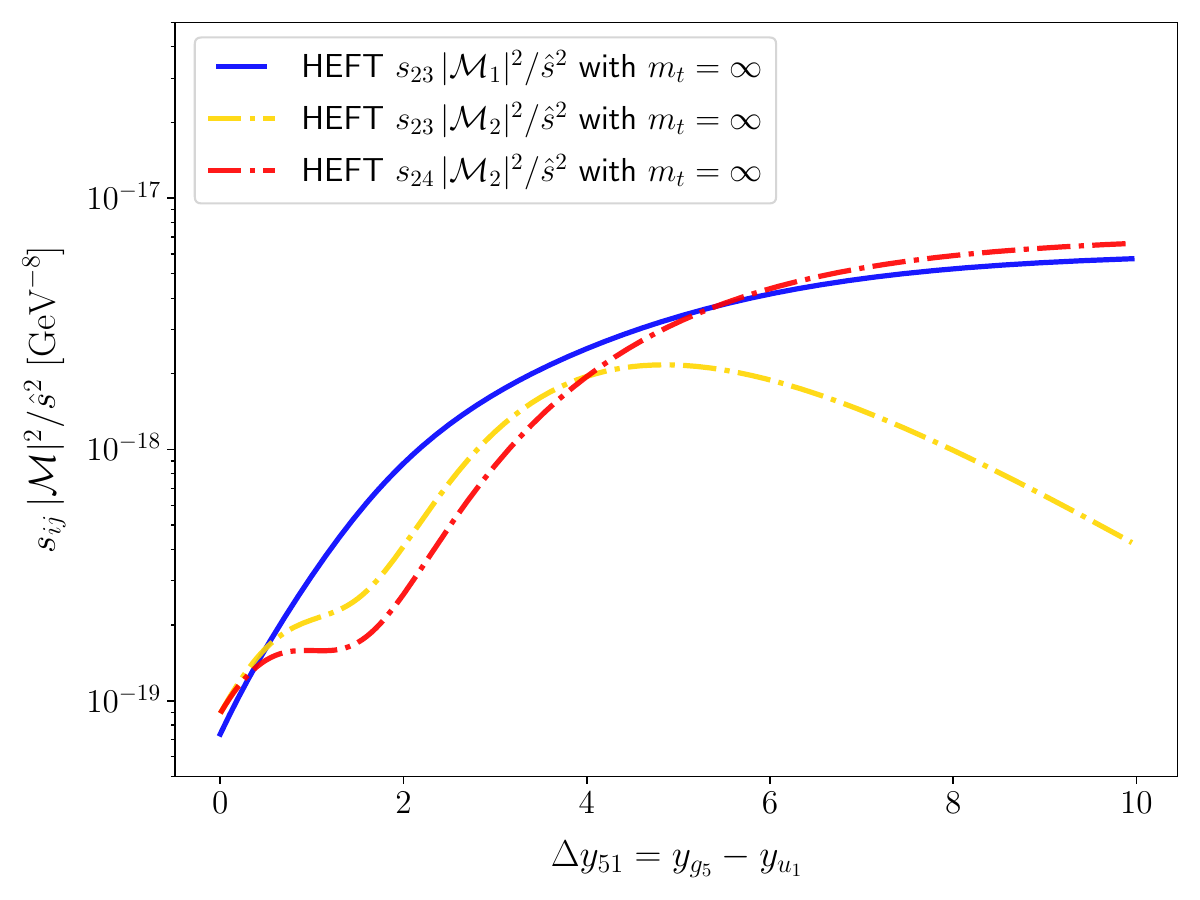}
        \caption{MRK scaling for $ ug \to u s H \bar{s} g$.}
    \label{fig:sub_qhqx_c}
    \end{subfigure}

    \vspace{1em}

    \begin{subfigure}[c]{0.49\textwidth}
        \centering
        \includegraphics[width=\textwidth]{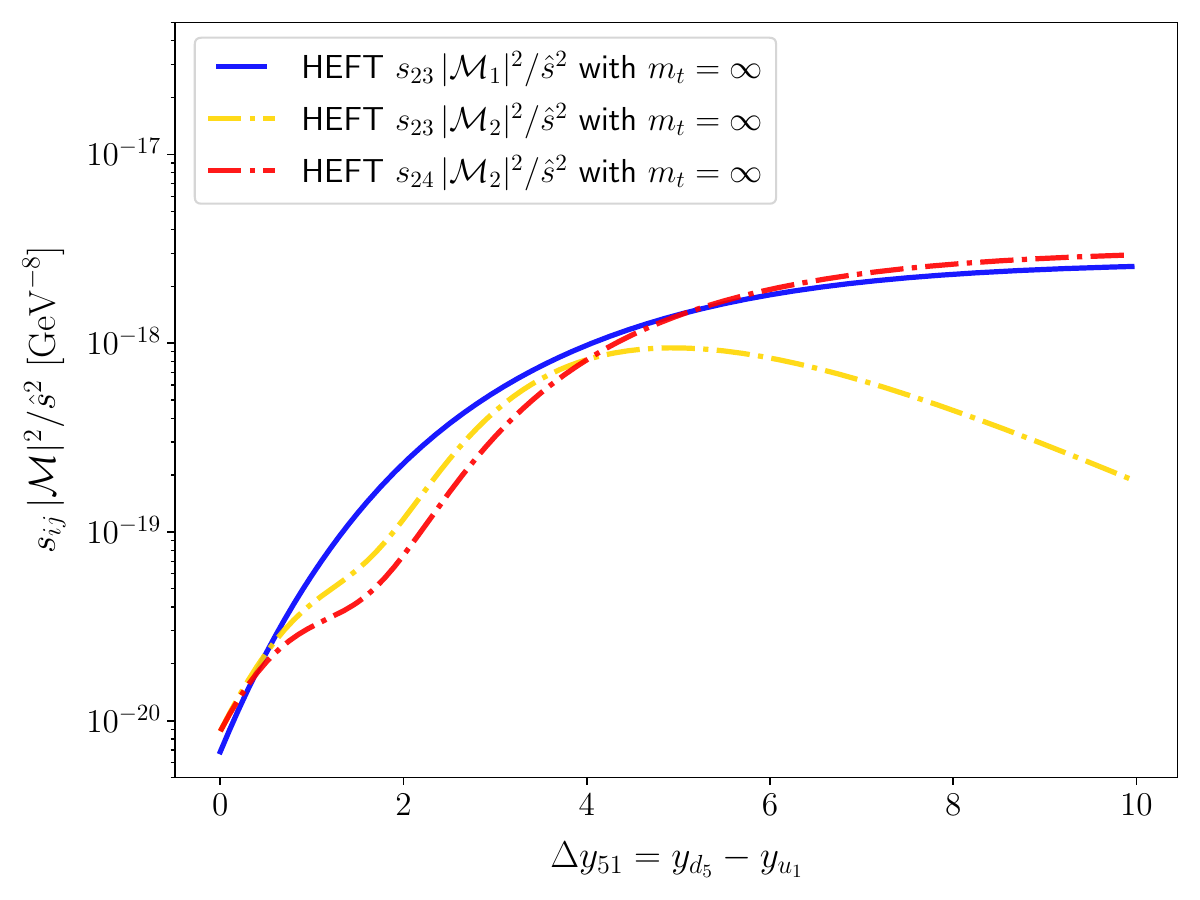}
        \caption{MRK scaling for for $ ud \to u s H \bar{s} d$.}
    \label{fig:sub_qhqx_e}
    \end{subfigure}

\caption{
HEFT matrix element squared as a function of the rapidity separation between the most forward and the most backward partons at leading order in the infinite top-quark mass limit for the three processes $gg \to g u H \bar{u} g$, $ug \to u s H \bar{s} g$, and $ud \to u s H \bar{s} d$.
The solid blue curves correspond to the scaling of $\vert \mathcal{M}_{1} \vert^2$, describing configurations in which the Higgs boson is emitted centrally and adjacent to the central quark-pair emission.
The dot-dashed yellow lines and dot-dashed red curves instead show the scaling of $\vert \mathcal{M}_{2} \vert^2$, corresponding to configurations where the Higgs boson is emitted centrally and between the quark-pair emission.
The quantities used to parametrise the momenta of the outgoing particles are reported in table~\ref{tab:explorer_cenqqx} for $\mathcal{M}_1$ and in table~\ref{tab:explorer_cenqhqx} for $\mathcal{M}_2$.
}
\label{fig:sub_qhqx}
\end{figure}

The asymptotic behaviour of the two configurations for increasing rapidity separation is shown in the panels of figure~\ref{fig:sub_qhqx} for three different processes.
The solid blue curves correspond to the Regge scaling of the first configuration, in which the Higgs boson is emitted centrally and adjacent to the quark pair, while the dot-dashed red curves display the Regge scaling of the second configuration, in which the Higgs boson is emitted centrally between the two quarks.
The dot-dashed yellow curves illustrate the exponential suppression of $|\mathcal{M}_{2}|^2$ relative to $|\mathcal{M}_{1}|^2$.

The suppression therefore depends only on the rapidity hierarchy of the final-state particles and becomes increasingly pronounced as the MRK limit is approached.
Consequently, configurations in which the Higgs boson is emitted between the quark pair are logarithmically subleading and contribute at NNLL accuracy, whereas configurations with the Higgs boson adjacent to the quark pair contribute already at NLL accuracy.

\begin{table}[!htbp]
\begin{center}
\begin{tabular}{ |c||l| }
\hline
Process & Momenta configuration \\
\hline
$gg \to guH\bar{u}Hg$ &
\(
\begin{cases}
      y_g = -\Delta, y_u=-\frac{\Delta}{2}, y_{\bar{u}} = 0, y_{H}=\frac{\Delta}{2} \text{ and } y_g = \Delta &\\
      \phi_u = \frac{6}{11}\pi, \phi_{\bar{u}} = \frac{3}{4}\pi, \phi_{H} = \frac{8}{5}\pi \text{ and } \phi_g = \frac{11}{9}\phi & \\
     p_{u\perp} = 40 \, \text{GeV}, p_{\bar{u}\perp} = 40 \, \text{GeV},p_{H\perp} = 40 \, \text{GeV} \text{ and } p_g = 40 \, \text{GeV} &
    \end{cases}
\)  \\
$ug \to usH\bar{s}Hg$ &
\(
\begin{cases}
      y_u = -\Delta, y_s=-\frac{\Delta}{2}, y_{\bar{s}} = 0, y_{H}=\frac{\Delta}{2} \text{ and } y_g = \Delta &\\
      \phi_s = \frac{6}{11}\pi, \phi_{\bar{s}} = \frac{3}{4}\pi, \phi_{H} = \frac{8}{5}\pi \text{ and } \phi_g = \frac{11}{9}\phi & \\
     p_{s\perp} = 40 \, \text{GeV}, p_{\bar{s}\perp} = 40 \, \text{GeV},p_{H\perp} = 40 \, \text{GeV} \text{ and } p_g = 40 \, \text{GeV} &
    \end{cases}
\)  \\
$ud \to usH\bar{s}Hd$ &
\(
\begin{cases}
      y_u = -\Delta, y_s=-\frac{\Delta}{2}, y_{\bar{s}} = 0, y_{H}=\frac{\Delta}{2} \text{ and } y_d = \Delta &\\
      \phi_s = \frac{6}{11}\pi, \phi_{\bar{s}} = \frac{3}{4}\pi, \phi_{H} = \frac{8}{5}\pi \text{ and } \phi_d = \frac{11}{9}\phi & \\
     p_{s\perp} = 40 \, \text{GeV}, p_{\bar{s}\perp} = 40 \, \text{GeV},p_{H\perp} = 40 \, \text{GeV} \text{ and } p_d = 40 \, \text{GeV} &
    \end{cases}
\)  \\
\hline
\end{tabular}
\end{center}
\caption{The momenta configuration used for the solid blue curves in figure~\ref{fig:sub_qhqx}.}
\label{tab:explorer_cenqqx}
\end{table}

\begin{table}[!htbp]
\begin{center}
\begin{tabular}{ |c||l| }
\hline
Process & Momenta configuration \\
\hline
$gg \to guH\bar{u}Hg$ &
\(
\begin{cases}
      y_g = -\Delta, y_u=-\frac{\Delta}{2}, y_{H} = 0, y_{\bar{u}}=\frac{\Delta}{2} \text{ and } y_g = \Delta &\\
      \phi_u = \frac{6}{11}\pi, \phi_{H} = \frac{3}{4}\pi, \phi_{\bar{u}} = \frac{8}{5}\pi \text{ and } \phi_g = \frac{11}{9}\phi & \\
     p_{u\perp} = 40 \, \text{GeV}, p_{\bar{u}\perp} = 40 \, \text{GeV},p_{H\perp} = 40 \, \text{GeV} \text{ and } p_g = 40 \, \text{GeV} &
    \end{cases}
\)  \\
$ug \to usH\bar{s}Hg$ &
\(
\begin{cases}
      y_u = -\Delta, y_s=-\frac{\Delta}{2}, y_{H} = 0, y_{\bar{s}}=\frac{\Delta}{2} \text{ and } y_g = \Delta &\\
      \phi_s = \frac{6}{11}\pi, \phi_{H} = \frac{3}{4}\pi, \phi_{\bar{s}} = \frac{8}{5}\pi \text{ and } \phi_g = \frac{11}{9}\phi & \\
     p_{s\perp} = 40 \, \text{GeV}, p_{\bar{s}\perp} = 40 \, \text{GeV},p_{H\perp} = 40 \, \text{GeV} \text{ and } p_g = 40 \, \text{GeV} &
    \end{cases}
\)  \\
$ud \to usH\bar{s}Hd$ &
\(
\begin{cases}
      y_u = -\Delta, y_s=-\frac{\Delta}{2}, y_{H} = 0, y_{\bar{s}}=\frac{\Delta}{2} \text{ and } y_d = \Delta &\\
      \phi_s = \frac{6}{11}\pi, \phi_{H} = \frac{3}{4}\pi, \phi_{\bar{s}} = \frac{8}{5}\pi \text{ and } \phi_d = \frac{11}{9}\phi & \\
     p_{s\perp} = 40 \, \text{GeV}, p_{\bar{s}\perp} = 40 \, \text{GeV},p_{H\perp} = 40 \, \text{GeV} \text{ and } p_d = 40 \, \text{GeV} &
    \end{cases}
\)  \\
\hline
\end{tabular}
\end{center}
\caption{The momenta configuration used for the dot-dashed curves in figure~\ref{fig:sub_qhqx}.}
\label{tab:explorer_cenqhqx}
\end{table}

\section{Numerical Impact of Quark Pair Component}\label{app:qqbar}

\begin{figure}[htbp]
    \centering

    \begin{subfigure}[b]{0.49\textwidth}
        \centering
        \includegraphics[width=\textwidth]{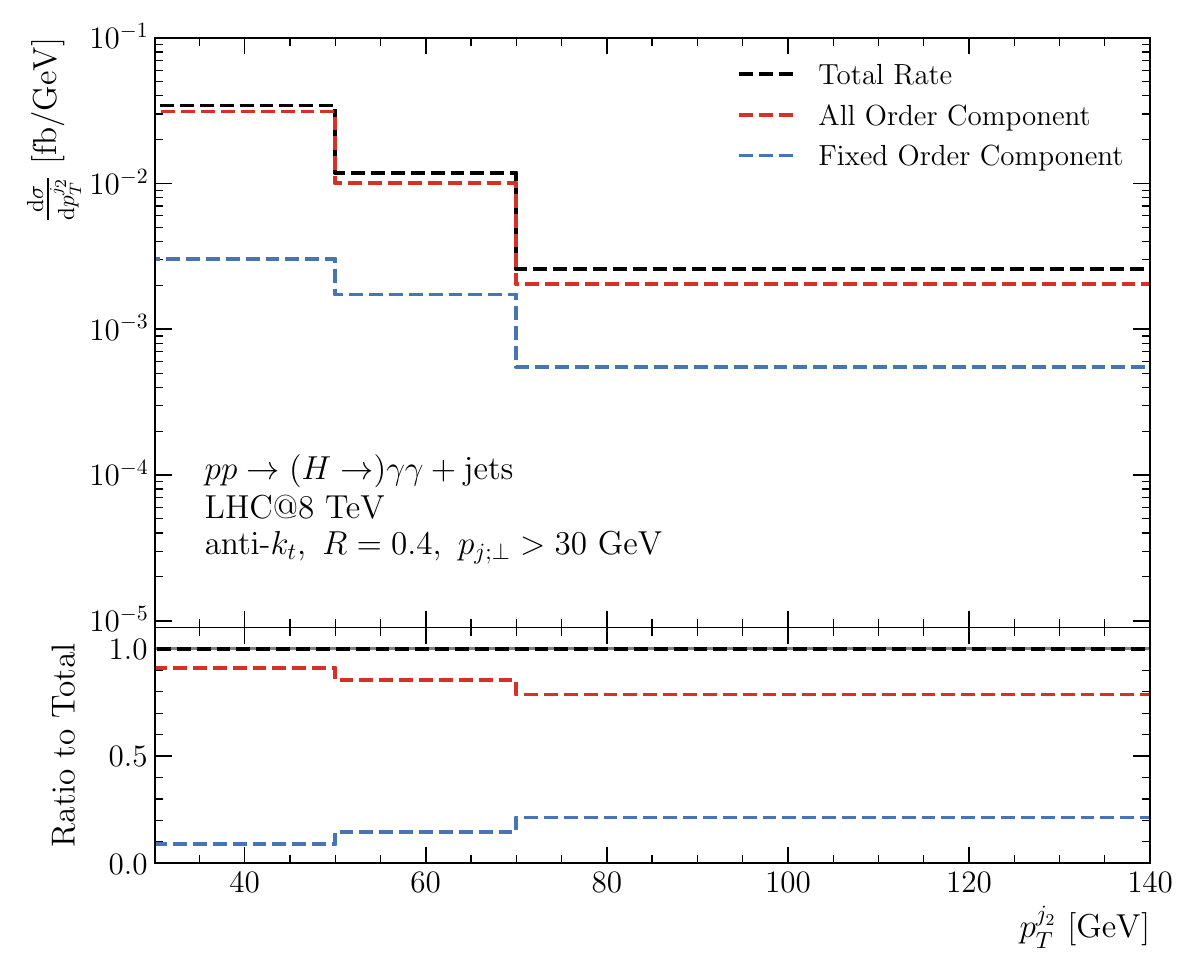}
        \caption{Case 1: LL + NLL (unordered).}
        \label{fig:components_pTj2_a}
    \end{subfigure}
    \hfill
    \begin{subfigure}[b]{0.49\textwidth}
        \centering
        \includegraphics[width=\textwidth]{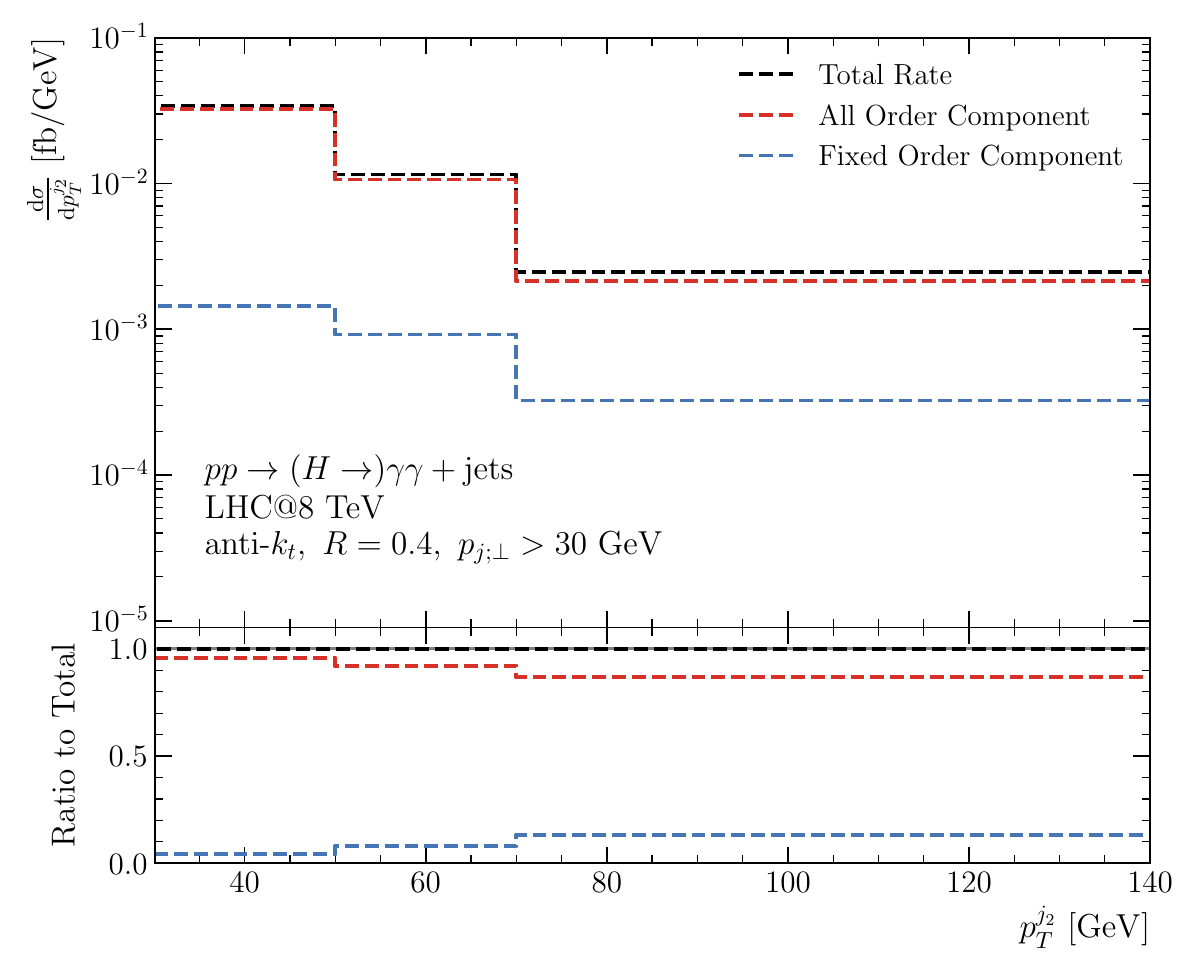}
        \caption{Case 2: LL + NLL (unordered + $q\bar{q}$).}
        \label{fig:components_pTj2_b}
    \end{subfigure}

    \vspace{1em}

    \begin{subfigure}[b]{0.48\textwidth}
        \centering
        \includegraphics[width=\textwidth]{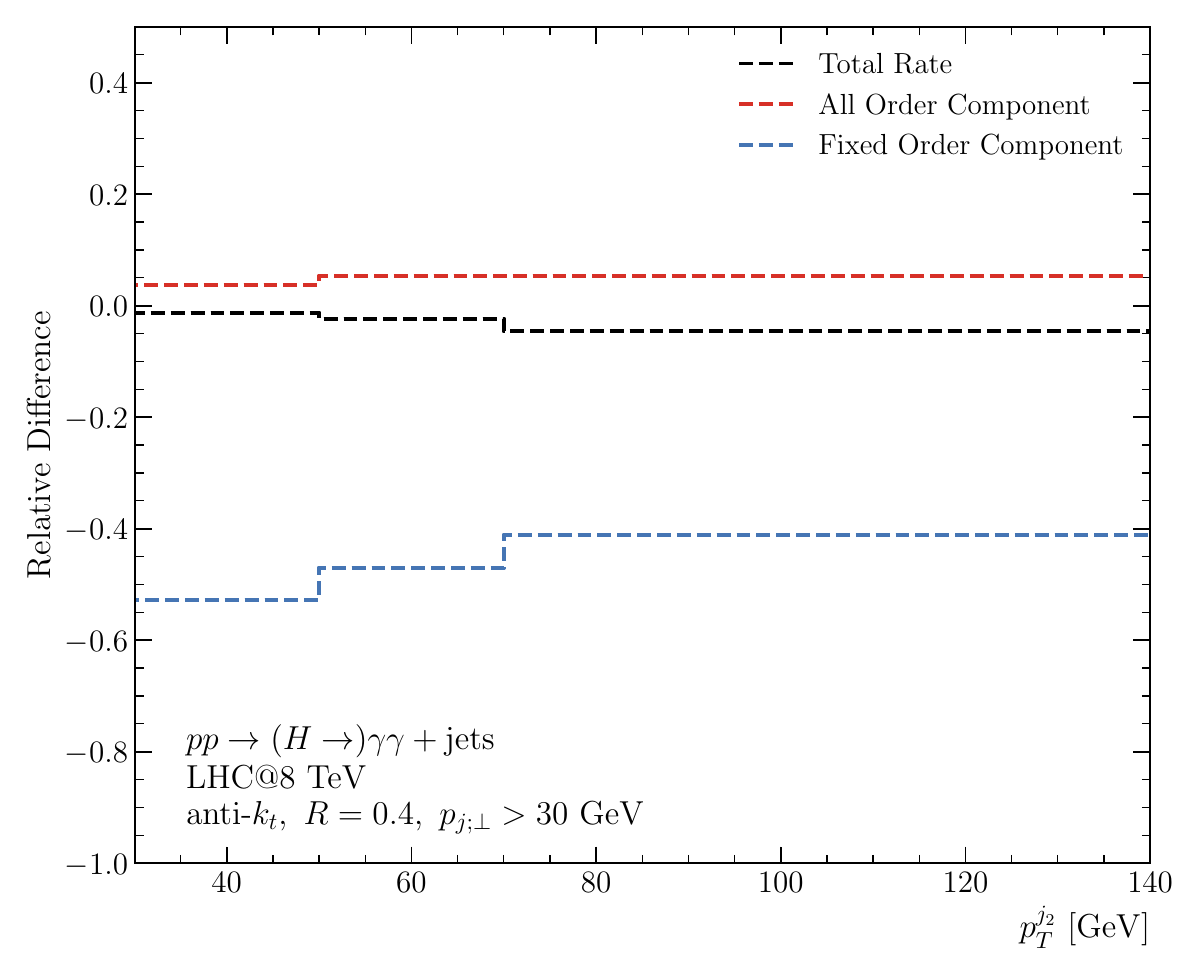}
        \caption{Relative difference.}
        \label{fig:components_pTj2_c}
    \end{subfigure}

\caption{
The differential distribution (black, dashed) in the transverse momentum of the subleading jet, $p_T^{j_2}$, in $pp \to (H \to \gamma \gamma) + \geq 2j$.
Panel~\ref{fig:components_pTj2_a} shows when the resummation is applied only to all LL states and a class of NLL states, i.e. the unordered emissions.
The case where the resummation is applied to all LL and NLL states, thus including quark-pair emissions as well, is shown in panel~\ref{fig:components_pTj2_b}.
For each case is also shown the split into the components where all-order resummation is applied (red, dashed) and the component which remains described at fixed-order only (blue, dashed).
The relative difference in each line is displayed in panel~\ref{fig:components_pTj2_c}, as explained in the text.
}
\label{fig:components_pTj2}
\end{figure}

\begin{figure}[htbp]
    \centering

    \begin{subfigure}[b]{0.49\textwidth}
        \centering
        \includegraphics[width=\textwidth]{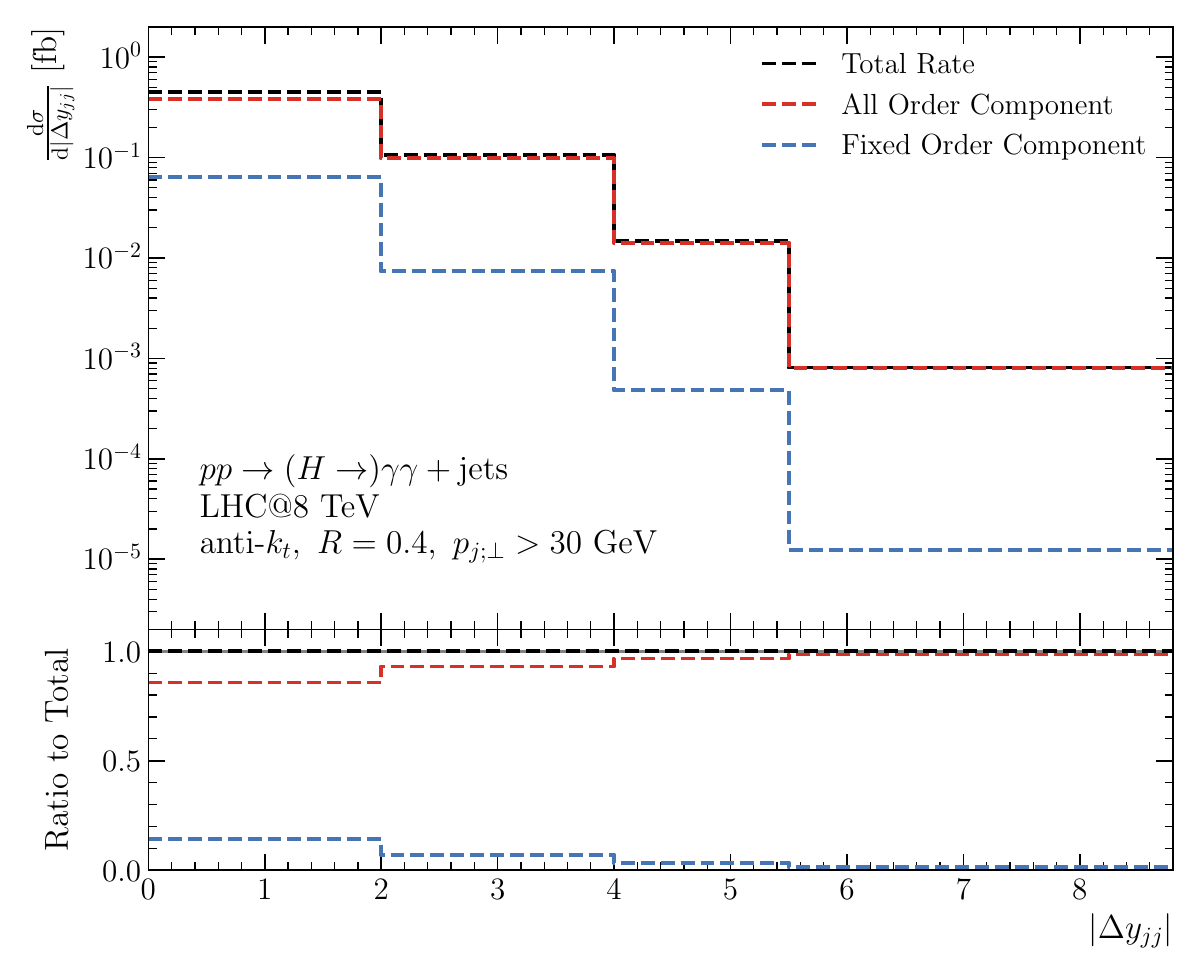}
        \caption{Case 1: LL + NLL (unordered).}
        \label{fig:components_dyjj_a}
    \end{subfigure}
    \hfill
    \begin{subfigure}[b]{0.49\textwidth}
        \centering
        \includegraphics[width=\textwidth]{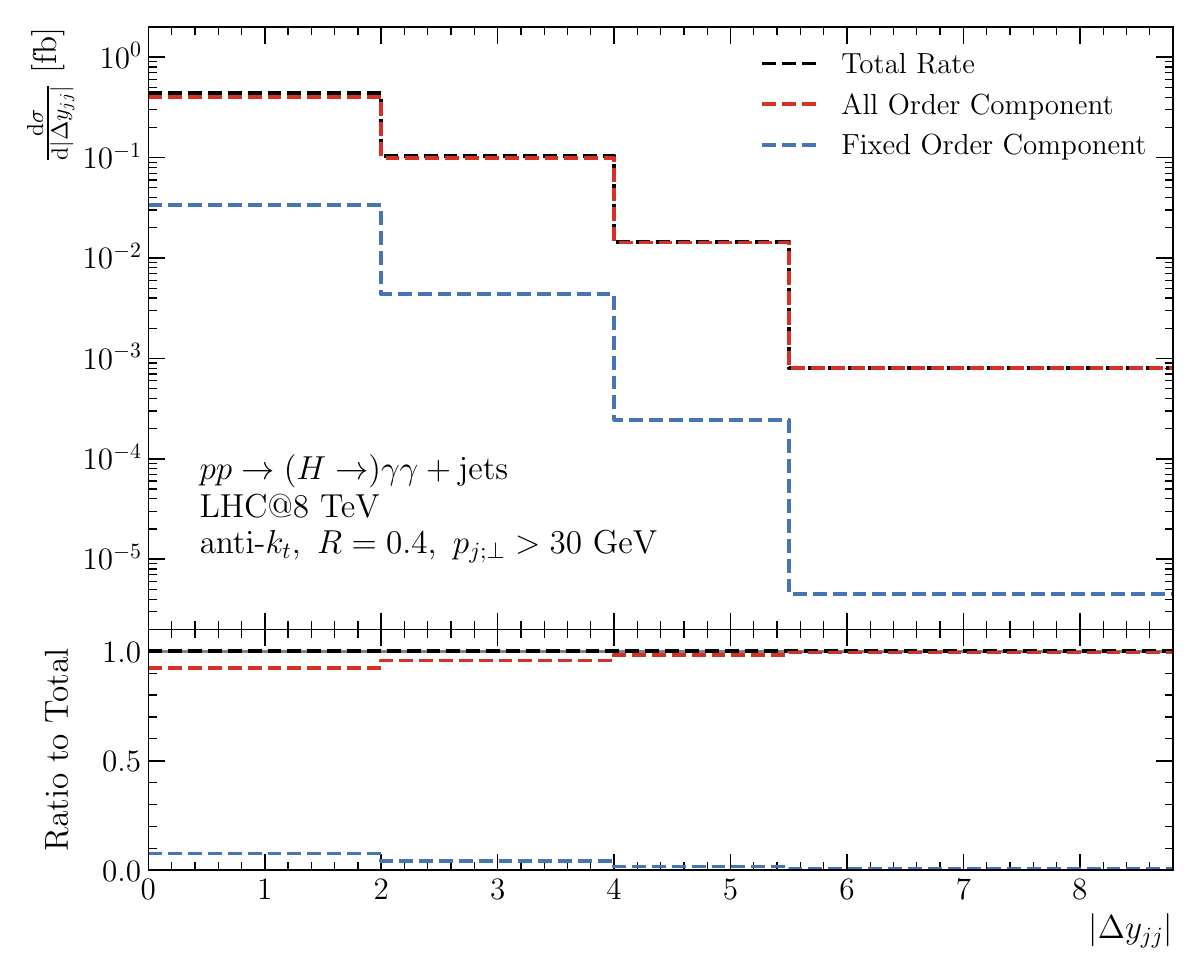}
        \caption{Case 2: LL + NLL (unordered + $q\bar{q}$).}
        \label{fig:components_dyjj_b}
    \end{subfigure}

    \vspace{1em}

    \begin{subfigure}[b]{0.48\textwidth}
        \centering
        \includegraphics[width=\textwidth]{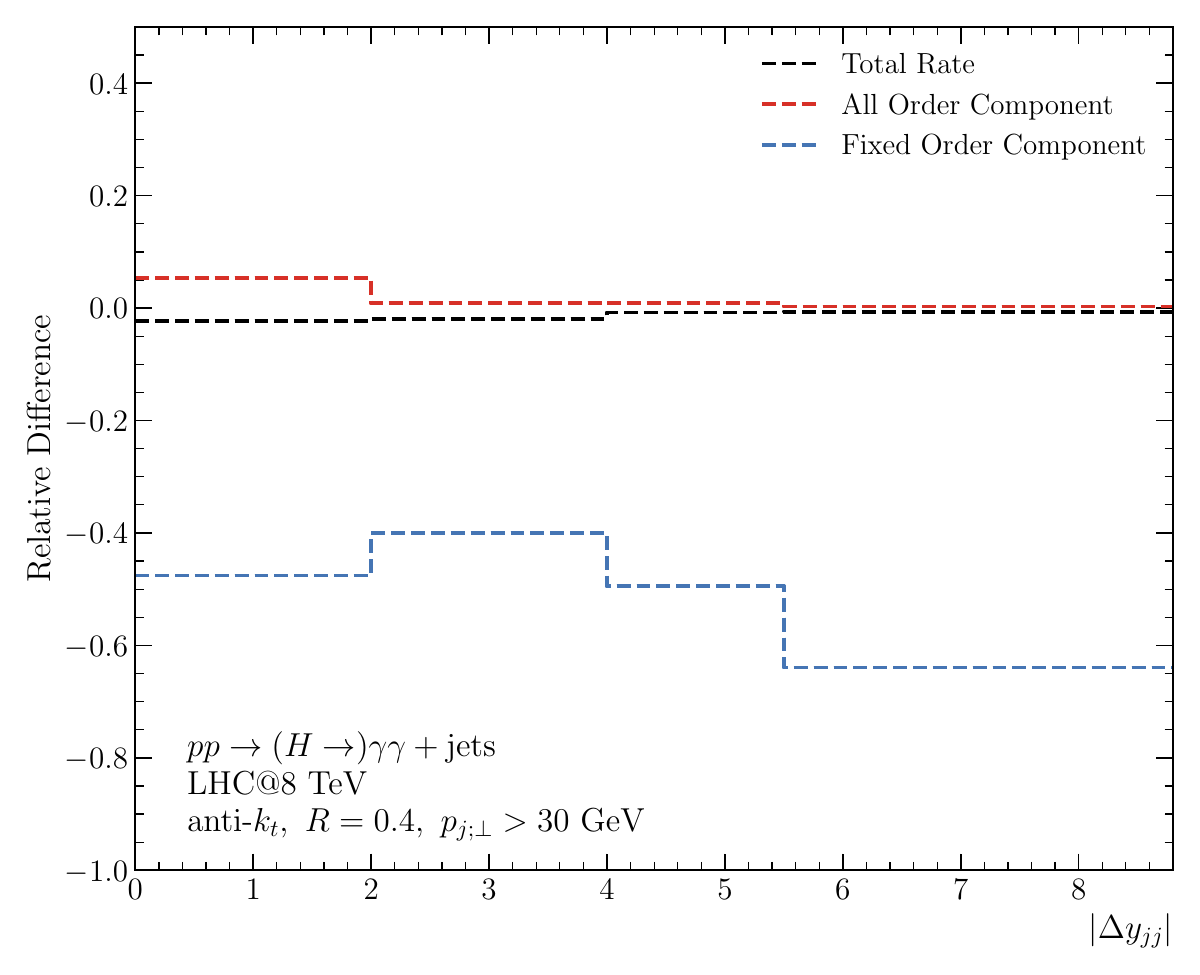}
        \caption{Relative difference.}
        \label{fig:components_dyjj_c}
    \end{subfigure}

\caption{
The differential distribution (black, dashed) in the dijet rapidity separation, $\vert \Delta y_{jj} \vert$, in $pp \to (H \to \gamma \gamma) + \geq 2j$, with and without resummation applied to quark-pair configurations.
The panels and lines are as in figure~\ref{fig:components_pTj2}.
}
\label{fig:components_dyjj}
\end{figure}

\begin{figure}[htbp]
    \centering

    \begin{subfigure}[b]{0.49\textwidth}
        \centering
        \includegraphics[width=\textwidth]{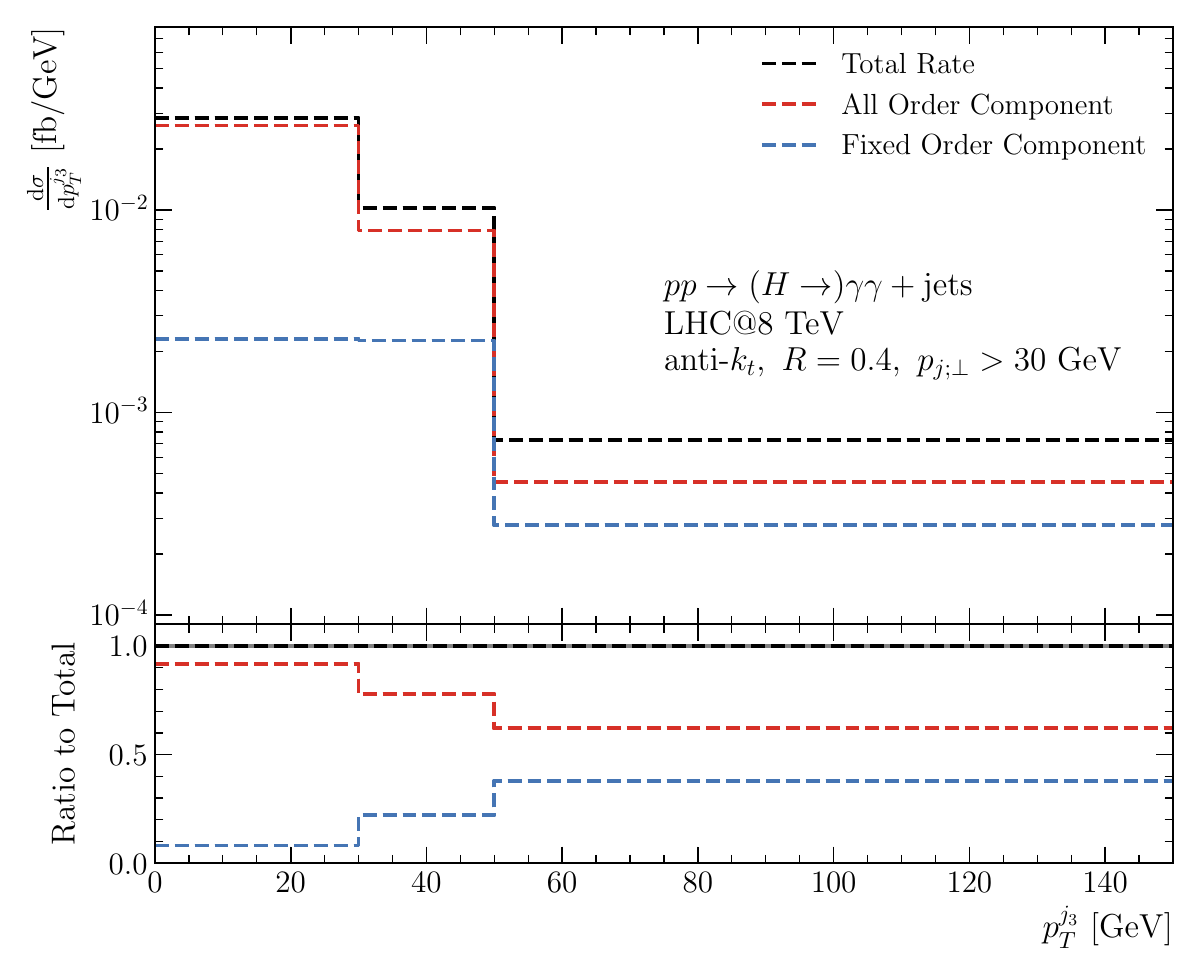}
        \caption{Case 1: LL + NLL (unordered).}
        \label{fig:components_pTj3_a}
    \end{subfigure}
    \hfill
    \begin{subfigure}[b]{0.49\textwidth}
        \centering
        \includegraphics[width=\textwidth]{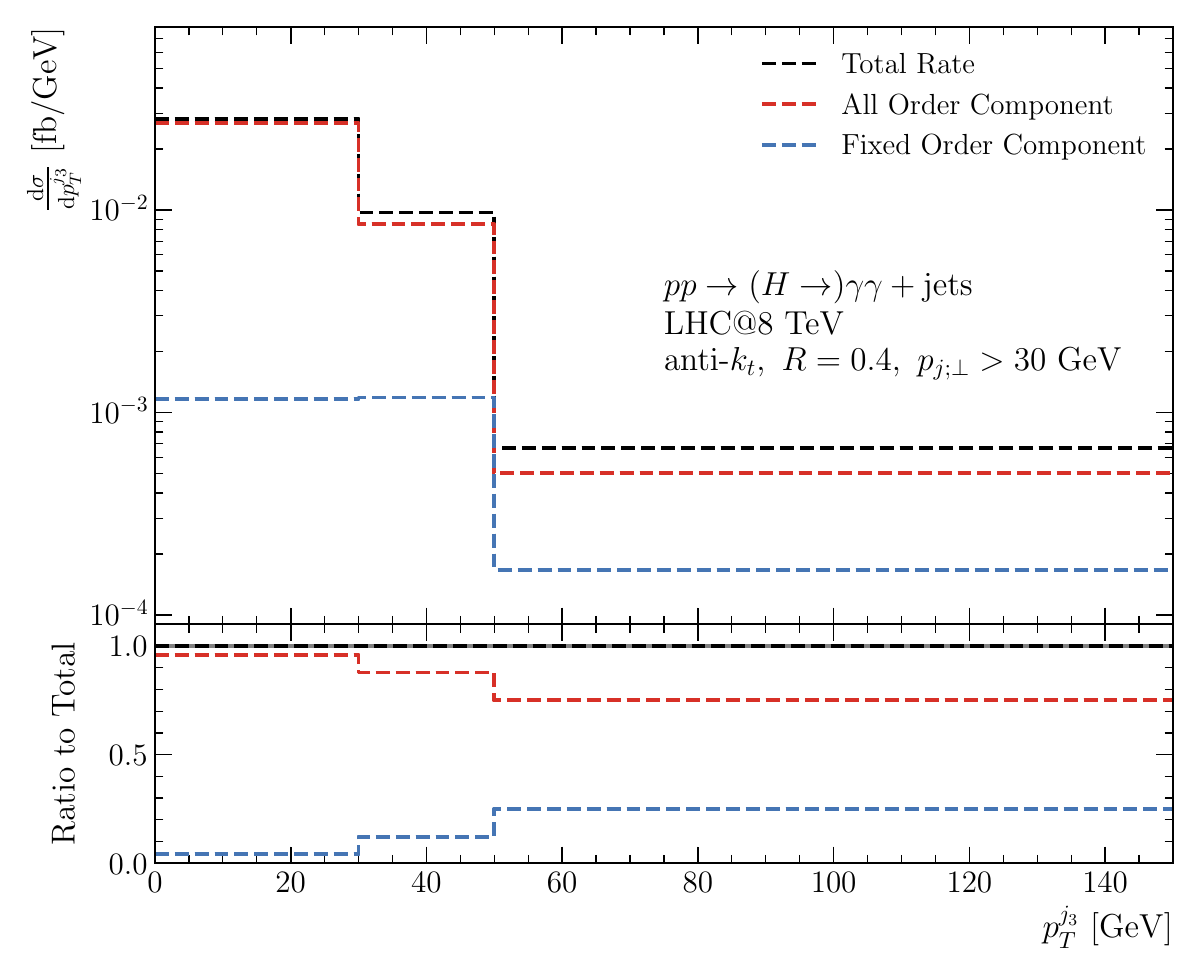}
        \caption{Case 2: LL + NLL (unordered + $q\bar{q}$).}
        \label{fig:components_pTj3_b}
    \end{subfigure}

    \vspace{1em}

    \begin{subfigure}[b]{0.48\textwidth}
        \centering
        \includegraphics[width=\textwidth]{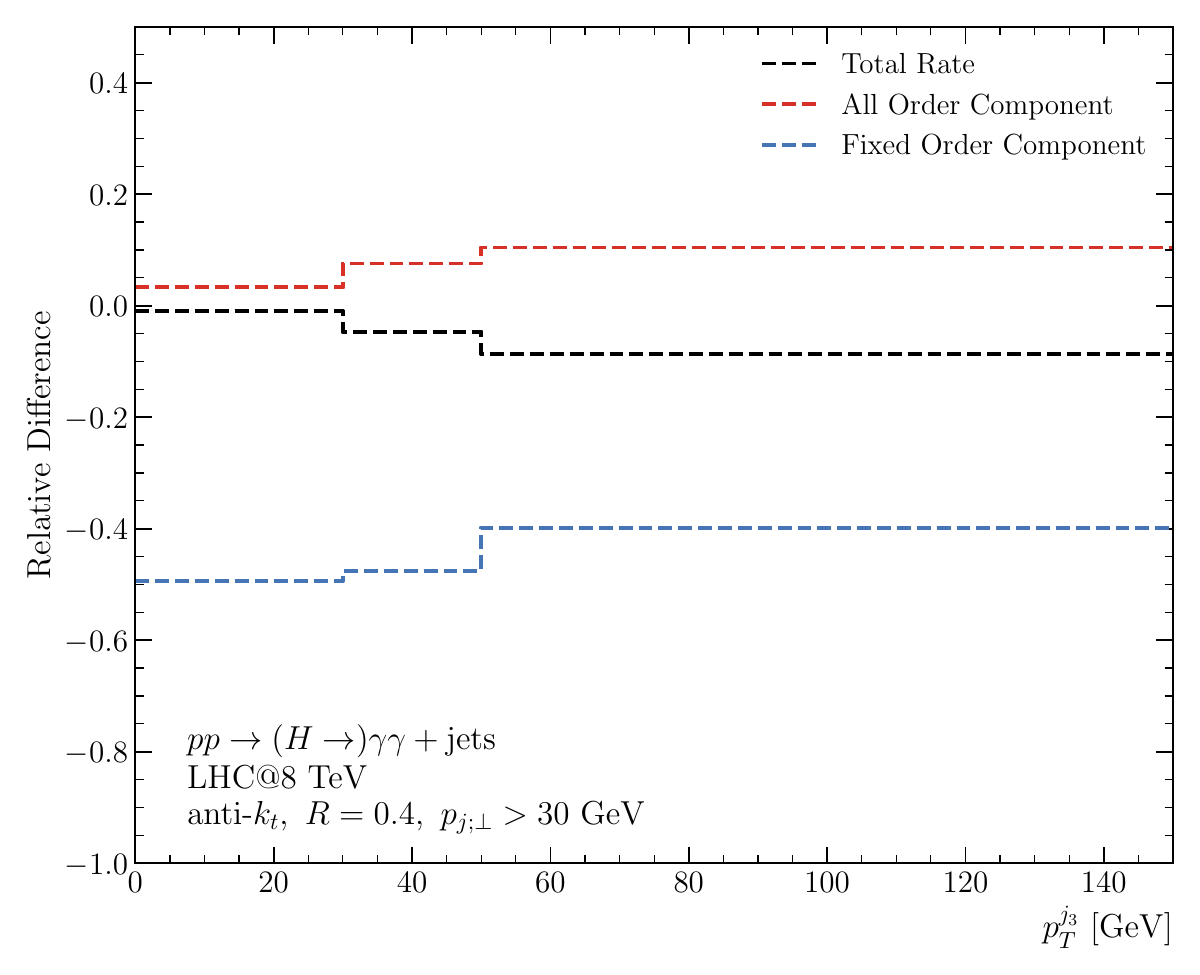}
        \caption{Relative difference.}
        \label{fig:components_pTj3_c}
    \end{subfigure}

\caption{
The differential distribution (black, dashed) in the third-leading jet transverse momentum, $p_{T}^{j_3}$, in $pp \to (H \to \gamma \gamma) + \geq 2j$, with and without resummation applied to quark-pair configurations.
The panels and lines are as in figure~\ref{fig:components_pTj2}.
}
\label{fig:components_pTj3}
\end{figure}

In section~\ref{sec:HEJFORHIGGS}, we discussed the main aspects of matrix elements and classification of events within \HEJ.
The individual contributions were separated into leading-log (LL), next-to-leading-log (NLL) and other (i.e.~further suppressed in the MRK limit) configurations.
Here, we want to illustrate the numerical impact of the new quark-antiquark component in figures~\ref{fig:components_pTj2}-\ref{fig:components_dyjj}-\ref{fig:components_pTj3}.
The results presented in this section are obtained using the same analysis setup described in section~\ref{sec:8TeV}, which follows the ATLAS measurement~\cite{ATLAS:2014yga}.
We plot the total differential cross section for the inclusive Higgs boson plus dijets process and show its all-order and fixed-order subcontributions as follows:

\begin{itemize}
\item Case~1: The LO result plus all LL corrections ($\alpha_{s}^{2+k} \, \log^k (\hat{s}/p_T^2)$) and the class of NLL corrections ($\alpha_{s}^{3+k} \, \log^k (\hat{s}/p_T^2)$) arising from \textit{unordered} gluon and Higgs emissions.
This is plotted in panel (a) of figures~\ref{fig:components_pTj2}-\ref{fig:components_dyjj}-\ref{fig:components_pTj3}.
The impact of the resummation is shown by the red dashed line marked ``All Order Component''.
The remaining configurations are described at fixed order only and enter the ``Fixed Order Component''.
\item Case~2: The LO result plus the LL and NLL corrections of case 1 are included plus the new NLL contributions from the \textit{quark-antiquark} configurations.
This is plotted in panel (b) of figures~\ref{fig:components_pTj2}-\ref{fig:components_dyjj}-\ref{fig:components_pTj3}.
Similarly to Case~1, the resummation contribution is given by the red dashed line.
On the other hand, all other subprocesses are described at fixed order only and enter the blue dashed line.
\item Relative difference: the difference between Case~2 and~1 divided by the results of Case~1.
This is plotted in panel (c) of figures~\ref{fig:components_pTj2}-\ref{fig:components_dyjj}-\ref{fig:components_pTj3}.
\end{itemize}
The first observable used to assess the numerical impact of the quark-pair resummation is the transverse momentum of the second-hardest jet, shown in figure~\ref{fig:components_pTj2}.
Once the resummation is extended to quark-pair configurations, the contribution labelled ``Fixed Order Component'' is reduced by approximately a factor of two across the entire spectrum, as illustrated by the dashed blue curve in panel~\ref{fig:components_pTj2_c}.
This indicates that roughly half of the configurations that were not resummed in Case~1 are now incorporated into the all-order description.
Correspondingly, the ``All Order Component'' increases by about $5\%$ throughout the distribution, as shown by the dashed red curve.
As a result, the resummed contribution accounts for nearly the entire cross section in the first bin and still covers close to $90\%$ of the total rate in the last bin.
Although no direct logarithmic enhancement is associated with the transverse momentum itself, the inclusion of quark-pair resummation leads to a visible redistribution between the fixed-order and all-order components.

The suppression of the new NLL contribution in the approach to the MRK limit is more clearly illustrated by the dijet rapidity separation shown in figure~\ref{fig:components_dyjj}.
For $|\Delta y_{jj}| \gtrsim 4$, corresponding to the last two bins of the distribution, the all-order component remains essentially unchanged when moving from Case~1 to Case~2.
This behaviour reflects the expected dominance of the LL configurations in the asymptotic high-energy regime.
At the same time, the fixed-order component is significantly reduced, dropping to approximately $50$-$60\%$ in the same region, as shown by the dashed blue line in panel~\ref{fig:components_dyjj_c}.
The numerical impact of the quark-pair resummation is therefore concentrated at smaller rapidity separations, where the all-order contribution receives an enhancement of roughly $5\%$ in the first bin.
More generally, this behaviour illustrates the expected hierarchy of logarithmic contributions: configurations entering at $\mathrm{N}^n\mathrm{LL}$ accuracy become progressively more relevant as one moves away from the strict MRK limit, while LL configurations dominate asymptotically.

Finally, figure~\ref{fig:components_pTj3} presents the transverse-momentum distribution of the third-hardest jet.
A pattern similar to that observed for $p_T^{j_2}$ emerges.
The fixed-order component is reduced by approximately $40$-$50\%$ with respect to Case~1, while the fraction described by the all-order resummation increases by about $5$-$10\%$, as illustrated by the dashed blue and dashed red curves in~\ref{fig:components_pTj3_c}, respectively.
In general, for increasing jet multiplicity, there is a richer event topology associated to many more possible final states which are either LL or $\mathrm{N}^n\mathrm{LL}$.
Therefore, the fraction of the cross section to which all-order corrections are applied is less than in figure~\ref{fig:components_pTj2}.

The relative small change in the total rate (black dashed line) across all distributions is a strong indication of the perturbative stability of the \HEJ\ framework.
In summary, the inclusion of quark-pair resummation has a moderate impact on the overall size of the all-order contribution.
Its most significant effect is instead the substantial reduction of configurations that could only be described at fixed order, thereby extending the reach of the resummed calculation to a considerably larger fraction of the total cross section.

\clearpage

\bibliographystyle{JHEP}
\bibliography{bibliography.bib}

\end{document}